\documentclass[prd,nofootinbib,twocolumn,preprintnumbers,balancelastpage,superscriptaddress,longbibliography]{revtex4-1}
\usepackage{tabularx}
\usepackage{array}
\newcolumntype{P}[1]{>{\centering\arraybackslash}p{#1}}
\usepackage{color}
\usepackage{hepunits}
\usepackage{amsmath}
\usepackage{booktabs}
\usepackage{array}
\usepackage{makecell}
\usepackage[framemethod=TikZ]{mdframed}
\usepackage[colorlinks=true
,urlcolor=blue
,anchorcolor=blue
,citecolor=blue
,filecolor=blue
,linkcolor=red
,menucolor=blue
,linktocpage=true
,pdfproducer=medialab
,pdfa=true
]{hyperref}
\usepackage[utf8]{inputenc}
\usepackage[english]{babel}

\usepackage{pifont}
\mathchardef\mhyphen="2D 

\newcolumntype{C}[1]{>{\centering\let\newline\\\arraybackslash\hspace{0pt}}m{#1}}

\newcommand{\es}[2] {\begin{equation} \label{#1} \begin{split} #2 \end{split} \end{equation}}

\newcommand{\ra}[1]{\renewcommand{\arraystretch}{#1}}
\newcommand{\be}{\begin{equation}}
\newcommand{\ee}{\end{equation}}

\begin{document}
\title{A deep search for decaying dark matter with {\it XMM-Newton} blank-sky observations}

\author{Joshua W. Foster}
\email{fosterjw@umich.edu}
\affiliation{Leinweber Center for Theoretical Physics, Department of Physics, University of Michigan, Ann Arbor, MI 48109, USA} 
\affiliation{Berkeley Center for Theoretical Physics, University of California, Berkeley, CA 94720, USA}
\affiliation{Theoretical Physics Group, Lawrence Berkeley National Laboratory, Berkeley, CA 94720, USA}
\author{Marius Kongsore}
\affiliation{Leinweber Center for Theoretical Physics, Department of Physics, University of Michigan, Ann Arbor, MI 48109, USA}
\author{Christopher Dessert}
\affiliation{Leinweber Center for Theoretical Physics, Department of Physics, University of Michigan, Ann Arbor, MI 48109, USA}
\affiliation{Berkeley Center for Theoretical Physics, University of California, Berkeley, CA 94720, USA}
\affiliation{Theoretical Physics Group, Lawrence Berkeley National Laboratory, Berkeley, CA 94720, USA}
\author{Yujin Park}
\affiliation{Leinweber Center for Theoretical Physics, Department of Physics, University of Michigan, Ann Arbor, MI 48109, USA}
\affiliation{Berkeley Center for Theoretical Physics, University of California, Berkeley, CA 94720, USA}
\affiliation{Theoretical Physics Group, Lawrence Berkeley National Laboratory, Berkeley, CA 94720, USA}
\author{Nicholas L. Rodd}
\affiliation{Berkeley Center for Theoretical Physics, University of California, Berkeley, CA 94720, USA}
\affiliation{Theoretical Physics Group, Lawrence Berkeley National Laboratory, Berkeley, CA 94720, USA}
\author{Kyle Cranmer}
\affiliation{Center  for  Cosmology  and  Particle  Physics,  New  York  University,  New  York,  NY  10003,  USA}
\author{Benjamin R. Safdi}
\email{brsafdi@lbl.gov}
\affiliation{Berkeley Center for Theoretical Physics, University of California, Berkeley, CA 94720, USA}
\affiliation{Theoretical Physics Group, Lawrence Berkeley National Laboratory, Berkeley, CA 94720, USA}

\preprint{}

\begin{abstract}
Sterile neutrinos with masses in the keV range are well-motivated  extensions to the Standard Model that could explain the observed neutrino masses while also making up the dark matter (DM) of the Universe.  If sterile neutrinos are DM then they may slowly decay into active neutrinos and photons, giving rise to the possibility of their detection through narrow spectral features in astrophysical X-ray data sets.  In this work, we perform the most sensitive search to date for this and other decaying DM scenarios across the mass range from 5 to 16 keV using archival {\it XMM-Newton} data.  We reduce 547 Ms of data from both the MOS and PN instruments using observations taken across the full sky and then use this data to search for evidence of DM decay in the ambient halo of the Milky Way.  We determine the instrumental and astrophysical baselines with data taken far away from the Galactic Center, and use Gaussian Process modeling to capture additional continuum background contributions.  No evidence is found for unassociated X-ray lines, leading us to produce the strongest constraints to date on decaying DM in this mass range.
\end{abstract}

\maketitle

Sterile neutrino dark matter (DM) is a well-motivated DM candidate that may give rise to observable nearly monochromatic X-ray signatures~\cite{Dodelson:1993je,Shi:1998km,Kusenko:2006rh}.  In this scenario the DM has a mass in the keV range and may decay into an active neutrino and an X-ray, with energy set by half the rest mass of the sterile neutrino~\cite{PhysRevD.25.766}.  Sterile neutrino DM is motivated in part by the seesaw mechanism for explaining the active neutrino masses~\cite{Yanagida:1980xy,Mohapatra:1979ia}.  In this work we present one of the most sensitive searches for sterile neutrino DM, along with other DM candidates that may decay to monochromatic X-rays, over the mass range $m_\chi \in [5, 16]$ keV.  We do so by searching for DM decay from the ambient halo of the Milky Way using all archival data from the {\it XMM-Newton} telescope collected from its launch until September 5, 2018.

This work builds heavily off the method developed in \textcite{Dessert:2018qih}, which used {\it XMM-Newton} blank-sky observations (BSOs) to strongly disfavor the decaying DM explanation of the previously-observed 3.5 keV unidentified X-ray line (UXL).
This UXL was found in nearby galaxies and clusters~\cite{Bulbul:2014sua,Boyarsky:2014jta,Urban:2014yda,Jeltema:2014qfa,Cappelluti:2017ywp}.  However the analysis performed in~\textcite{Dessert:2018qih} was able to robustly rule out the DM decay rate required to explain the previous 3.5 keV UXL signals~\cite{Dessert:2020hro}. (For additional non-observations, see Refs.~\cite{Horiuchi:2013noa,Malyshev:2014xqa,Anderson:2014tza,Tamura:2014mta,Jeltema:2015mee,Aharonian:2016gzq,Gewering-Peine:2016yoj}.)  We extend the search in~\textcite{Dessert:2018qih} to the broader mass range $m_\chi \in [5,16]$, and in doing so implement the following notable differences: (i) we use a data-driven approach to construct stacked, background-subtracted data sets in rings around the Galactic Center (GC), while Ref.~\cite{Dessert:2018qih} performed a joint-likelihood analysis at the level of individual exposures, and (ii) we use Gaussian Process (GP) modeling to describe continuum residuals, instead of parametric modeling as used in~\cite{Dessert:2018qih}. 

As demonstrated in~\textcite{Dessert:2018qih}, BSO searches for DM decaying in the Milky Way halo can be both more sensitive and more robust than extra-galactic searches, because (i) the expected DM flux, even at angles $\sim$45$^\circ$ away from the GC, rivals the expected flux from the most promising extra-galactic objects, such as M31 and the Perseus cluster; (ii) promising extra-galactic targets have continuum and line-like X-ray features that are confounding backgrounds for DM searches (dwarf galaxies being an exception~\cite{Jeltema:2015mee,Ruchayskiy:2015onc}), while BSOs instead focus on the lowest-background regions of the sky; (iii) extra-galactic targets require pointed observations, while in principle any observation collected by {\it XMM-Newton} is sensitive to DM decay in the Milky Way, opening up considerably more exposure time.

The limits presented in this work represent the strongest found using the {\it XMM-Newton} instrument over the energy range $\sim$2.5--8 keV.
At higher energies our limits are superseded with those found using the NuSTAR satellite~\cite{Neronov:2016wdd,Perez:2016tcq,Roach:2019ctw,Riemer-Sorensen:2015kqa,Ng:2019gch}.  Ref.~\cite{Roach:2019ctw} performed a search similar in spirit to that in this work (though with NuSTAR) in that they looked for DM decay from the Milky Way halo near the GC ($\sim$$10^\circ$ away in their case), while Ref.~\cite{Ng:2019gch} searched for DM decay from M31 with NuSTAR. 

Our results put in tension efforts to explain the abundance of DM with sterile neutrinos.  For example, in the Neutrino Minimal Standard Model ($\nu$MSM)~\cite{Asaka:2005an,Asaka:2005pn,Canetti:2012kh}, which may simultaneously explain the observed neutrino masses, DM density, and baryon asymmetry,   
the Standard Model is supplemented by three heavier sterile neutrino states, the lightest of which is the DM candidate.  The DM abundance is generated through the mixing of sterile and active neutrinos~\cite{Dodelson:1993je}, which can further be resonantly enhanced by a finite lepton chemical potential~\cite{Shi:1998km,Dolgov:2002ab,Serpico:2005bc,Boyarsky:2009ix,Laine:2008pg,Canetti:2012kh,Venumadhav:2015pla,Cherry:2017dwu}, though other production mechanisms are also possible~\cite{Kusenko:2006rh,Petraki:2007gq,Abazajian:2019ejt}. 
DM models such as axion-like-particle DM~\cite{Higaki:2014zua} and moduli DM~\cite{Kusenko:2012ch} predict similar UXL signatures from DM decay.

\noindent
{\bf Data reduction and processing.} We process and analyze all publicly-available data collected before 5 September 2018 by the metal oxide semiconductor (MOS) and positive-negative (PN) cameras on board {\it XMM-Newton}.
We subject each exposure to a set of quality cuts, which are described shortly. 
Those exposures satisfying the quality cuts are included in our angularly-binned data products.  In particular, we divide the sky into 30 concentric annuli centered around the GC, each with a width of $6^{\circ}$ in angular radius from the GC, $r_{\rm GC}$, where $\cos (r_{\rm GC}) = \cos (l) \cos (b)$ in terms of the Galactic longitude, $l$, and latitude, $b$. We label these from 1 to 30, starting from the innermost ring. We further mask the Galactic Plane such that we only include the region $|b| \geq 2^\circ$.  In each ring we then produce stacked spectra where, in each energy bin, we sum over the counts from each exposure whose central position lies within that annulus.  We produce separate data sets for the MOS and PN cameras, which have 2400 and 4096 energy channels, respectively.  In addition to stacking the counts in each ring and energy channel, we also construct the appropriately weighted detector response matrices in every ring for forward modeling an incident astrophysical flux. The full-sky maps and associated modeling data are provided as Supplementary Data~\cite{supp-data-bso} in both the annuli and in finer-resolution \texttt{HEALPix} binning~\cite{Gorski:2004by}.  We analyze the MOS data from 2.5 to 8 keV and the PN data from 2.5 to 7 keV, in order to exclude intervals containing large instrumental features. 

\begin{figure}[!t]
\includegraphics[width = 0.47\textwidth]{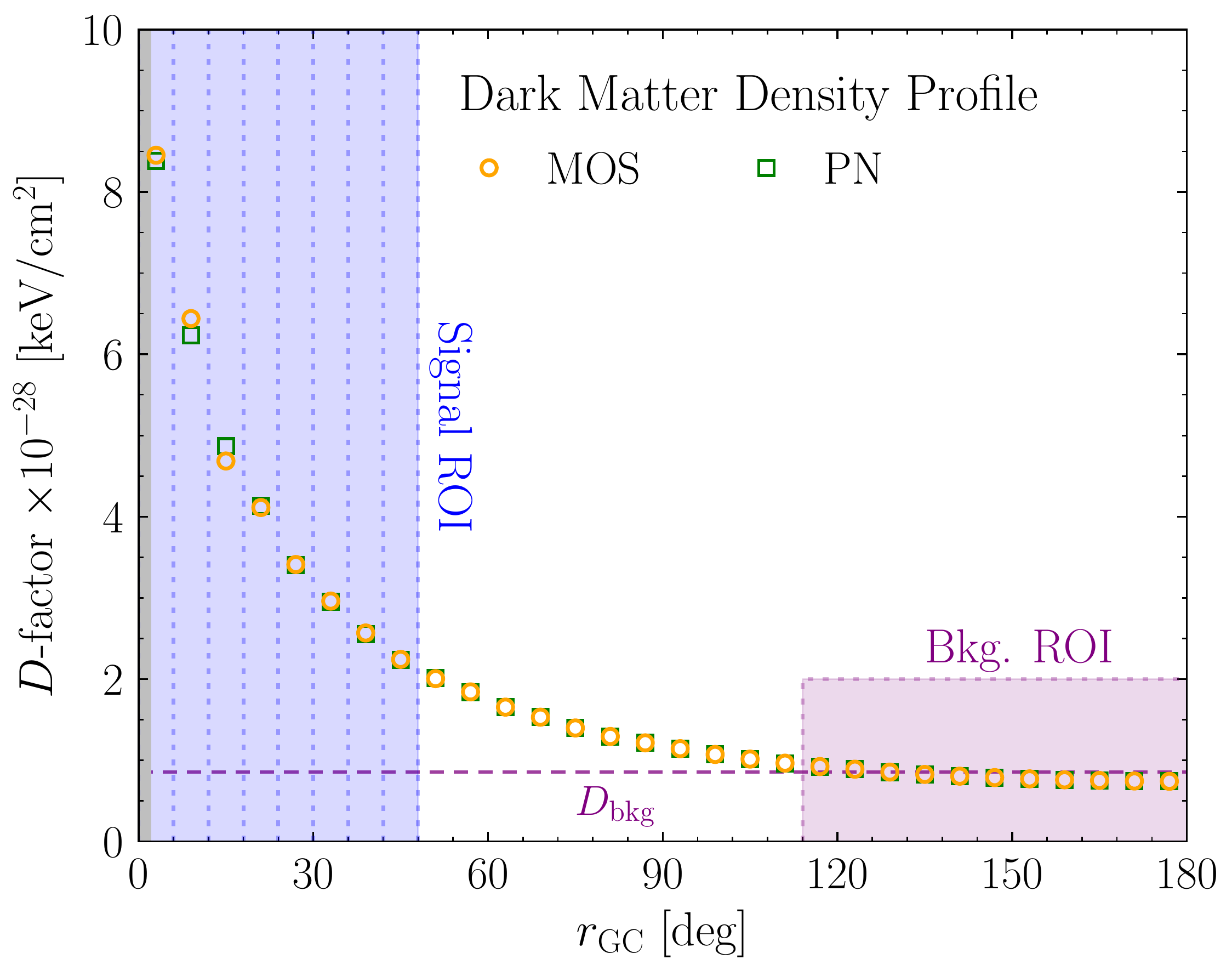}
\caption{Our fiducial $D$-factor, which is proportional to the expected DM signal flux.
Values are given in all 30 annuli, which are $6^{\circ}$ wide in angular distance from the GC (with $|b|>2^{\circ}$), and we define a signal and background ROI as shown. 
In each ring, we compute the $D$-factor of all MOS or PN exposures, weighted according to the observation time and field of view.
The horizontal line indicates $D_{\rm bkg}$, the mean $D$-factor in the background ROI.
}
\label{fig:ROI_ill}
\end{figure}

\noindent
{\bf Data analysis.}
Having constructed our data in all 30 rings, we divide the full sky into two regions of interest (ROI): a signal ROI, consisting of annuli 1 through 8 ($0^{\circ} \leq r_{\rm GC} \leq 48^{\circ}$ with $|b| \geq 2^\circ$), inclusive, and the background ROI, consisting of annuli 20 through 30 ($114^{\circ} \leq r_{\rm GC} \leq 180^{\circ}$ with $|b| \geq 2^\circ$).
The regions are illustrated in Fig.~\ref{fig:ROI_ill}. The MOS (PN) exposure time in the signal ROI is 25.27 Ms (5.56 Ms), whereas in the background ROI it is 62.51 Ms (17.54 Ms).  The signal flux is proportional to the $D$-factor, which is defined by the line-of-sight integral of the Galactic DM density $\rho_{\rm DM}$, $D \equiv \int ds \,\rho_{\rm DM}$.  In Fig.~\ref{fig:ROI_ill} we show the appropriately weighted $D$-factor in each annuli.
The motivation for the two ROIs is that the signal should dominate in the inner regions of the Galaxy and become progressively weaker further away from the GC.  The background ROI is chosen to be large enough to have significantly more exposure time than the signal ROI, so that using the background-subtracted data does not significantly broaden the statistical uncertainties.
We stack the data over the full background ROI, which has $D$-factor $D_{\rm bkg}$, and use this as an estimate of the instrumental and astrophysical baseline fluxes by subtracting this data from the data in each ring of the signal ROI.  This subtraction mostly removes large instrumental lines, as illustrated in Supplementary Material (SM) Fig. S1.  

We analyze the background-subtracted data in each annulus for evidence of a UXL. The data is modeled as a combination of narrow spectral features at the locations of known astrophysical and instrumental lines, and a continuum flux which we account for using GP modeling. Note that the instrumental lines need not be completely removed by the data-subtraction procedure, leaving a residual flux or flux deficit that must be modeled.  Astrophysical emission lines from the Milky Way plasma should be brighter in the signal ROI, and so are also expected to appear in the background-subtracted data. For both astrophysical and instrumental lines, the lines are modeled using the forward modeling matrices for MOS and PN. We allow the instrumental lines to have either positive or negative normalizations, while the astrophysical lines are restricted to have positive normalizations. 
To decide which lines to include in our residual background model we start with an initial list of known instrumental and astrophysical lines.
The instrumental lines are determined from an analysis of the background ROI data, while the astrophysical lines are those expected to be produced by the Milky Way.
In each ring, and for MOS and PN independently, we then determine the significance of each emission line, keeping those above $\sim$2$\sigma$. As a result, every ring has a different set of lines included in the analysis. 
We note that it is conceivable that a UXL might be inadvertently removed by an overly-subtracted instrumental line at the same energy; however, it would be highly unlikely for such a conspiracy to occur in every ring, given the varying $D$-factor.
The effects of sub-threshold instrumental lines are mitigated through a {\it spurious-signal} nuisance parameter~\cite{Aad:2014eha}, as discussed in the SM.

The unprecedented data volume incorporated into this analysis necessitates a flexible approach to modeling the residual continuum emission, which is accomplished with GP modeling, in order to minimize background mismodeling. 
As opposed to parametric modeling, where the model is specified by a specific functional form and associated list of model parameters, GP modeling is non-parametric: the model expectations for the data at two different energies, $E$ and $E'$, are assumed to be normally distributed with non-trivial covariance.  Taking the model expectation to have zero mean, the GP model is then fully specified by the covariance kernel, $K(E,E')$.  We model the mean-subtracted data using the non-stationary kernel $K(E,E') = A_{\rm GP}  \exp \left[{- (E-E')^2 / (2 E E'\sigma_E^2)} \right]$, implemented in \texttt{george}~\cite{hodlr},
where $\sigma_E$ is the correlation-length hyperparameter and $A_{\rm GP}$ is the amplitude hyperparameter. 
We fix $\sigma_E$ such that it is larger than the energy resolution of the detector, which is $\delta E / E \sim 0.03$ across most energies for MOS and PN, while ensuring $\sigma_E$ is kept small enough to have the flexibility to model real variations in the data. The goal is to balance two competing effects. If $\sigma_E$ approaches the lower limit imposed by the energy resolution of the detector, then the GP model would have the flexibility to account for line-like features, which would reduce our sensitivity when searching for such features over the continuum background. On the other hand, if $\sigma_E$ is too large then the GP continuum model may not accurately model real small-scale variations in the data.
In our fiducial analysis we fix $\sigma_E = 0.3$, though in the SM we show that our results are robust to variations not only in this choice, but also to modifications to the form of the kernel itself. In contrast, the hyperparameter $A_{\rm GP}$ is treated as a nuisance parameter that is profiled over when searching for UXLs.  

We then follow the statistical approach developed in~\textcite{Frate:2017mai}, which used GP modeling to perform an improved search for narrow resonances over a continuum background in the context of the Large Hadron Collider.  In particular, we construct a likelihood ratio $\Lambda$ between the model with and without the signal component, where the signal is the UXL line at fixed energy $E_{\rm sig}$. The null model is as above, the combination of a GP model with a single nuisance parameter $A_{\rm GP}$, and a set of background lines, whose amplitudes are treated as nuisance parameters. We use the marginal likelihood from the GP fit in the construction of the likelihood ratio~\cite{Frate:2017mai}.  Note that as the number of counts in all energy bins is large ($\gg$ 100), we are justified in assuming normally-distributed errors in the context of the GP modeling. 
We then profile over all nuisance parameters.
Finally, the discovery significance is quantified by the test statistic (TS) $t = - 2 \ln \Lambda$.  We verify explicitly in the SM that under the null hypothesis $t$ follows a $\chi^2$-distribution.
The 95\% one-sided upper limits are constructed from the profile likelihood, as a function of the signal amplitude.

\begin{figure}[!t]
\includegraphics[width = 0.48\textwidth]{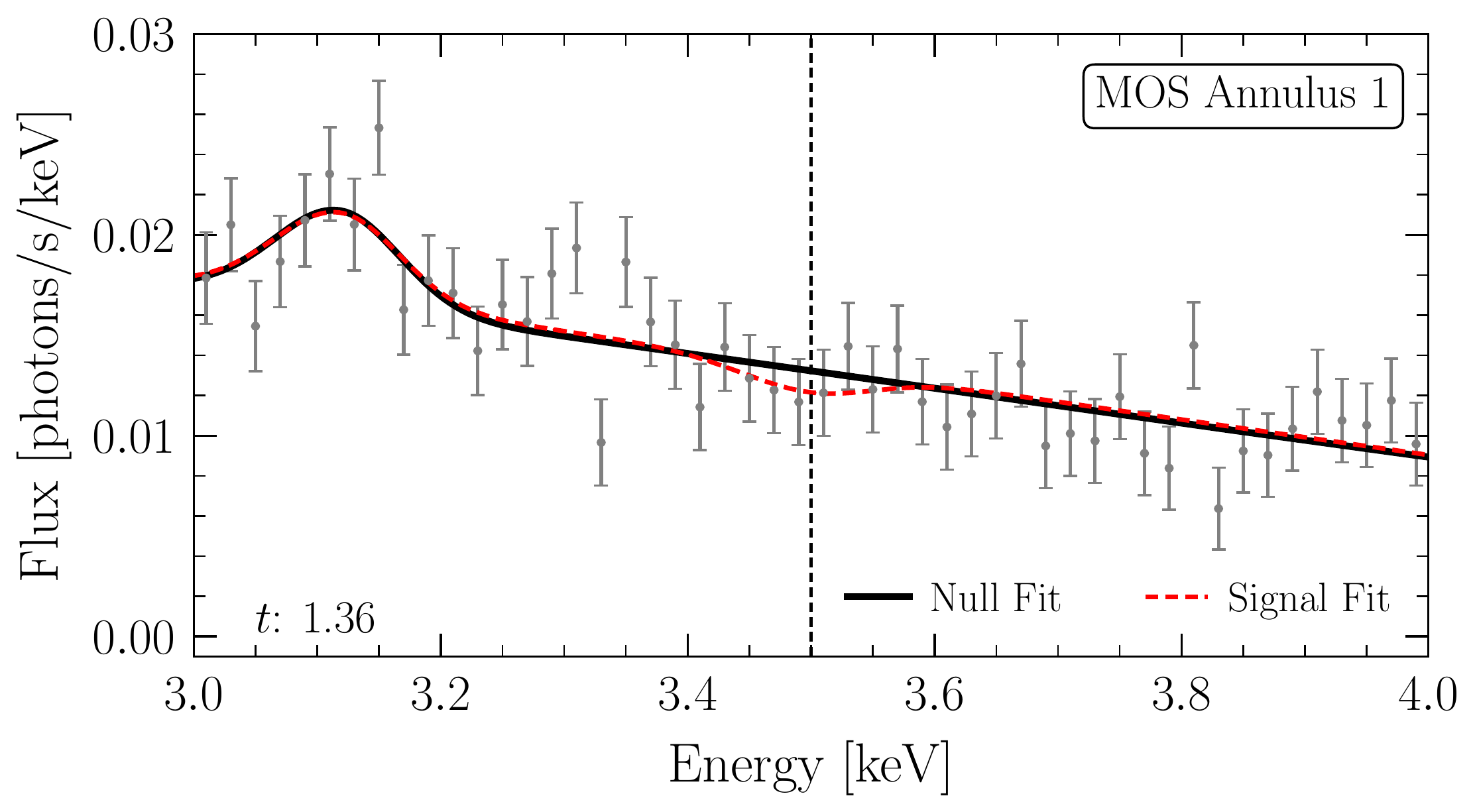}
\caption{
The background-subtracted MOS data for the innermost annulus, downbinned by a factor of four for presentation purposes.
The indiciated best fit null and signal models, for a 3.5 keV UXL, are constructed using the GP modeling described in the text.
}
\label{fig:fit_example_mos_1}
\end{figure}

We implement this procedure and scan for a UXL from 2.5 to 8 keV in 5 eV intervals.
At each test point we construct profile likelihoods for signal flux independently for each ring using the background-subtracted MOS and PN data.
We then combine the likelihoods between rings -- and eventually cameras -- in a joint likelihood in the context of the DM model, as discussed shortly. As an example, Fig.~\ref{fig:fit_example_mos_1} illustrates the signal and null model fits to the innermost MOS background-subtracted signal-annulus data for a putative UXL at 3.5 keV (indicated by the vertical dashed line).  Note that while the fit is performed over the full energy range (2.5$-$8 keV) for clarity we show the data zoomed in to the range 3 to 4 keV.  
In this case the data have a deficit, which manifests itself as a signal with a negative amplitude.

\noindent
{\bf DM interpretation.} We combine together the profile likelihoods from the individual annuli to test the decaying DM model.  In the context of sterile neutrino DM with mass $m_\chi$ and mixing angle $\theta$, the DM decay in the Galactic halo produces an X-ray flux at energy $m_\chi/2$ that scales as $\Phi \propto m_\chi^4 D \sin^2(2\theta)$~\cite{Pal:1981rm}.
Note that the $D$-factors, appropriately averaged over observations in the individual annuli, are illustrated in Fig.~\ref{fig:ROI_ill}.
Thus, at fixed DM mass $m_\chi$ we may construct profile likelihoods as functions of $\sin^2 (2 \theta)$ to appropriately combine the profile likelihoods as functions of flux in the individual annuli.  
We subtract off $D_{\rm bkg}$ from the $D$-factors in each signal ring since any UXL would also appear in the background ROI and thus be included in the background subtraction.

The $D$-factors may be computed from the DM density profile of the Milky Way.
Modern hydrodynamic cosmological simulations indicate that the DM density profile in Milky Way mass halos generally have a high degree of spherical symmetry (for a review, see Ref.~\cite{Vogelsberger:2019ynw}). Further, the presence of baryons contracts the inner $\sim$10 kpc of the profile away from the canonical Navarro, Frenk, and White (NFW) DM distribution~\cite{Navarro:1995iw,Navarro:1996gj}, so that there is an enhancement of the DM density at smaller radii versus the NFW expectation~\cite{Gnedin:2004cx,Schaller:2014uwa,Zhu:2015jwa,Dutton:2016okm,Hopkins:2017ycn,Lovell:2018amb}, though cores could develop on top of this contraction at radii $\lesssim 2$ kpc~\cite{Chan:2015tna,Mollitor:2014ara,2017MNRAS.465.1621P,2020MNRAS.497.2393L}.  For example, in Milky Way analogue halos within the \texttt{Fire-2} simulations the DM-only and hydrodynamic simulations produce DM density profiles that agree within $\sim$25\% at 10 kpc, but with baryons the density profiles are typically around twice as large as the NFW DM-only expectation at distances $\sim$1 kpc away from the GC~\cite{Hopkins:2017ycn}.  To be conservative we assume the canonical NFW density profile for all radii, though in the SM we discuss how our results change for alternate density profiles.

The NFW  profile is specified by a characteristic density $\rho_0$ and a scale radius $r_s$: $\rho_{\rm DM}(r) = \rho_0 /(r/r_s) / (1 + r/r_s)^2$.  We use the recent results from~\textcite{2020MNRAS.494.4291C}, who combined {\it Gaia} DR2 Galactic rotation curve data~\cite{2019ApJ...871..120E} with total mass estimates for the Galaxy from satellite observations~\cite{2019ApJ...873..118W,2019MNRAS.484.5453C}.  These data imply, in the context of the NFW model, a virial halo mass $M_{200}^{\rm DM} = 0.82_{-0.18}^{+0.09} \times 10^{12}$ $M_\odot$ and a concentration $c = r_{200}/r_s = 13.31_{-2.68}^{+3.60}$, with a non-trivial covariance between $M_{200}^{\rm DM}$ and $c$~\cite{2020MNRAS.494.4291C} such that lower concentrations prefer higher halo masses.  Within the 2D 68\% containment region for $M_{200}^{\rm DM}$ and  $c$ quoted in Ref.~\cite{2020MNRAS.494.4291C}, the lowest DM density at $r \approx 0.5$ kpc, and thus the most conservative profile for the present analysis, is obtained for $\rho_0 = 6.6 \times 10^6$ $M_\odot / {\rm kpc}^3$ and $r_s = 19.1$ kpc.  We adopt these values for our fiducial analysis.
With our choice of NFW DM parameters the local DM density, at the solar radius, is $\sim$$0.29$ GeV$/$cm$^3$, which is consistent with local measurements of the DM density using the vertical motion of tracer stars perpendicular to the Galactic plane, see, \emph{e.g.}, Refs.~\cite{Read:2014qva,deSalas:2020hbh}.  

\begin{figure}[!t]
\includegraphics[width = 0.49\textwidth]{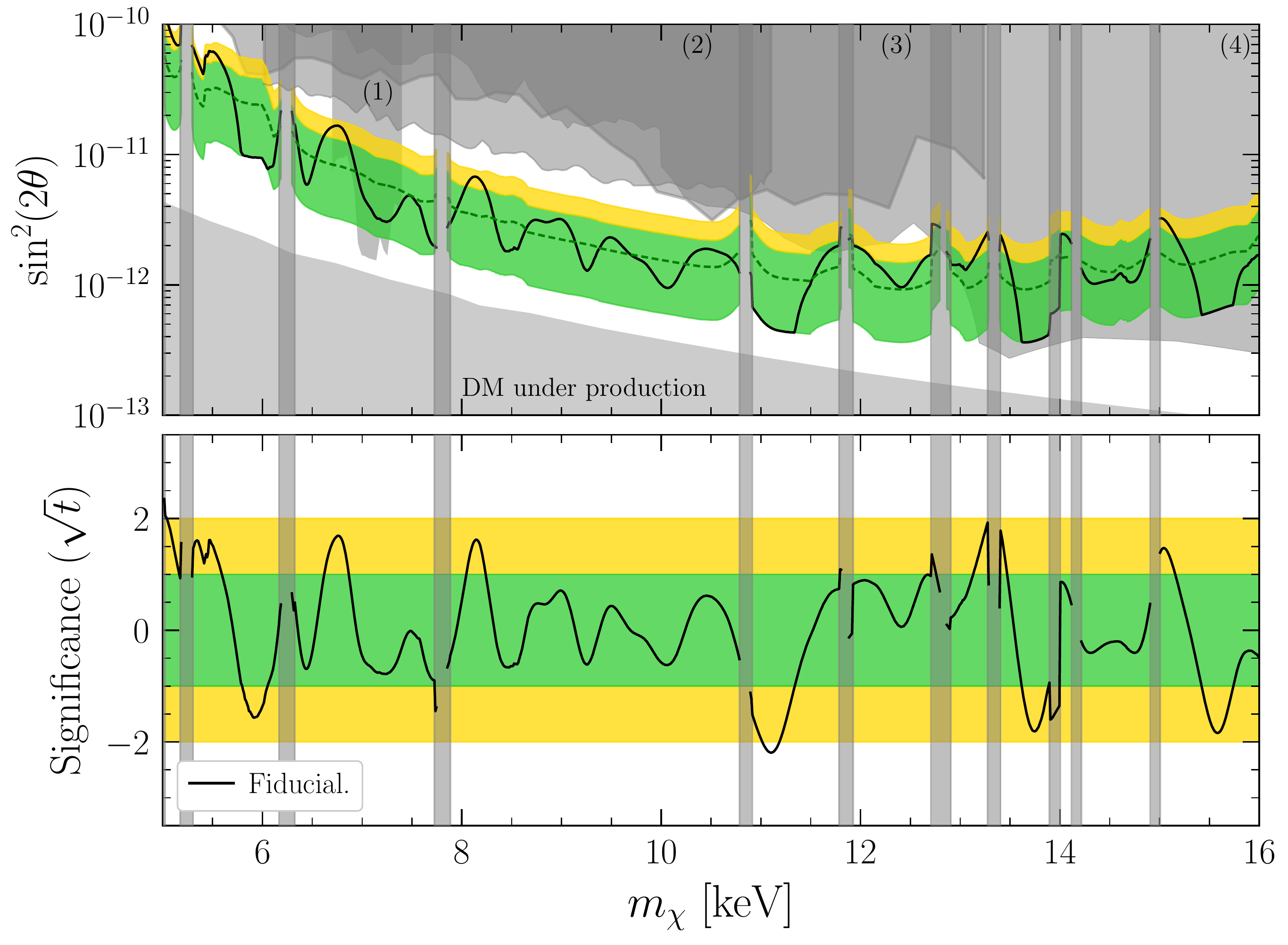}
 \vspace{-0.7cm}
\caption{
(Upper) The power-constrained 95\% upper limit on the DM lifetime from this work, presented in the context of the sterile-neutrino mixing angle $\sin^2(2\theta)$, as a function of the DM mass $m_\chi$. The dark grey regions correspond to theoretical bounds for DM underproduction in the $\nu$MSM or bounds from previous X-ray searches (1)--(5); see text for details.
(Lower) The associated sign-weighted significance for the UXL.  Vertical grey regions denote background lines and are at least partially masked. Green and gold regions indicate 1/2$\sigma$ expectations under the null hypothesis.  These results are shown in the context of more general DM models as constraints on the DM lifetime in SM Fig. S6. 
}
\label{fig:final_limit}
\end{figure} 

We search for evidence of decaying DM in 10 eV intervals in mass between 5$-$16 keV, masking 0.1 keV windows around the locations of known lines, as indicated in Fig.~\ref{fig:final_limit}.  We construct the joint likelihoods for the MOS and PN data sets.  
We test and account for additional background mismodeling
in the MOS and PN analyses by looking at the distribution of best-fit mixing angles in the energy side-bands, using a technique similar to the ``spurious signal" used by ATLAS in the search for the Higgs boson~\cite{Aad:2014eha}. This procedure is described in the SM and only has a small effect at low masses.  We then combine, at a given mass, the resulting MOS and PN profile likelihoods to obtain the final profile likelihood used to construct the limit and discovery significance shown in Fig.~\ref{fig:final_limit}.  In that figure we show the one-sided 95\% upper limit on $\sin^2(2\theta)$ in the upper panel, along with the 1 and 2$\sigma$ expectations for the power-constrained upper limit~\cite{Cowan:2011an} under the null hypothesis (shaded green and gold, respectively). 

We find no evidence for decaying DM signals above our pre-determined significance threshold of 5$\sigma$ global significance (corresponding to $\sim$6$\sigma$ local significance), as shown in the bottom panel. 
In that figure we compare our upper limit to previous limits in the literature, adjusted to our fiducial DM model for the Milky Way where appropriate. In the context of the $\nu$MSM it is impossible to explain all of the observed DM in the region marked ``DM under production" because of the big bang nucleosynthesis  bound on the lepton chemical potential~\cite{Dolgov:2002ab,Serpico:2005bc,Boyarsky:2009ix}.  Note that the $\nu$MSM also predicts that the DM becomes increasingly warm for decreasing $m_\chi$, which leads to tension with Milky Way satellite galaxy counts for low $m_\chi$: data from the Dark Energy Survey and other Galactic satellite surveys~\cite{Drlica-Wagner:2019vah} constrains $m_\chi$ greater than $\sim$15--20 keV in the $\nu$MSM~\cite{Nadler:2020prv} (which can be strengthened further when combined with strong lensing measurements~\cite{Nadler:2021dft}), though we note that our results apply to more general DM production mechanisms that do not predict modifications to small-scale structure.  In Fig.~\ref{fig:final_limit} we also show previous X-ray limits from 
(1) \textcite{Dessert:2018qih},
(2) a {\it Chandra} search for DM decay in the Milky Way~\cite{Sicilian:2020glg}, 
(3) a {\it Chandra} search for DM decay in M31~\cite{Horiuchi:2013noa}, and 
(4) combined {\it NuSTAR} searches for DM decay: in the Milky Way~\cite{Neronov:2016wdd,Perez:2016tcq,Roach:2019ctw}, the Bullet Cluster~\cite{Riemer-Sorensen:2015kqa}, and M31~\cite{Ng:2019gch}. 
Note that the results from Milky Way searches have been adjusted to use the same DM density profile as in our fiducial analysis.

\noindent
{\bf Discussion.} 
We find no significant evidence for decaying DM, which leads us to set some of the strongest constraints to-date on the DM lifetime.  {We confirm the results of Dessert {\it et al.}~\cite{Dessert:2018qih} for the non-observation of a DM decay line near 3.5 keV using a more robust and flexible analysis strategy, leaving little room for a decaying DM explanation of the previously-observed 3.5 keV anomalies~\cite{Bulbul:2014sua,Boyarsky:2014jta,Urban:2014yda,Jeltema:2014qfa,Cappelluti:2017ywp}. (See the SM for further discussion.)}

Given the data volume incorporated into this analysis it is unlikely that further analyses of {\it XMM-Newton} data, or {\it Chandra} data, could produce qualitatively stronger results on the DM lifetime in the mass range considered here.  However, the approach taken in this work may lead to a powerful advancement in discovery power with future data sets from surveys such as those by the upcoming {\it Athena}~\cite{Barcons:2015dua} and {\it XRISM}~\cite{XRISMScienceTeam:2020rvx} telescopes.  A combination of the data collected by those missions and the analysis framework introduced in this work may lead to the discovery of decaying DM in the few-keV mass range at lifetimes beyond those probed in this work.

\section*{Acknowledgments}
{\it 
We thank Kerstin Perez and Christoph Weniger for useful conversations. This work was supported  in  part  by  the  DOE  Early Career  Grant  DESC0019225. This research used resources of the National Energy Research Scientific Computing Center (NERSC) and the Lawrencium computational cluster provided by the IT Division at the Lawrence Berkeley National Laboratory, supported by the Director, Office of Science, and Office of Basic Energy Sciences, of the U.S. Department of Energy under Contract No.  DE-AC02-05CH11231. 
NLR is supported by the Miller Institute for Basic Research in Science at the University of California, Berkeley. 
KC is partially supported by NSF grant PHY-1505463m, NSF awards ACI-1450310, OAC-1836650, and OAC-1841471, and the Moore-Sloan Data Science Environment at NYU.
}

\bibliography{xmm}


\clearpage

\onecolumngrid
\begin{center}
  \textbf{\large Supplementary Material for A deep search for decaying dark matter with {\it XMM-Newton} blank-sky observations}\\[.2cm]
  \vspace{0.05in}
  { Joshua W. Foster, Marius Kongsore, Christopher Dessert, Yujin Park, Nicholas L. Rodd, Kyle Cranmer, and Benjamin R. Safdi}
\end{center}

\twocolumngrid
\setcounter{equation}{0}
\setcounter{figure}{0}
\setcounter{table}{0}
\setcounter{section}{0}
\setcounter{page}{1}
\makeatletter
\renewcommand{\theequation}{S\arabic{equation}}
\renewcommand{\thefigure}{S\arabic{figure}}
\renewcommand{\thetable}{S\arabic{table}}

\onecolumngrid

This Supplementary Material (SM) is organized as follows.  In Sec.~\ref{app:data} we provide additional details behind our data reduction and analysis procedure.  In Sec.~\ref{app:extended} we provide additional results from the main analysis presented in this Letter.  Sec.~\ref{app:injection} presents non-trivial checks of our analysis procedure using synthetic signals.  Lastly, in Sec.~\ref{app:syst} we perform multiple analysis variations to demonstrate the robustness of our main results.

\section{Data Reduction and Analysis}
\label{app:data}

In this section, we detail our process for data reduction and analysis.

\subsection{Data Reduction}

We selected all {\it XMM-Newton} observations performed until September 5, 2018. For each of these observations, we retrieved the raw data products from the \href{http://nxsa.esac.esa.int/}{XMM-Newton Science Archive}. For data reduction, we used the {\it XMM-Newton} Extended Source Analysis Software (ESAS) package, which is a part of the Science Analysis System~\cite{SAS} (SAS) version 17.0, and used for modeling sources covering the full {\it XMM-Newton} field-of-view and diffuse backgrounds.

The data reduction process is described in detail in Ref.~\cite{Dessert:2018qih}; here, we summarize the important steps and point out any differences. To reduce a given observation, we obtain the list of science exposures (i.e. pointings taken in a mode usable for scientific purposes) from the summary files. For each exposure (independent of camera), we generate an event list and filter this list to only include events which were recorded during a period of low-background, which cuts contamination from soft-proton flares. We then mask point sources in the field of view which contribute in any energy range (c.f. Ref.~\cite{Dessert:2018qih} where we only masked point sources in the 3-4 keV range). We also mask data from CCDs operating in anomalous states. From the filtered and masked data products we create the photon-count data, the ancillary response file (ARF), and the redistribution matrix file (RMF).

The reduced data contains 11,805 observations, with 21,388 and 8,190 individual MOS and PN exposures, totaling 438 Ms and 109 Ms of data. Given our focus is on searching for DM emission in otherwise dark regions of the sky, we place a cut on these data sets to isolate the astrophysically quietest amongst them. In particular, we construct the integrated flux from $2-10$ keV in all exposures, and determine the median value for MOS and PN separately as 0.09 photons/keV/s and 0.39 photons/keV/s. All observations with integrated fluxes higher than these median values are excluded. This cut will remove observations with above average astrophysical emission, but also those where there is large instrumental or quiescent particle background (QPB)  emission (c.f. Ref.~\cite{Dessert:2018qih} where a separate cut on the QPB emission was performed). 
For regions of the sky that are not focused on a bright Galactic or extra-galactic source, the QPB counts should dominate over the extra-galactic X-ray background~\cite{Lumb:2002sw}.  However, the QPB is time-dependent and will vary over exposures because of {\it e.g.} flaring activity (which, as described in the SM, we filter for).
Further, we emphasize that even in the most optimistic scenario, a DM UXL will only provide an exceptionally small contribution to the total integrated flux, and thus this cut will not bias against a potential signal. In addition to removing these bright exposures, we place two additional cuts. Firstly, all exposures with less than 500s of data are removed, as the flux in such short exposures can be poorly characterized. Finally, we exclude all observations within $2^{\circ}$ of the plane of the Milky Way, which excises only a small amount of the expected DM signal, but a much larger fraction of the expected astrophysical emission associated with emission from our own galaxy.  The cuts leave 215 Ms (57 Ms) of the total 438 Ms (109 Ms) of full-sky ($|b| \geq 2^\circ$) exposure time for MOS (PN).

Exposure passing all three cuts are then divided into 30 rings, each of width $6^{\circ}$ from the GC as described in the main body. The rings, numbered 1-30 starting from the GC, are used to form our signal ROI (rings 1-8) and background ROI (rings 20-30).

\subsection{Public Data Products}
\label{sec:public_data}

The processed data used to perform the analysis in this work is made fully publicly available at \href{https://github.com/bsafdi/XMM_BSO_DATA}{github.com/bsafdi/XMM\_BSO\_DATA}. There we provide all the data required to reproduce our results. In particular, we provide the data after the cuts described above in each ring for the MOS and PN cameras separately. The instrument response files, appropriately weighted across the exposures in each ring, are provided.

\subsection{Analysis}
\label{sec:sup_analysis}

In this section we provide additional details behind the analysis framework used to interpret the data products described above in the context of the decaying DM model.  First, we describe how we analyze the flux data in the individual annuli, and then we detail how those results are joined together to constrain the DM lifetime.  Lastly, we describe how we test for and incorporate systematic uncertainties.  

\subsubsection{Construction of the profile likelihood}

\begin{figure}[htb]
\includegraphics[width = 0.95\textwidth]{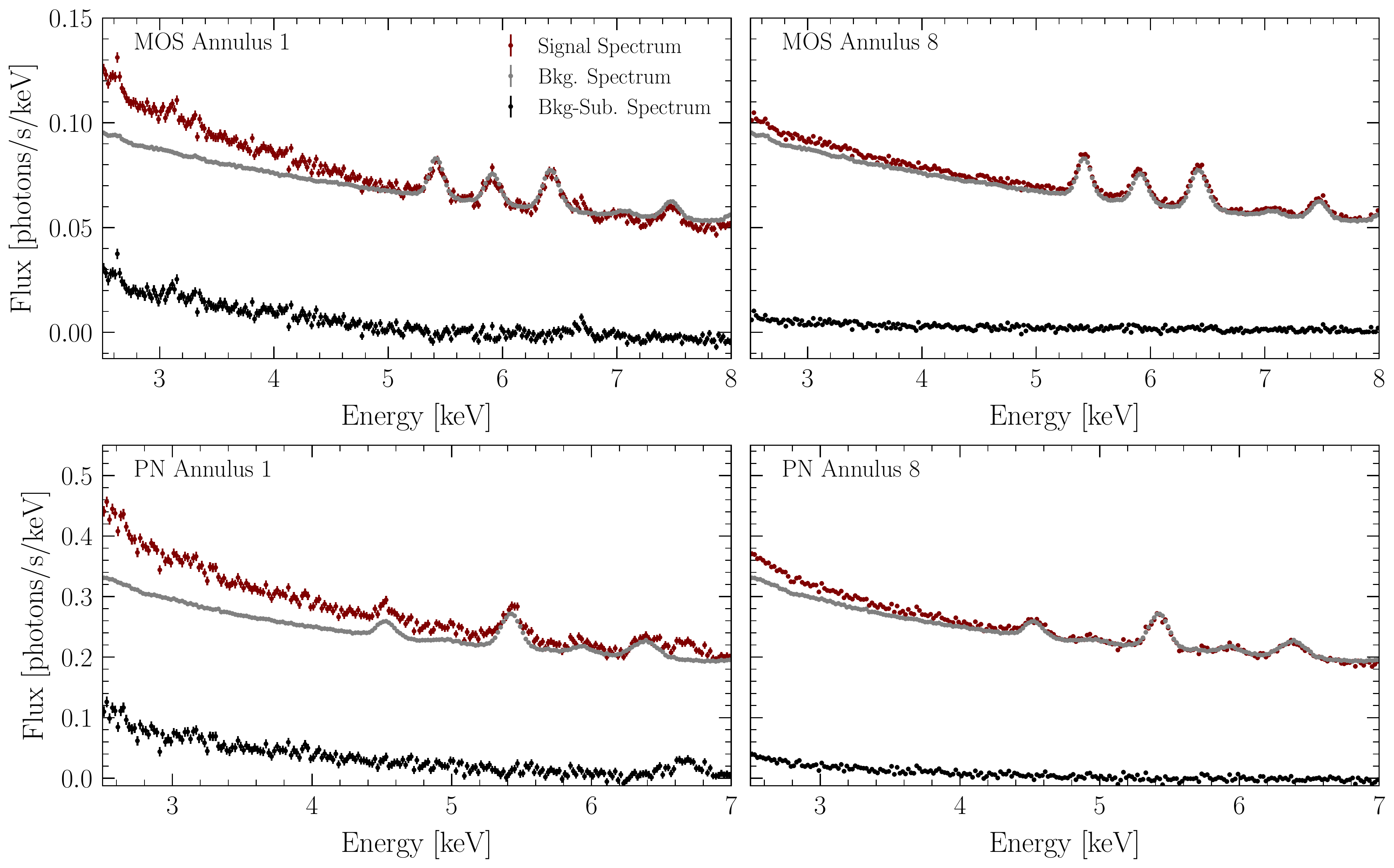}
\caption{Examples of the signal region spectra for MOS (top panels) and PN (bottom panels) in Ring 1 (left panels) and Ring 8 (right panels) with and without background subtraction in red and black, respectively. The background-region spectra are shown in grey. Many of the large instrumental features that are removed when looking at the background-subtracted data.  Note that for visual clarity these spectra have been down-binned by a factor of 4.}
\label{fig:back_sub_spectra}
\end{figure}

Let us first focus on the analysis of the (either MOS or PN) data in an individual ring $k \in [1,8]$.  The data set $d^k$ in this ring consists of background subtracted count rates $d^k_{i}$ in each energy channel $i$.  The count rates have units of cts$/$s$/$keV, as illustrated in {\it e.g.} Fig.~\ref{fig:back_sub_spectra}, with Poisson counting uncertainties $\sigma^k_i$ that arise from combining the statistical uncertainties in the signal and background data sets in the large-count limit, where the uncertainties become normally distributed.  Our goal is then to compute the log-likelihood $\log p(d_k | {\bf \theta})$ as a function of the model parameters ${\bf \theta} = \{A_{\rm sig}, {\bf \theta}_{\rm nuis}\}$, which consist of our parameter of interest, $A_{\rm sig}$, and our nuisance parameters ${\bf \theta}_{\rm nuis}$.  The nuisance parameters include background line amplitudes, $A_j$, with $j$ indexing the different lines at energies $E_j$, and also hyperparameters for the GP model.  In our fiducial analysis the only GP model hyperparameter is the amplitude of the double-exponential kernel $A_{\rm GP}$.  Note that, as described shortly, in determining the instrumental line list we also assign nuisance parameters to the locations of the lines.  Our goal is to construct the profile likelihood $\log p(d_k | A_{\rm sig}) = {\rm max}_{ {\bm \theta}_{\rm nuis}} \log p(d_k | {\bm \theta}) $.

Before describing the log-likelihood function in detail, we note that because we are in the large-count limit, so that the statistical fluctuations are normally distributed, we may interchangeably use the concept of modeling the data as the sum of model components and subtracting model components from the data and considering the residuals. 
In the small-count limit, where the Poisson fluctuations are not nearly Gaussian, this approach would not be appropriate.

The log-likelihood function that we use is a modification of the zero-mean GP marginal likelihood.  The modification that we implement incorporates the background lines and the signal line of interest.  For a given set of model parameters $\{A_{\rm sig},A_j\}$ we construct the modified data vector\footnote{Note that the line energies $E_j$ are fixed in all analyses except those of the background ROI data for constructing our lists of instrumental lines; in those analyses only, the $E_j$ are also model parameters.} 
\es{eq:y_deff}{
y^k_i({\bm \theta}) \equiv d^k_i -A_{\rm sig} \mu_{\mathrm{sig}, i}^k - \sum_j A_j \mu_{j,i}^k  -\big\langle d^k_i -A_{\rm sig} \mu_{\mathrm{sig}, i}^k - \sum_j A_j \mu_{j,i}^k  \big\rangle_i \,,
}
where $\mu_{\rm sig}$ is the spectral template of the signal line of interest, with fixed normalization, as obtained by appropriately summing the forward modeling matrices of the individual exposures that compose the observations within the ring of interest, $k$.  Similarly, $\mu_{j,i}^k$ denotes the fixed-normalization spectral template of the $j^{\rm th}$ background line, at energy $E_j$, in ring $k$ (recall that $i$ labels the detector energy channel).  The quantity $\langle \cdots \rangle_i$ in~\eqref{eq:y_deff} denotes the average over energy bins $i$, which implies that by construction the $y^k_i({\bm \theta})$ have zero mean when averaged over the full energy range of the analysis.
We postulate that the $y^k_i$ are described by GP models, so that we may use the GP marginal likelihood to compute the hyperparameter $A_{\rm GP}$:
\es{eq:GP_marginal_likelihood}{
\log p(d_k | {\bm \theta}) = - {1 \over 2} {{\bf y}^k}^T \left[ {\bf K} + (\sigma^k)^2 {\bf I}\right]^{-1} {\bf y}^k - {1 \over 2} \log | {\bf K} + (\sigma^k)^2 {\bf I} | - {n \over 2} \log(2 \pi) \,.
}
Above, $n$ is the number of energy channels, and all matrix operations are taken in the space of energy channels, with $(\sigma^k)^2 {\bf I}$ denoting the diagonal matrix with entries $(\sigma^k_i)^2$.  The matrix ${\bf K}$ denotes the GP kernel.  We implement the non-stationary kernel 
\es{eq:kernel}{
K(E,E') = A_{\rm GP}  \exp \left[{- {(E-E')^2 \over  2 E E'\sigma_E^2 }} \right] \,,
}
which has the hyperparameters $A_{\rm GP}$ and $\sigma_E$.  Note that later in the SM we show that similar results are obtained using the more standard double exponential kernel, but we chose the form of the kernel in~\eqref{eq:kernel} for reasons discussed below.

It is worth emphasizing that we have made the choice to describe the residuals of the background-subtracted data, after also subtracting the contributions from the instrumental lines, by a zero-mean GP model.  An alternative strategy would be to allow the GP model to have an energy-dependent mean.  Equivalently, we could include a parametric model component (such as a power-law or exponential) to model the clear upward trend in the data at low energies observed in {\it e.g.} Fig.~\ref{fig:back_sub_spectra}, with the GP model then describing fluctuations about that parametric component.  Such an approach would likely result in smaller values of the hyperparameter $A_{\rm GP}$ and, potentially, increased sensitivity.  Such a hybrid parametric plus GP modeling approach could be explored in future work.

Our goal is to look for narrow lines on top of a smooth continuum flux. We know that even a narrow line will manifest itself as a broader feature in the detector-level data due to the detector response. So the correlation-length of the GP kernel has a lower-bound set by the detector resolution. Because the energy resolution $\delta E$ of {\it XMM-Newton} increases linearly with energy ({\it i.e.}, $\delta E / E$ is roughly constant), a stationary kernel with a fixed correlation length is not adequate and a kernel of the form in~\eqref{eq:kernel} is more natural. However, we expect the continuum to be much smoother, even before the smearing induced by the detector resolution.  
A common approach in GP literature when the hyperparameters are not motivated from some other considerations is to fit them to the data. This approach leads to the best-fit values $\sigma_E \approx 0.608$ ($\sigma_E \approx 0.77$) for MOS (PN) in the first annulus, with comparable results in the annuli further from the Galactic Center.  However, we chose to fix $\sigma_E = 0.3$ because this is an intermediate value between the lower-bound of a narrow line given by the energy resolution and the best-fit result reflecting the smoothness of the observed continuum. This choice leads to more conservative limits, since for smaller values of $\sigma_E$ the GP model is able to capture smaller-scale fluctuations in the data, absorbing what would otherwise be attributed to narrow lines.

Lastly, note that while the marginal likelihood in~\eqref{eq:GP_marginal_likelihood} is defined within the context of Bayesian statistics, as it is obtained by integrating the likelihood times prior distribution for the formal GP model parameters, we will use the likelihood to perform frequentist parameter inference.  This approach is called the ``hybrid approach" in~\cite{Frate:2017mai}.  As noted in~\cite{Frate:2017mai}, the asymptotic expectations for the distribution of the TS constructed from the marginal likelihood may differ from the frequentist expectations~\cite{Cowan:2010js}, because of the use of the Bayesian marginal likelihood, and so in principle the $p$-values and upper-limit criteria should be calibrated on Monte Carlo (MC).  However, as we show below, we find through MC simulations that in our examples the TS statistical distributions follow the asymptotic frequentist expectations to high accuracy.  With that in mind, we briefly review the asymptotic expectations for translating discovery TS values to $p$-values and forming 95\% one-sided upper limits.

As discussed in the main Letter, the TS in favor of the signal model is given by 
\es{}{
t = -2 \big[  \max_{\bm \theta}\log p(d_k | {\bm \theta}) - \max_{\bf \theta_{\rm nuis}}\log p(d_k | \{A_{\rm sig} = 0,{\bm \theta}_{\rm nuis}\} \big] \,,
}
where the second term is the maximum marginal likelihood for the null model without a signal line.  When searching for evidence of DM, the discovery TS is set to zero for unphysical model parameters (in that case, $\sin^2(2\theta) < 0$), but for the purpose of testing for systematic uncertainties it is useful to allow for both positive and negative signal amplitudes.  The discovery TS is asymptotically $\chi^2$ distributed with one degree of freedom under the null hypothesis (see, {\it e.g.},~\cite{Cowan:2010js}).  In addition to searching for evidence of the signal model over the null hypothesis using $t$, we also set 95\% one-sided upper limits using the likelihood ratio.  We define the profile likelihood ratio $q(A_{\rm sig})$ by  
\es{eq:prof_LL}{
q(A_{\rm sig}) = -2 \big[  \max_{\bm \theta_{\rm nuis}}\log p(d_k | \{A_{\rm sig},{\bm \theta}_{\rm nuis}\}) - \max_{\bm \theta}\log p(d_k | {\bm \theta}) \big] \,,
}
where in the first term we maximize the log-likelihood over the nuisance parameters ${\bm \theta}_{\rm nuis}$ at fixed signal parameter $A_{\rm sig}$.  Let $\hat A_{\rm sig}$ be the best-fit signal parameter; {\it i.e.}, $q({\hat A}_{\rm sig}) = 0$.  Then, the 95\% one-sided upper limit $A_{\rm sig}^{95\%}$ is given by the value $A_{\rm sig}^{95\%} > {\hat A}_{\rm sig}$ which satisfies $q(A_{\rm sig}^{95\%}) \approx -2.71$~\cite{Cowan:2010js}.

Note that the profile likelihood in~\eqref{eq:prof_LL} is computed as a function of the signal normalization $A_{\rm sig}$ at fixed UXL energy (or, equivalently, fixed DM mass).  All of the analyses presented in the main Letter are performed in this way ({\it i.e.}, we have a grid of different UXL energies to probe and then for each fixed energy we compute the profile likelihood as a function of the signal-strength amplitude).  In SM Sec.~\ref{sec:sig_inj_real}, however, we consider our ability to localize a putative signal in $m_a$ and $\sin^2(2\theta)$ using synthetic data.  In that analysis, and that analysis only, we simultaneously constrain the mass and the signal strength.   

\subsubsection{Instrumental and astrophysical background lines}

Several instrumental and astrophysical background spectral lines are expected to contribute to the observed flux spectra. 
Here, we outline the procedure by which we obtain our candidate instrumental and astrophysical background lines, with the complete list of included lines for MOS and PN presented in Tab.~\ref{tab:MOS_Lines} and  Tab.~\ref{tab:PN_Lines}, respectively.

\begin{table}[htb]
\ra{1.3}
\centering
\begin{tabularx}{\textwidth}{p{0.1\textwidth}*{10}{P{0.08\textwidth}}}
\hline
 Energy [keV] & Origin & Type & Ring 1 & Ring 2 & Ring 3 & Ring 4 & Ring 5 & Ring 6 & Ring 7 & Ring 8  \\ \hline
 2.46 & S  & Astro.  & \textbf{12.5} & 0.7 & \textbf{7.2} & 0.7 & 2.0 & \textbf{5.5} & 0.9 & \textbf{5.4}  \\ \hline
 2.62 & S  & Astro.  & \textbf{36.8} & \textbf{7.6} & \textbf{4.3} & 1.9 & 0.0 & 2.1 & 0.0 & \textbf{6.1}  \\ \hline
 3.12 & Ar  & Astro. & \textbf{15.0} & 1.1 & \textbf{8.8} & 2.8 & 0.0 & \textbf{3.5} & 0.0 & 0.9  \\ \hline
 3.90 & Ca  & Astro. & 0.3 & 0.0 & 0.0 & \textbf{9.8} & 0.0 & 0.2 & \textbf{4.1} & 0.0 
 \\ \hline
 5.42 & Cr  & Inst.  & \textbf{8.9} & \textbf{7.7} & \textbf{22.1} & 0.1 & 0.0 & 1.2 & \textbf{7.2} & 0.0 \\ \hline
 5.92 & Mn  & Inst.  & 1.8 & 1.0 & \textbf{3.6} & 0.3 & 0.4 & 2.5 & \textbf{10.0} & 1.7  \\ \hline
 6.42 & Fe  & Inst.  & 0.1 & \textbf{7.1} & \textbf{82.0} & 0.3 & 1.5 & 0.0 & 2.7 & \textbf{7.2} \\ \hline
 6.67 & Fe  & Astro. & \textbf{55.4} & 1.7 & 0.0 & \textbf{4.3} & \textbf{5.9} & 2.5 & 0.0 & 0.0 \\ \hline
 6.97 & Fe  & Astro. & \textbf{5.9} & 0.0 & 0.5 & \textbf{4.3} & 0.3 & 0.2 & 0.4 & 0.2 \\ \hline
 7.08 & Fe  & Inst.  & 1.5 & 0.0 & \textbf{4.1} & 0.2 & 2.1 & 1.3 & 0.0 & 2.2 \\ \hline
 7.48 & Ni  & Inst.  & 2.0 & 0.1 & \textbf{11.9} & 0.1 & 0.0 & 0.1 & \textbf{3.1} & \textbf{4.7}  \\ \hline
  8.06 & Cu  & Inst.  & 1.1 & 0.9 & 0.3 & 0.1 & 0.0 & 0.2 & 0.4 & 0.1  \\ \hline
\end{tabularx}
\caption{The list of spectral lines of instrumental and astrophysical origins which are included in our background model for the MOS camera. For the line in each ring, we provide the value of $\Delta \chi^2$ associated with the addition/removal of the line from the best-fit background model which is obtained after our line-dropping procedure. Bolded values indicate the inclusion of a line in a ring's background model.}
\label{tab:MOS_Lines}
\end{table}

\begin{table}[htb]
\ra{1.3}
\centering
\begin{tabularx}{\textwidth}{p{0.10\textwidth}*{10}{P{0.08\textwidth}}}
\hline
 Line Energy [keV] & Origin & Type & Ring 1 & Ring 2 & Ring 3 & Ring 4 & Ring 5 & Ring 6 & Ring 7 & Ring 8  \\ \hline
 2.46 & S   & Astro. & \textbf{15.3} & 0.7 & \textbf{12.5} & \textbf{3.9} & 0.0 & \textbf{3.5} & 1.1 & \textbf{6.8} \\ \hline
 2.62 & S   & Astro. & \textbf{19.0} & \textbf{4.5} & \textbf{9.1} & \textbf{5.1} & 0.0 & 0.0 & 0.1 & \textbf{4.4} \\ \hline
 3.12 & Ar  & Astro. & \textbf{6.4} & 2.8 & \textbf{6.5} & \textbf{13.7} & 0.0 & 1.1 & 0.0 & 0.0 \\ \hline
 3.90 & Ca  & Astro. & \textbf{3.9} & 0.0 & 0.5 & \textbf{3.8} & 0.0 & \textbf{6.0} & 0.2 & 0.5 \\ \hline
 4.52 & Ti  & Inst.  & 0.6 & 0.7 & 0.8 & 0.6 & 0.1 & 0.3 & 0.5 & 0.0  \\ \hline
 5.42 & Cr  & Inst.  & 0.7 & 0.0 & 0.4 & \textbf{7.3} & 0.7 & 1.9 & 2.4 & \textbf{6.6}  \\ \hline
 5.93 & Cr  & Inst.  & 0.8 & 1.0 & 0.5 & \textbf{4.7} & 0.1 & 1.7 & \textbf{3.6} & 0.7 \\ \hline
 6.39 & Fe  & Inst.  & 0.0 & 0.1 & \textbf{3.2} & 1.2 & 0.0 & 1.6 & \textbf{9.3} & 0.0  \\ \hline
 6.67 & Fe  & Astro. & \textbf{79.2} & \textbf{5.5} & \textbf{8.9} & 2.8 & 0.8 & 2.3 & 0.3 & 2.3 \\ \hline
 6.97 & Fe  & Astro. &0.0 & 0.0 & 0.4 & 0.0 & 0.3 & 0.0 & 0.7 & 0.0 \\ \hline
\end{tabularx}
\caption{The list of spectral lines of instrumental and astrophysical origins which are included in our background model for the PN camera. For the line in each ring, we provide the value of $\Delta \chi^2$ associated with the addition/removal of the line from the best-fit background model which is obtained after our line-dropping procedure. Bolded values indicate the inclusion of a line in a ring's background model.}
\label{tab:PN_Lines}
\end{table}

We adopt an initial instrumental line list for PN and MOS from Refs.~\cite{Plaa:2010, Leccardi:2008}.
We then analyze the stacked data, for MOS and PN independently, in the background ROI in order to test for the presence of each candidate line.  We use an analysis framework analogous to that we use in the background-subtracted signal ROI data: in particular, our analysis of the background ROI data incorporates GP modeling for the continuum emission, in addition to the set of putative instrumental lines. We test for known instrumental lines in the vicinity of: 4.51 keV (Ti $K\alpha$), 5.41 keV (Cr K$\alpha$), 5.90 keV (Mn K$\alpha$), 5.95 keV (Cr K$\beta$), 6.40 keV (Fe K$\alpha$), 6.49 keV (Mn K$\beta$), 7.06 keV (Fe K$\beta$), 7.47 keV (Ni K$\alpha)$, 8.04 keV (Cu K$\alpha$). During this process, we allow the central location of the background lines to float by up to $25$ eV. Lines which are detected with $t > 16$ ($4\sigma$ local significance) in the background data analysis are accepted at their best-fit energy as a new component of our residual background model. In MOS, we accept instrumental lines at energies: 5.42 keV, 5.915 keV, 6.425 keV, 7.07 keV, 7.485 keV, and 8.06 keV. In PN, we accept instrumental lines at 4.52 keV , 5.42 keV, 5.95 keV, and 6.39 keV.

After constructing our list of instrumental background lines we include them in our analyses of the signal-ROI background-subtracted data sets.
In particular, each line is given an intensity nuisance parameter in each ring.
Given our procedure of subtracting the background flux from the signal region, variability in the instrumental lines between observations can result in the best fit instrumental line intensity in our data set having a positive or negative normalization.
Accordingly, we allow the normalization of the instrumental lines to be either positive or negative.

We also develop an initial list of candidate astrophysical background lines following~\cite{Bulbul:2014sua}, by selecting those with emissivities greater than $5\times 10^{-19}$ photons/cm$^{3}$/s at a temperature of 1 keV, which is the approximate temperature of the hot component of the Galactic Center, using the AtomDB database \cite{2016AAS...22721108F}. We include additional iron lines that are known to produce emission in the inner Galaxy~\cite{Koyama:2006pb}.
Taking this preliminary list, we then inspect the innermost ring and determine all lines which appear with TS $t > 3$ in either PN or MOS.  If such a line meets this criteria in either PN or MOS then we add it to our list of putative astrophysical lines for both instruments.
As with their instrumental counterparts, the astrophysical lines are treated with independent nuisance parameters describing their intensity in each annuli. 
However, for astrophysical background lines, we restrict intensities to values greater than or equal to zero.

The procedure described above leads to a list of astrophysical and instrumental lines, which are shown in Tabs.~\ref{tab:MOS_Lines} and~\ref{tab:PN_Lines}.  However, this does not mean that we included all of the those lines in every ring when performing our UXL search.  In each annulus we analyze the background-subtracted data to determine which of the lines in Tabs.~\ref{tab:MOS_Lines} and~\ref{tab:PN_Lines} are detected with moderate significance (we use the criteria $t > 3$) in the background-subtracted data set.  Note that in Tabs.~\ref{tab:MOS_Lines} and~\ref{tab:PN_Lines} we indicate whether the line is included in each annulus. To determine the significance of a given line we proceed iteratively, starting with the full list of lines and then calculating the change in the maximum likelihood when a given line is removed from the model.

\begin{figure*}[htb]
\includegraphics[width = \textwidth]{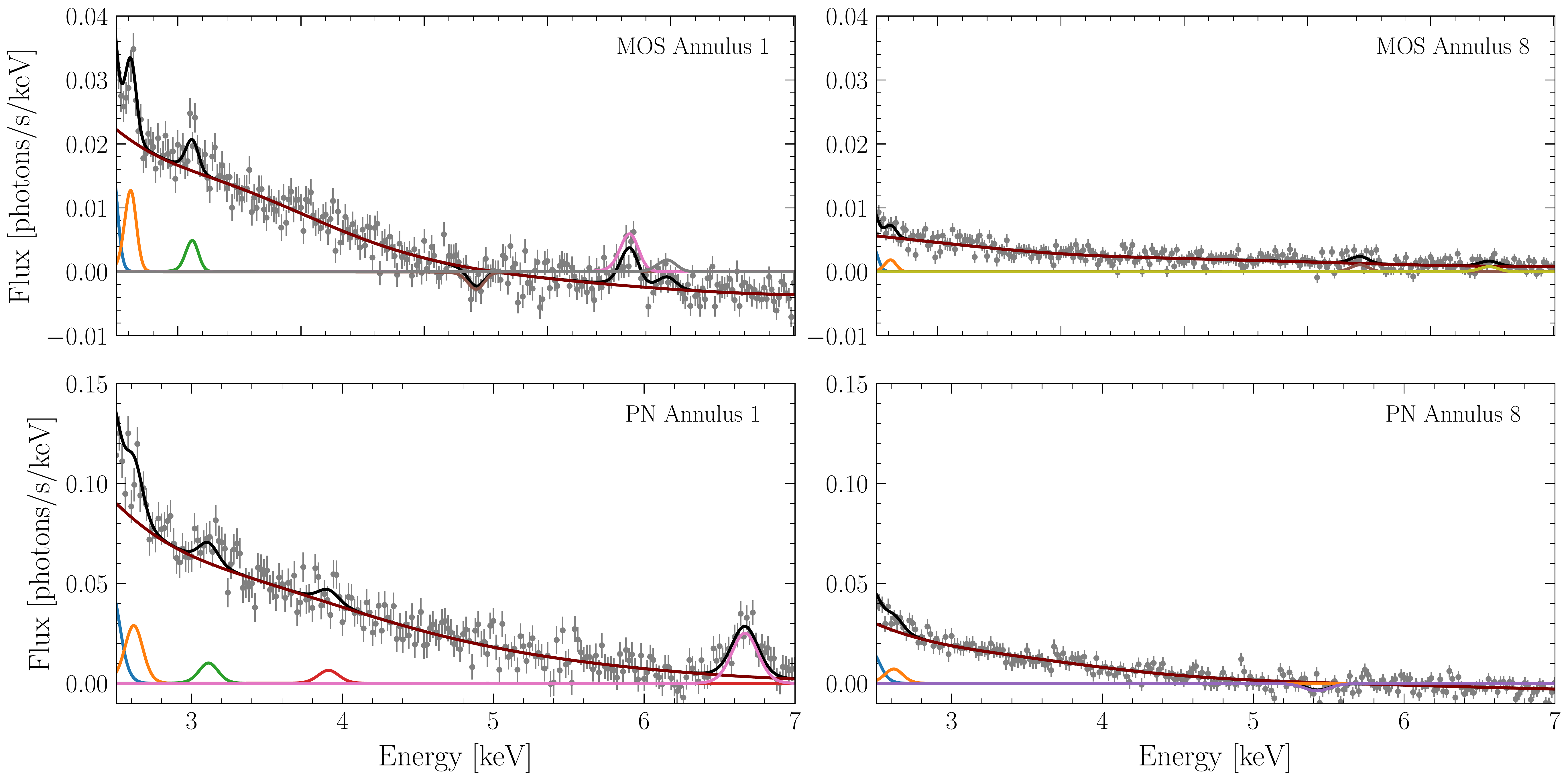}
\caption{The same background-subtracted data sets illustrated in Fig.~\ref{fig:back_sub_spectra} (also down-binned), but now shown along with their best-fits under the null hypothesis.  The best-fit model prediction is shown in black, which may be decomposed into the contribution from the GP model (dark red) and the contributions from the individual background lines (colored curves).  Note that the background lines to include in the analysis are determined independently in each annulus, as described in the text.  }
\label{fig:Example_Fits}
\end{figure*}

In Fig.~\ref{fig:Example_Fits} we illustrate example fits for our fiducial analyses to the data without the inclusion of an UXL.  These fits are to the same background-subtracted data as illustrated in Fig.~\ref{fig:back_sub_spectra}, as labeled. In black we show the combined best-fit model, which is the sum of the GP model contribution (dark red) and the contributions from the individual astrophysical and instrumental lines (colored curves).  Note that the number of background lines differs between each of the annuli because the important background lines are determined independently for each annulus. 

\subsubsection{The joint likelihood and background mismodeling}

After constructing the profile log likelihoods $q_k(A_{\rm sig})$ in each energy annulus ($k = 1,2, \cdots,8$) we then convert from $A_{\rm sig}$, which has units of cts$/$cm$^2/$s$/$sr, to $\sin^2(2\theta)$ using the relation
\es{eq:flux}{
\Phi &\approx 0.26\, {\rm photons}/{\rm cm}^2/{\rm s}/{\rm sr} \times \left( \frac{m_\chi}{7.0 \, {\rm keV}}  \right)^4 \left( \frac{D}{10^{29} \, {{\rm keV/cm}^2}} \right) \left( \frac{\sin^2 (2 \theta)}{10^{-10}} \right)\,.
}
To do so we use the background-subtracted $D$-factors, as discussed in the main Letter.  Then, at each test mass point for the DM model we construct the joint profile likelihood
\es{eq:q_joint}{
q_{\rm joint}\big(\sin^2(2\theta) \big) = \sum_{k=1}^8 q_k\big(\sin^2(2\theta) \big) \,,
}
for both MOS and PN independently. Later, we will also combine the MOS and PN profile likelihoods to construct our final joint profile likelihood that we use to search for evidence of decay DM.  First, however, we analyze the joint MOS and PN profile likelihoods independently for evidence of background mismodeling. 

We test and account for possible background mismodeling by extending the background model to include a component that is totally degenerate with the signal. This is a conservative approach that would remove all sensitivity to a UXL if the amplitude for this additional signal-like component were left free. Therefore we penalize the amplitude of such a signal like feature in the background model with a zero-mean Gaussian likelihood with variance hyperparameter $\sigma^2_{\rm spur}$ .
The approach we follow was developed and implemented in~\cite{Ackermann:2013uma,Albert:2014hwa,Ackermann:2015lka} within the context of searches for narrow spectral features in $\gamma$-ray astronomy and in the context of the Higgs boson search by the ATLAS experiment, where it is called the ``suprious signal"~\cite{Aad:2014eha}.  
We extend the likelihood to include two ``spurious signal" nuisance parameters, one for the MOS data and one for the PN data.  The MOS and PN likelihoods are then combined to produce the joint likelihood that we use for probing the DM model.

After extending the background model to include a signal-like component constrained by $\sigma^2_{\rm spur}$, the resulting profile likelihood (for either the MOS or PN data) is given by
\es{eq:joint_with_syst_profile}{
\tilde q_{\rm joint}\big( \sin^2(2\theta) \big) = {\rm max}_{A_{\rm spur}} \left[ q_{\rm joint}\big( \sin^2(2\theta) + A_{\rm spur} \big) - {\left( A_{\rm spur} \right)^2 \over \sigma^2_{\rm spur}} \right] \,,
}
where $q_{\rm joint}$ is defined in~\eqref{eq:q_joint}.
Note that the profile likelihood now depends on the hyperparameter $\sigma^2_{\rm spur}$, which determines the strength of the spurious-signal nuisance parameter
For example, in the limit $\sigma^2_{\rm spur} \to 0$ the nuisance parameter becomes fixed at zero ($A_{\rm spur} \to 0$) and
the modified profile likelihood $\tilde q$ approaches the un-modified likelihood $q$.  However, in the opposite limit $\sigma^2_{\rm spur} \to \infty$ we completely lose constraining power and $\tilde q_{\rm joint}\big( \sin^2(2\theta) \big) \to 0$ for all $\sin^2(2\theta) $.

In practice, we determine the value of the hyperparameter at each test mass point independently.
The philosophy is that if there is evidence that the background model is not properly describing the data in the immediate energy side-bands around a mass point of interest, then we should account for the possibility, through $A_{\rm spur}$, of similar background mismodeling at our mass point of interest.  Specifically, we implement the following approach.
At a given mass point $m_\chi^m$, where $m$ is the index that labels the mass point, we consider the subset of test mass points  in a 2 keV window around $m_\chi^m$, masking: (i) a 0.4 keV window in mass around $m_\chi^m$ and (ii) masking 0.1 keV windows around the locations are all background lines that were included in the analyses of the annuli.  Each test mass point within this side-band window has a best-fit $\sin^2(2\theta)$ from the likelihood analysis without the inclusion of the spurious-signal nuisance parameter.  The ensemble of best-fit points in the side-band window is denoted by $\{ \sin^2(2\theta) \}_m$.  We compute the variance over this ensemble of best-fit points, ${\rm Var}\left[ \{ \sin^2(2\theta) \}_m \right]_{\rm observed}$.  The observed variance is then compared to the expected variance ${\rm Var}\left[ \{ \sin^2(2\theta) \}_m \right]_{\rm expected}$, and specifically we set 
\es{eq:sigma_def}{\sigma^2_{\rm spur, m} = {\rm max}\left[ 0, {\rm Var}\left[ \{ \sin^2(2\theta) \}_m \right]_{\rm observed} - {\rm Var}\left[ \{ \sin^2(2\theta) \}_m \right]_{\rm expected}\right] \,,
}
where $\sigma^2_{\rm spur, m}$ denotes the hyperparameter at the mass point $m_\chi^m$.  The expected side-band best-fit variance ${\rm Var}\left[ \{ \sin^2(2\theta) \}_m \right]_{\rm expected}$ is computed from 500 MC simulations of the null hypothesis.  The null hypothesis model is that given by the fit of the background model to the data without any extra UXL signal components. 

We expect $\sigma^2_{\rm spur, m}$ to be non-zero if there is background mismodeling in the energy side-band, which increases the variance of observed best-fit points relative to the expectation under the null hypothesis.  However, sometimes $\sigma^2_{\rm spur, m}$  will be non-zero simply because of statistical fluctuations in the observed side-bands, in which case the nuisance parameter will weaken the limits more than intended.  However, this occasional weakening  of the limits is worth the advantage of having an analysis framework that is more robust to mismodeling.  Indeed, we know that there is an opportunity for some degree of background mismodeling because we have chosen to only include background lines that pass some significance threshold, and thus the aggregate effect of the sub-threshold lines could lead to mismodeling that could be partially mitigated by $A_{\rm spur}$.

\begin{figure*}[htb]
\includegraphics[width = .7\textwidth]{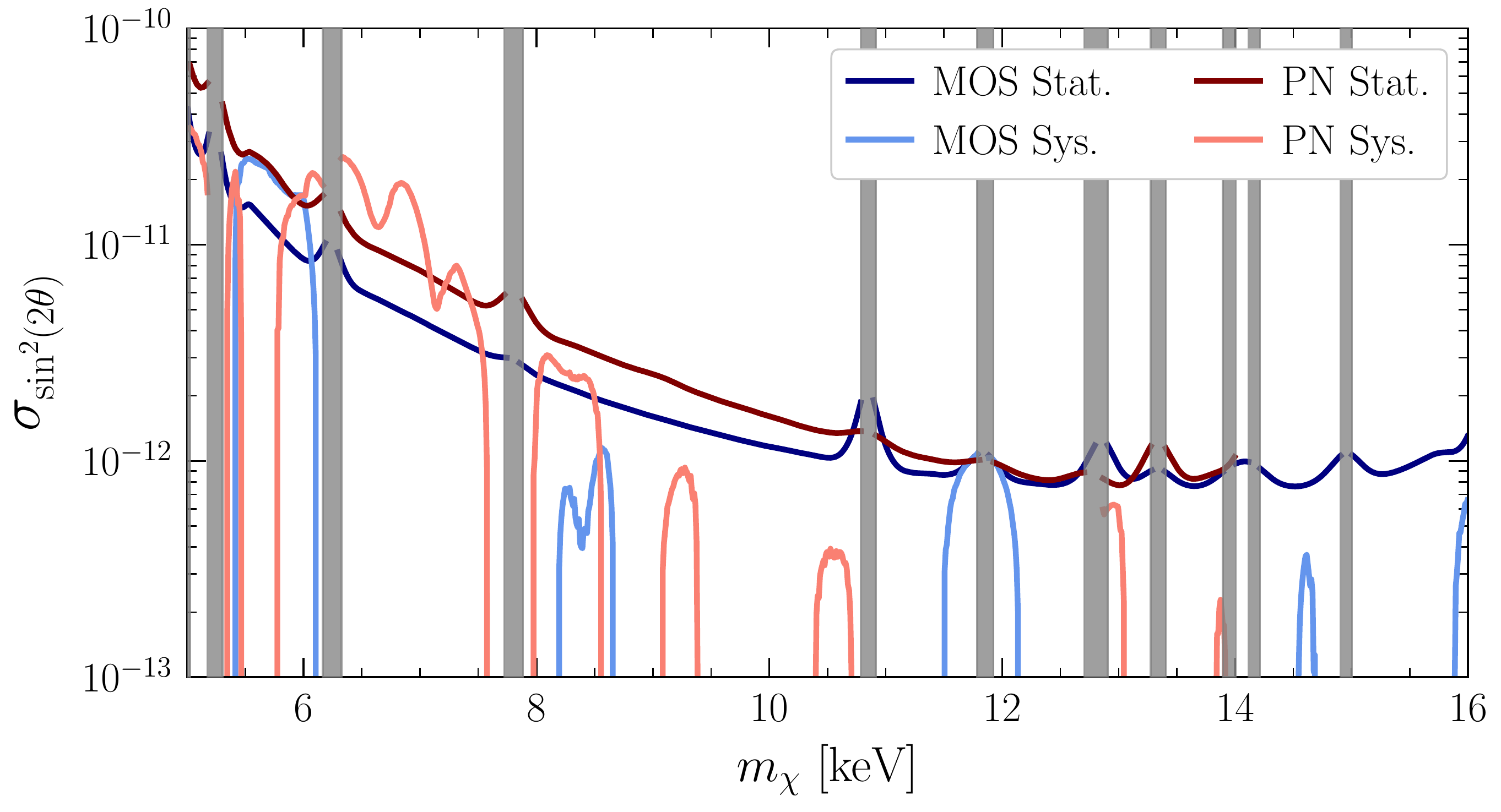}
\caption{The spurious-signal hyperparameter $\sigma^2_{\rm spur, m}$ (labeled MOS Sys. and PN Sys.), as computed in~\eqref{eq:sigma_def}, as a function of the DM mass.  For both MOS and PN the nuisance parameter $A_{\rm spur}$ is predominantly active at low energies, and it plays a more significant role in PN than in MOS.  We compare the hyperparameter to the statistical uncertainties (labeled MOS Stat. and PN Stat.), which are computed from the Hessian of the log-likelihood (without the spurious-signal) about the best-fit mixing angle at a fixed energy. {We note that several of the sharp variations of the expected sensitivity shown in Fig.~\ref{fig:final_limit} arise as a result of the variations of the spurious signal hyperparameter shown here.}
}
\label{fig:Nuisance_Params}
\end{figure*}

In Fig.~\ref{fig:Nuisance_Params} we illustrate the values of $\sigma^2_{\rm spur, m}$ (labeled MOS Syst. and PN Syst.) that we find from the data analyses of the MOS and PN data.  We compare the hyperparameter to the statistical uncertainty on $\sin^2(2\theta)$, labeled MOS Stat. and PN Stat.  Note that the statistical uncertainties are computed from the Hessian of the log-likelihood, for that data set, about the best-fit coupling at a fixed UXL energy, without the inclusion of $A_{\rm spur}$.  For both MOS and PN we see that the background mismodeling uncertainties, as captured by $A_{\rm spur}$, may dominate the statistical uncertainties at some low energies, though the nuisance parameter appears more important for PN than for MOS.  

\begin{figure*}[htb]
\includegraphics[width = .98\textwidth]{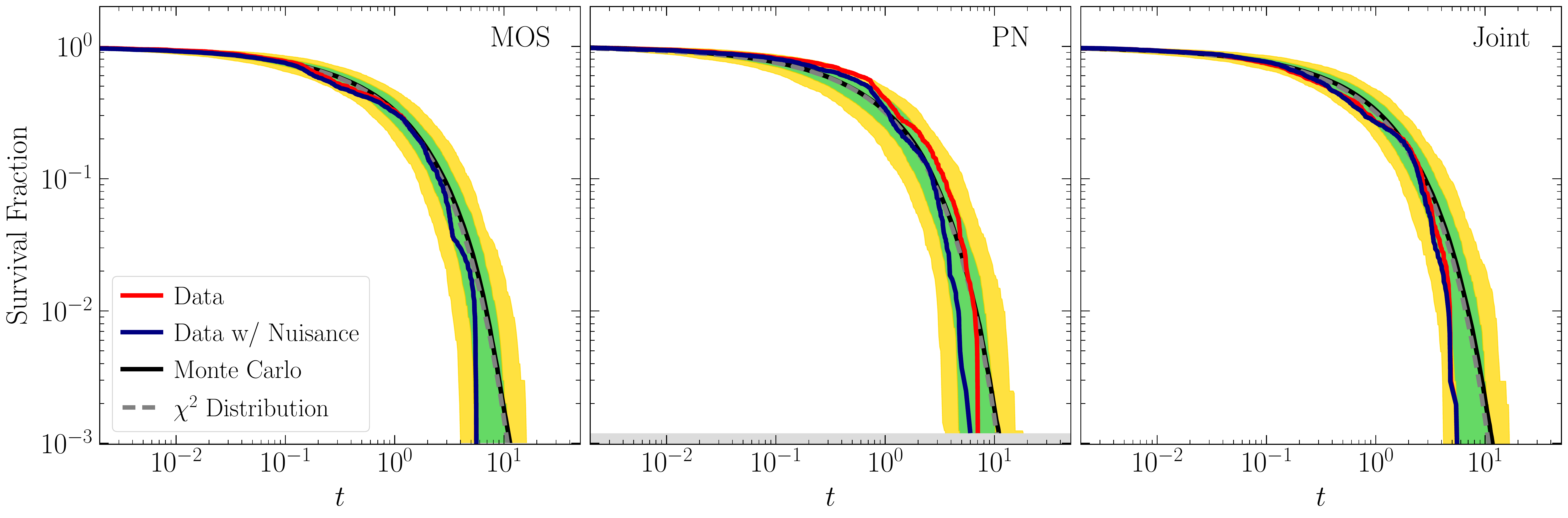}
\caption{(Left) The survival function of the test statistic for discovery in the analysis of the MOS data. Under the null hypothesis, and for a large number of samples, the survival fractions are expected to follow the $\chi^2$ distribution, as verified by MC (as labeled).  At a finite number of samples the expected chi-square distributions are found from MC to be expected to be contained within the green and gold shaded regions at 68\% and 95\% confidence, respectively. The negligible effect of the systematic nuisance parameter can be seen by comparing the survival function without the nuisance parameter (red, labelled ``Data") and with the nuisance parameter (blue, labeled ``Data w/ Nuisance Parameter"). (Center) As in the left panel, but for the PN analysis. (Right) The survival function for the joint analysis of MOS and PN data. In blue, the survival function for the joined PN and MOS analysis without systematic nuisance parameters; in red, the survival function for the joint analysis when the PN and MOS results are corrected by their independently-tuned systematic nuisance parameters prior to joining.
}
\label{fig:Joined_Survival}
\end{figure*}

It is interesting to consider the ensemble of discovery TSs in favor of the DM model across all tested mass points.  We denote this distribution of TSs without the spurious-signal by ${\bf T}$, while with the inclusion of the spurious-signal nuisance parameter we call this distribution ${\bf T}_{\rm sys}$.  We expect ${\bf T}_{\rm sys}$ to have fewer high-TS points than ${\bf T}$.
In the left and center panels of Fig.~\ref{fig:Joined_Survival} we illustrate the distributions of TSs for both ${\bf T}$ (labeled Data) and ${\bf T}_{\rm sys}$ (labeled Data w/ Nuisance Parameter) for MOS and PN, respectively.  More specifically, in that figure we illustrate the survival fractions for the distributions, which show the faction of TSs in ${\bf T}$ or ${\bf T}_{\rm sys}$ with a value above the TS indicated on the $x$-axis.  Asymptotically we expect, up to the caveat that we used the Bayesian marginal likelihood of the GP to define our TSs, that the TSs should be $\chi^2$ distributed~\cite{Cowan:2010js}.  The survival function of the $\chi^2$ distribution is shown in Fig.~\ref{fig:Joined_Survival}. We verify with a large number of MC simulations that the that the null-distribution of TSs is indeed $\chi^2$-distributed for both MOS and PN datasets.  The results of these tests are labeled ``Monte Carlo" in Fig.~\ref{fig:Joined_Survival} and overlap with the $\chi^2$ distribution, providing evidence that we are in the asymptotic regime~\cite{Cowan:2010js}.

\begin{figure}[htb]
\includegraphics[width = 0.98\textwidth]{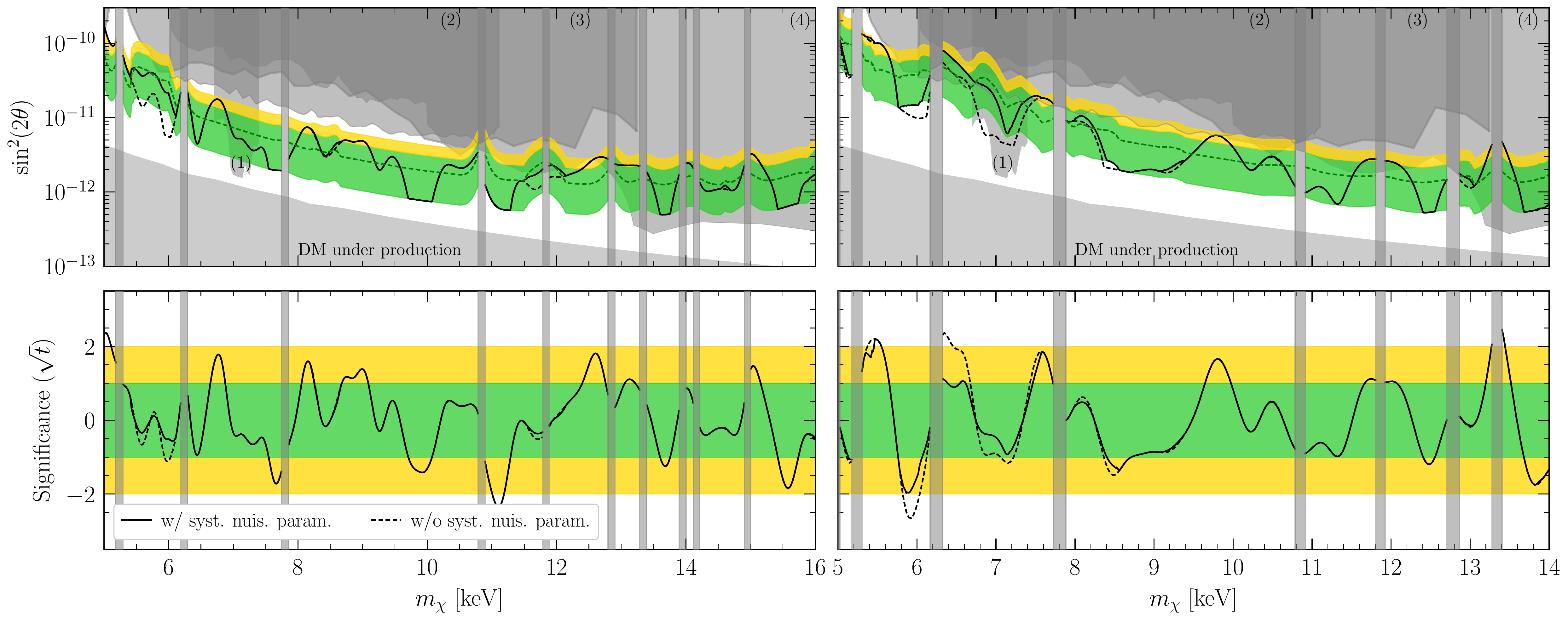}
\caption{As in Fig.~\ref{fig:final_limit}, but for the MOS (left panel) and PN (right panel) analyses individually and with and without the spurious-signal nuisance parameter.  The 1$\sigma$ and 2$\sigma$ expectations are shown only for the case with the spurious-signal nuisance parameter.  The limits without the nuisance parameter are slightly stronger at low masses. {The sharp variations in the expected sensitivity, especially visible in the PN results, arise from how the spurious-signal hyperparameter is determined through the sliding window procedure.} } 
\label{fig:final_limit_nuis}
\end{figure}

Because there are a finite number of samples in ${\bf T}$ and locations spaced within the detector energy resolution are correlated, the survival function for the observed data is not expected to follow the $\chi^2$-distribution exactly.  The green and gold bands in Fig.~\ref{fig:Joined_Survival} show the 68\% and 95\% containment regions for the survival fraction computed over 500 MC realizations of ${\bf T}$.  We expect that the data should fall within these bands if no signal is present, which is analogous to the green and gold bands for the significance in Fig.~\ref{fig:final_limit_nuis}.  In the left and center panels of Fig.~\ref{fig:Joined_Survival} we may see that the distributions of ${\bf T}$ for MOS and PN are broadly consistent with the MC expectations.  The distributions of ${\bf T}_{\rm sys}$, as expected, fall off slightly faster at large values of the TS.  The right panel of Fig.~\ref{fig:Joined_Survival} shows the survival fraction for the combined analysis where we combine the MOS and PN profile likelihoods, with and without the spurious-signal.   The most significant test point has a significance slightly above 2$\sigma$ local significance, which is less than 1$\sigma$ in global significance.  
Thus, we conclude that there is no evidence for decaying DM in our analysis above our 5$\sigma$ global predetermined detection threshold.  

The effect of the spurious-signal nuisance parameter on the individual MOS and PN limits is illustrated in Fig.~\ref{fig:final_limit_nuis}.  The inclusion of the nuisance parameter slightly decreases the discovery TSs at low masses and also causes a slight weakening of the limits.  Note that the expectations under the null hypothesis are indicated for the spurious-signal-corrected analysis in that figure.

\section{Extended Results}
\label{app:extended}

\begin{figure}[htb]
\includegraphics[width = 0.98\textwidth]{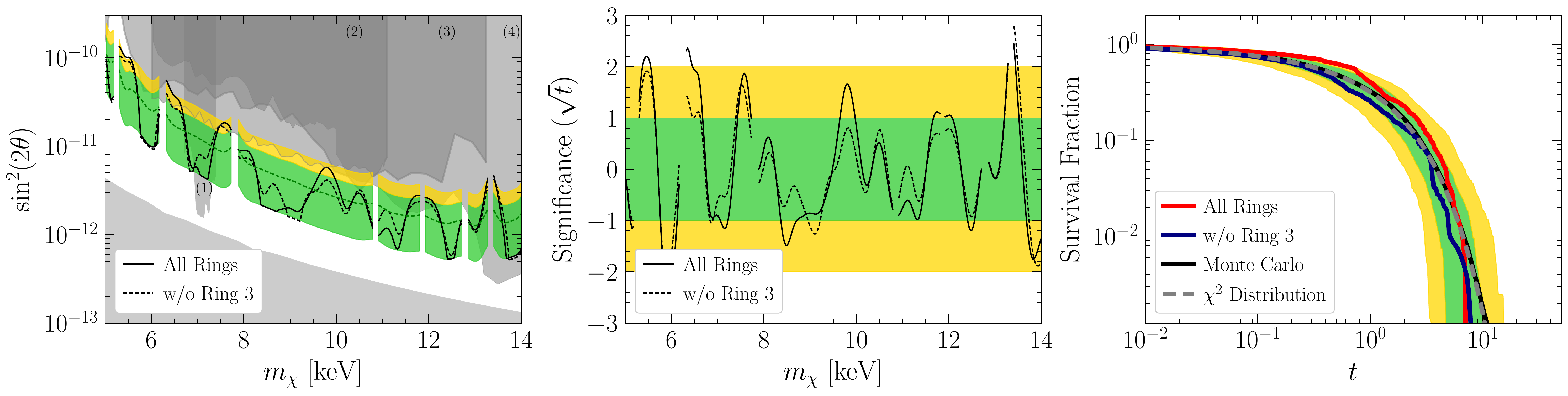}
\caption{{A comparison of all results obtained in the joint analysis of PN data with and without the inclusion of Ring 3, which may be subject to systematic mismodeling. Note that for this comparison we do not profile over the spurious-signal nuisance parameter.}} 
\label{fig:PN_Ring3_Masking}
\end{figure}

In this section we present extended results for the analyses that go into producing Fig.~\ref{fig:final_limit}. {First, we provide a measure of the goodness-of-fit of our null model to the data, quantified through the $\chi^2$ per degree of freedom (dof), in each annulus for the PN and MOS data sets in Tab.~\ref{tab:GoF}.  Note that we also quote the $p$-value associated with the $\chi^2$ per dof, with smaller numbers indicating a worse null-model fit. We observe acceptable $p$-values ($p \gtrsim 0.1$) in all rings except for Ring 3 of the PN data set, which realizes a $p$-value associated with the $\chi^2/\mathrm{dof}$ of $p \approx 5.7 \times 10^{-6}$. We would not expect to observe a $p$-value this small in any of the 16 rings. 
For example, Fig.~\ref{fig:Nuisance_Params} shows some evidence for mild systematic uncertainties at low energies in the PN data, though these are captured through our spurious-signal formalism. 
We also note that there is some indication that the poor $\chi^2 / \mathrm{dof}$ in PN Ring 3 arises from statistical fluctuations on scales much smaller than the detector energy resolution; for example, down-binning that data set to bins of width 45 eV, which is still smaller by a factor of a few relative to the energy resolution across the full energy range, improves the $p$-value associated with the $\chi^2 / \mathrm{dof}$ to $p \approx 4 \times 10^{-3}$.
As an additional test, we compare the results obtained without the spurious-signal formalism in the joint analysis of the PN data with and without the inclusion of the data in Ring 3. These results are presented in Fig.~\ref{fig:PN_Ring3_Masking}, and are qualitatively unchanged by the inclusion or exclusion of the PN Ring 3 data set.} We also provide best-fit normalizations for our GP kernels, presented in Tab.~\ref{tab:GP_Fit}. 

\begin{figure*}[htb]
\includegraphics[width = .65\textwidth]{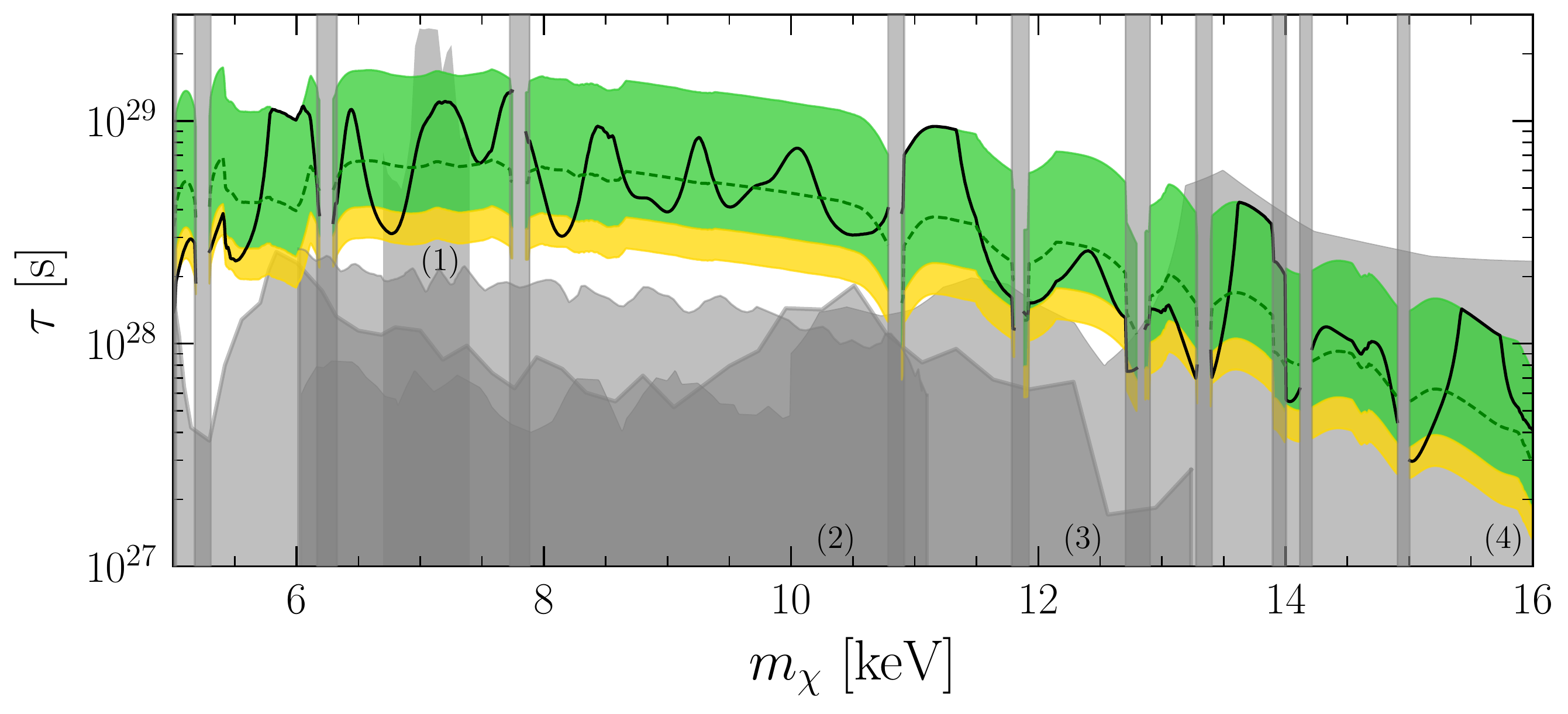}
\caption{As in Fig.~\ref{fig:final_limit}, but interpreted as limits on the DM lifetime. This figure applies for DM whose decays produce a single mono-energetic photon at energy $m_\chi/2$.  If the DM decay produces two photons (as in an axion model), then the lifetime limits are twice as strong. }
\label{fig:Lifetime}
\end{figure*}

In Fig.~\ref{fig:Lifetime} we present the main result in Fig.~\ref{fig:final_limit} in terms of the DM lifetime instead of in terms of $\sin^2(2\theta)$.  The result in Fig.~\ref{fig:Lifetime} is more general than in Fig.~\ref{fig:final_limit} since it holds for more general DM models beyond the sterile neutrino model.  Note, however, that this figure applies to DM whose decays produces one mono-energetic photon at energy $m_\chi/2$.  Axion-like models produce two photons during the decay, in which case the limits are twice as strong as those shown in Fig.~\ref{fig:Lifetime}. 

\begin{table}[htb]
\ra{1.3}
\centering
\begin{tabularx}{\textwidth}{p{0.15\textwidth}*{8}{P{0.095\textwidth}}}
\hline
  Instrument & Ring 1 & Ring 2 & Ring 3 & Ring 4 & Ring 5 & Ring 6 & Ring 7 & Ring 8  \\ \hline
  MOS [$\chi^2/\mathrm{dof}]$ & 1133.6/1093 & 1069.6/1096 & 1190.7/1091 & 1114.2/1096 & 1157.7/1098 & 1073.2/1097 & 1083.4/1095 & 1100.1/1095 \\ \hline
  MOS [$p$-value] & 0.19 & 0.71 & 0.02 & 0.34 & 0.10 & 0.69 & 0.59 & 0.45 \\ \hline
  PN [$\chi^2/\mathrm{dof}]$ & 860.7/894 & 845.2/897  & 1091.8/894 & 873.2/893 & 915.5/899 & 920.8/897 & 838.1/897 & 874.5/896  \\ \hline
  PN [$p$-value] & 0.77 & 0.89  & $5.7 \times 10^{-6}$ & 0.65 & 0.34 & 0.25 & 0.92 & 0.68  \\ \hline
\end{tabularx}
\caption{{The goodness-of-fit of the null model fit in each annulus for PN and MOS data sets as measured by the $\chi^2$ divided by the number of degrees of freedom (dof). With the exception of Ring 3 of the PN data set, this measure indicates an acceptable goodness-of-fit to the data under the null, as quantified through the $p$-value (see text for details).}}
\label{tab:GoF}
\end{table}

\begin{table}[htb]
\ra{1.3}
\centering
\begin{tabularx}{\textwidth}{p{0.10\textwidth}*{8}{P{0.105\textwidth}}}
\hline
 Instrument & Ring 1 & Ring 2 & Ring 3 & Ring 4 & Ring 5& Ring 6  & Ring 7 & Ring 8  \\ \hline
  MOS & $6.9 \times 10^{-3} $ & $1.6 \times 10^{-3} $ & $1.1 \times 10^{-3} $ & $3.4 \times 10^{-3} $ & $6.6 \times 10^{-4} $ & $1.9 \times 10^{-3} $ & $9.0 \times 10^{-4} $ & $1.5 \times 10^{-3} $  \\
  PN & $2.3 \times 10^{-2}$ & $5.3 \times 10^{-3}$ & $5.7 \times 10^{-3}$ & $9.1 \times 10^{-3}$ & $2.0 \times 10^{-3}$ & $4.2 \times 10^{-3}$  & $2.0 \times 10^{-4}$ & $1.1\times10^{-2}$  \\ \hline
\end{tabularx}
\caption{The best fit normalization of the GP kernel for each ring in PN and MOS. We present $\sqrt{A_\mathrm{GP}}$ in units of photons/cm$^{2}$/s/keV for $A_\mathrm{GP}$ defined in~\eqref{eq:kernel}. }
\label{tab:GP_Fit}
\end{table}

\begin{figure*}[htb]
\includegraphics[width = .95\textwidth]{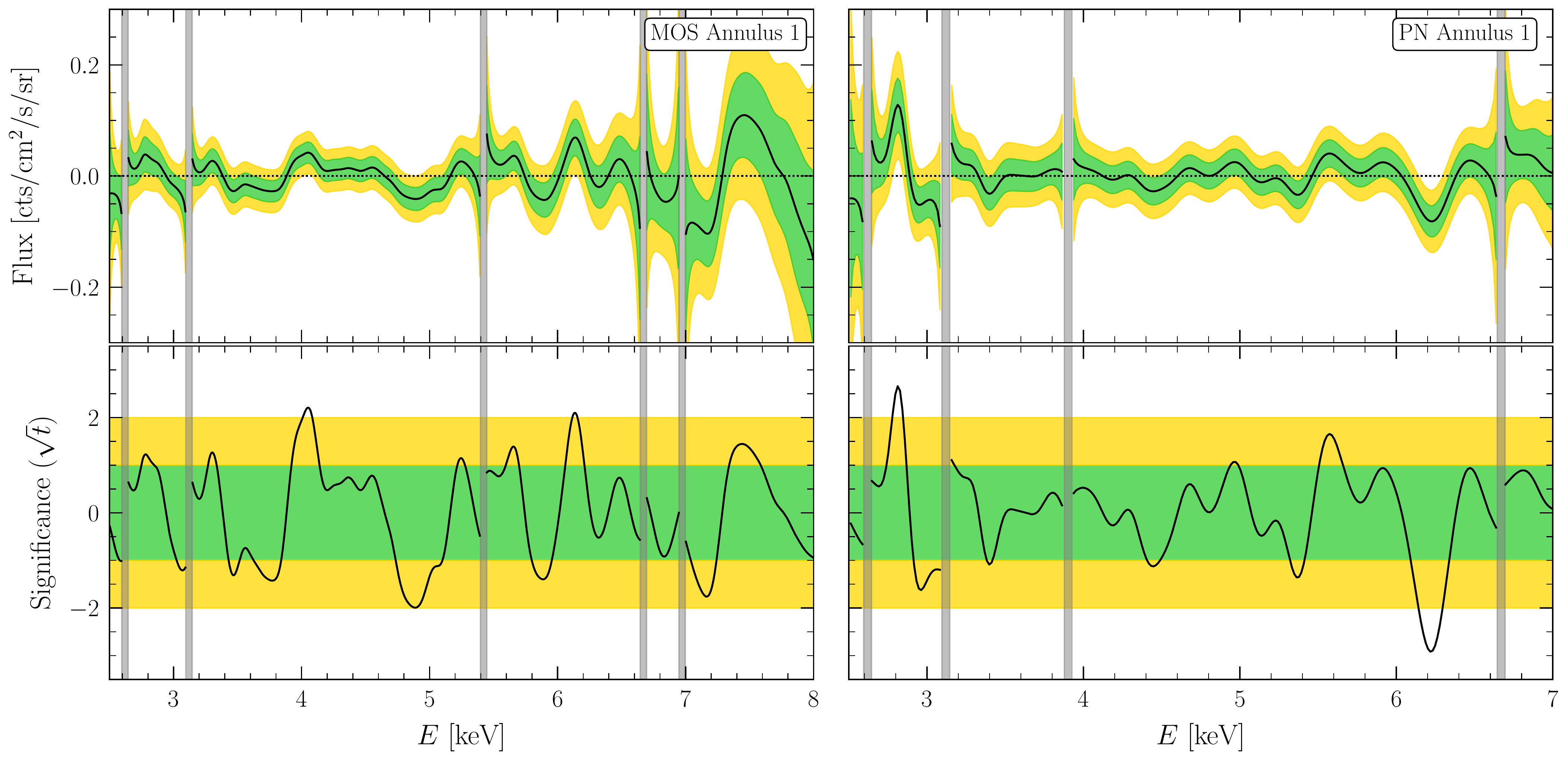}
\caption{
(Upper Left) The best-fit signal flux, and $1$ and $2\sigma$ uncertainties, as a function of the central UXL energy across our full energy range for the innermost MOS ring.  (Lower Left) The corresponding significance in favor of the signal model, multiplied by the sign of the best fit UXL normalization at that energy, along with the 1/$2\sigma$ expectations under the null hypothesis.
(Right Panel) As in the left panel but for the innermost PN annulus.}
\label{fig:Ring_0}
\end{figure*}

\begin{figure*}[htb]
\includegraphics[width = .95\textwidth]{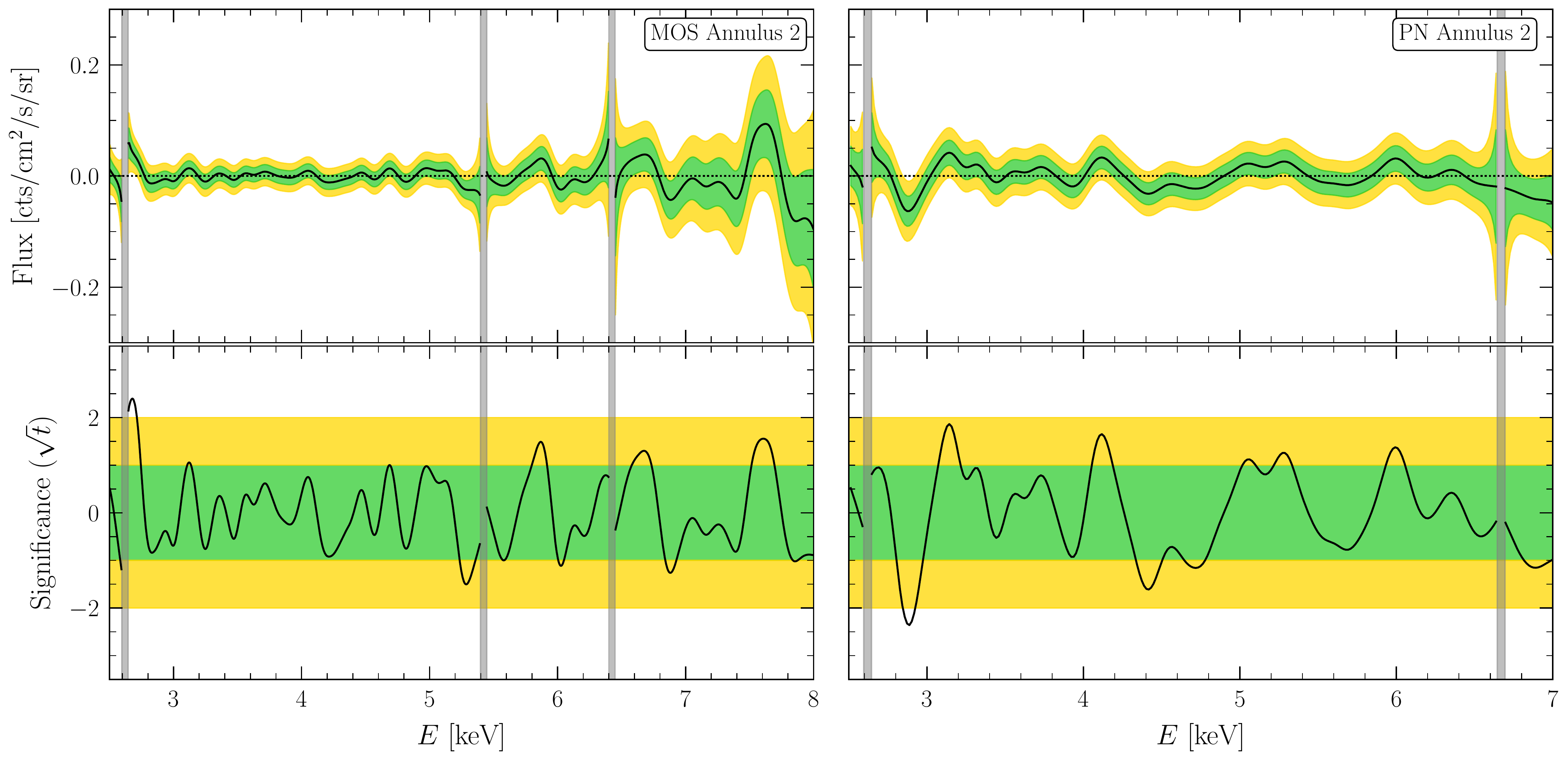}
\caption{As in Fig.~\ref{fig:Ring_0} but for annulus 2.}
\label{fig:Ring_1}
\end{figure*}

\begin{figure*}[htb]
\includegraphics[width = .95\textwidth]{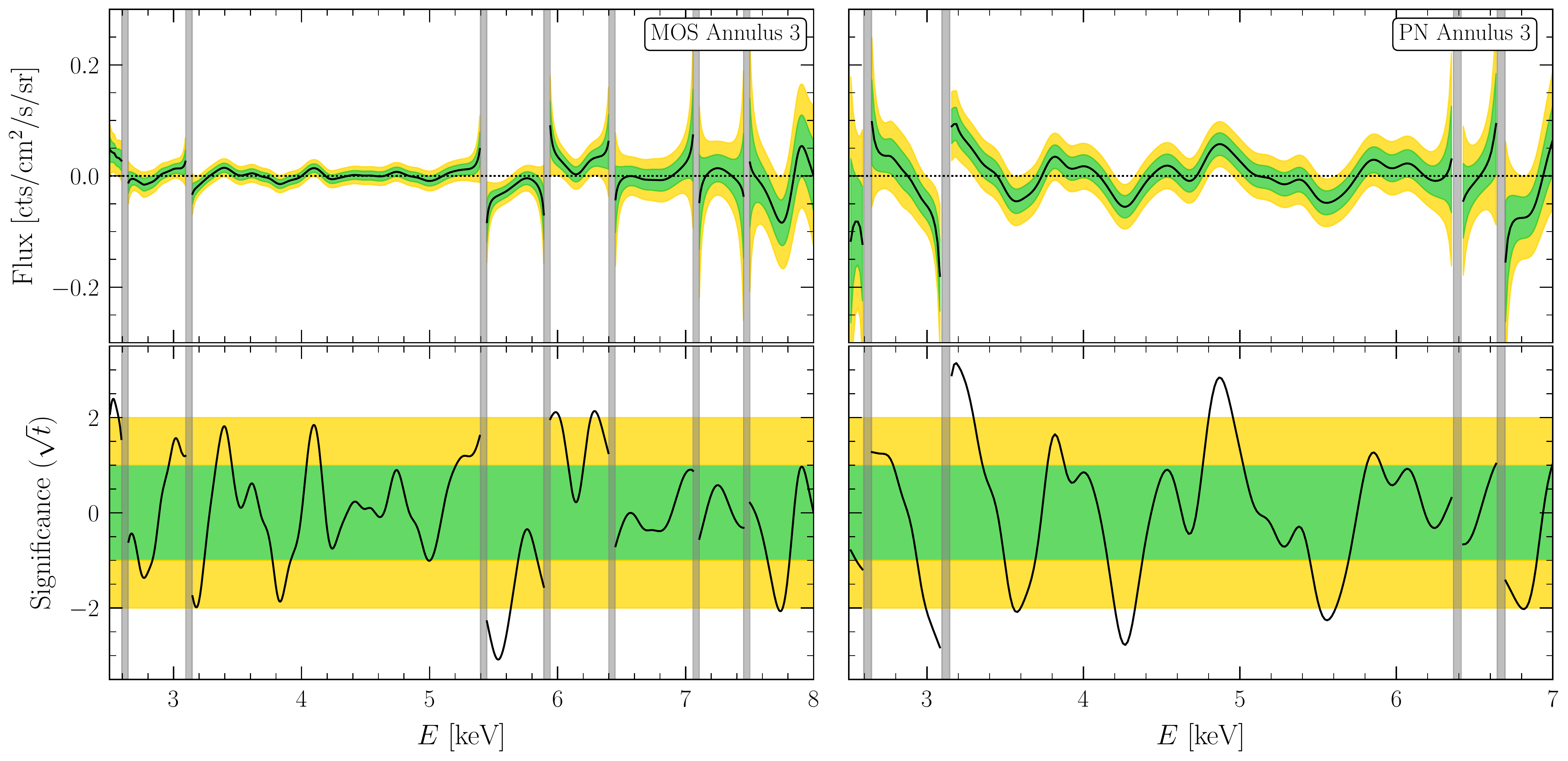}
\caption{As in Fig.~\ref{fig:Ring_0} but for annulus 3.}
\label{fig:Ring_2}
\end{figure*}

\begin{figure*}[htb]
\includegraphics[width = .95\textwidth]{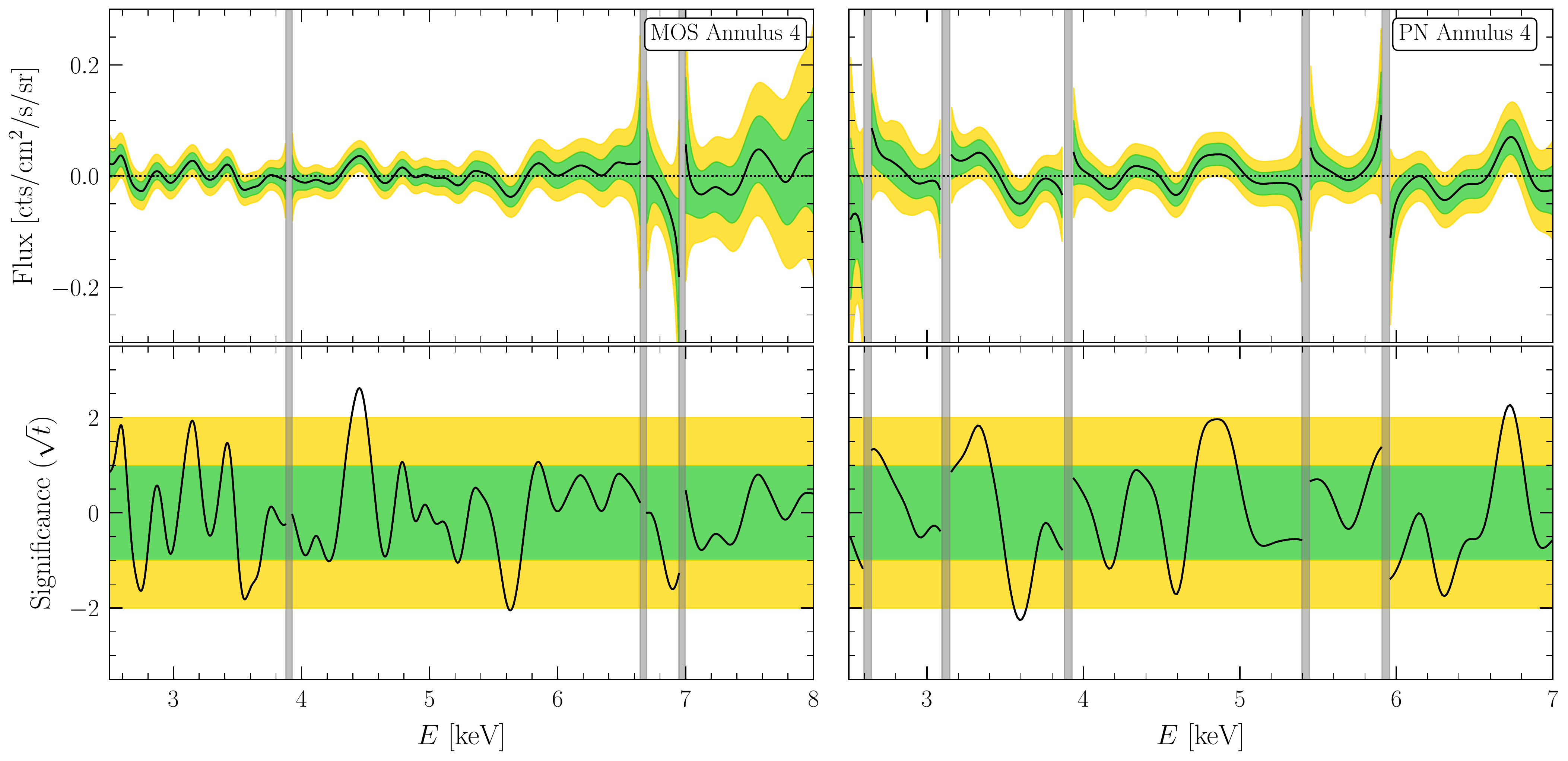}
\caption{As in Fig.~\ref{fig:Ring_0} but for annulus 4.}
\label{fig:Ring_3}
\end{figure*}

\begin{figure*}[htb]
\includegraphics[width = .95\textwidth]{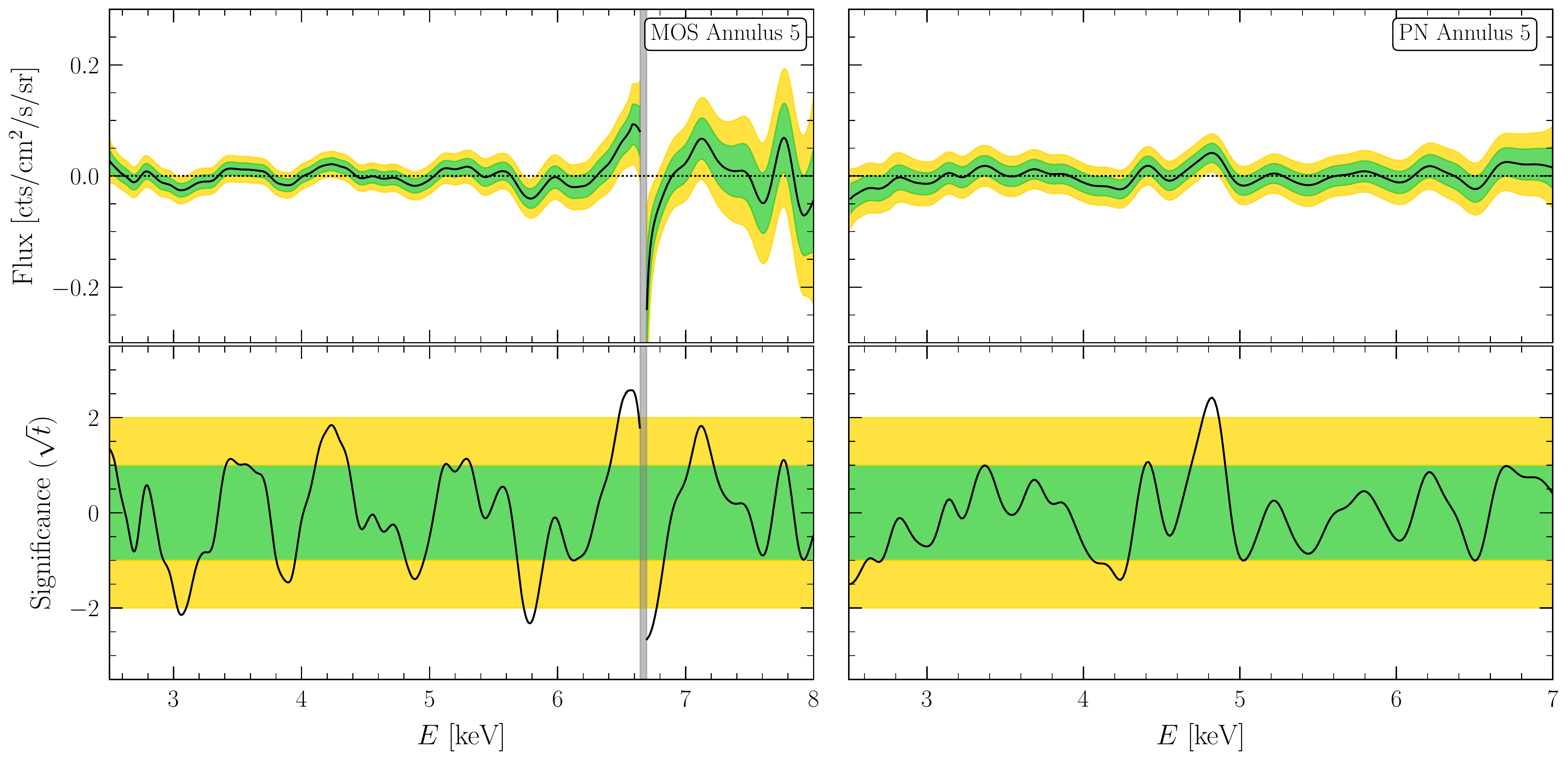}
\caption{As in Fig.~\ref{fig:Ring_0} but for annulus 5.}
\label{fig:Ring_4}
\end{figure*}

\begin{figure*}[htb]
\includegraphics[width = .95\textwidth]{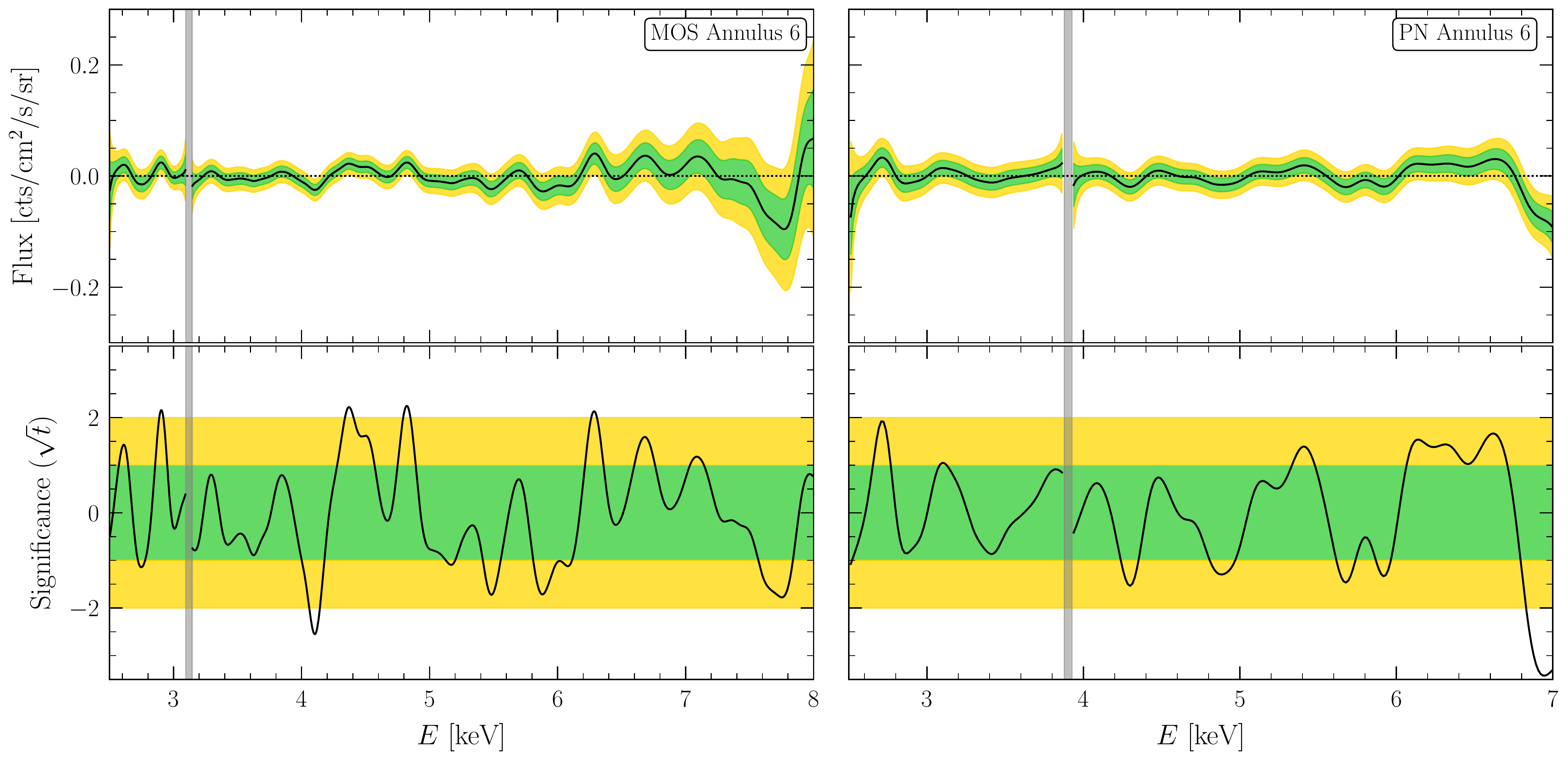}
\caption{As in Fig.~\ref{fig:Ring_0} but for annulus 6.}
\label{fig:Ring_5}
\end{figure*}

\begin{figure*}[htb]
\includegraphics[width = .95\textwidth]{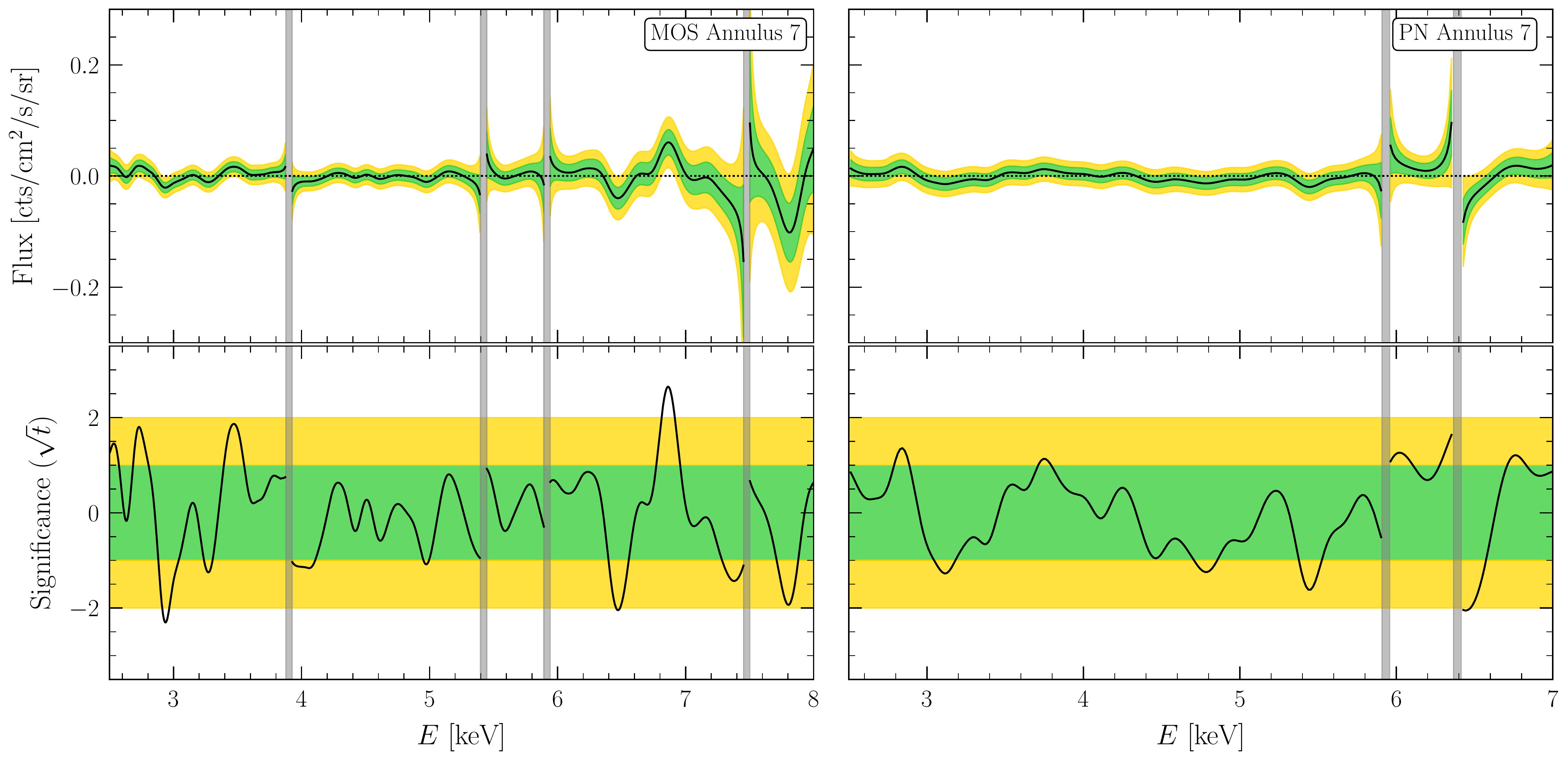}
\caption{As in Fig.~\ref{fig:Ring_0} but for annulus 7.}
\label{fig:Ring_6}
\end{figure*}

\begin{figure*}[htb]
\includegraphics[width = .95\textwidth]{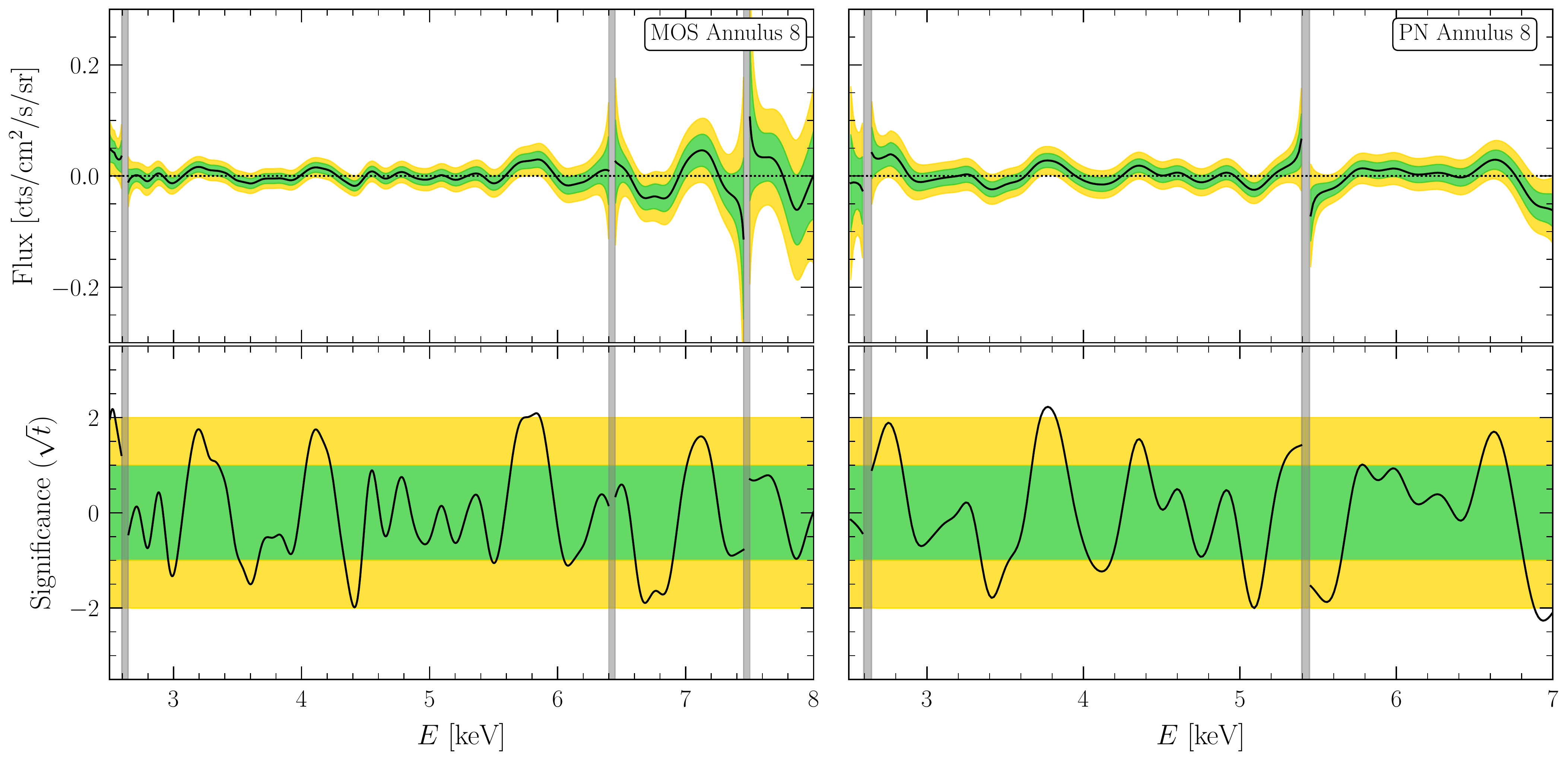}
\caption{As in Fig.~\ref{fig:Ring_0} but for annulus 8.}
\label{fig:Ring_7}
\end{figure*}

Next, we show our results from the analyses to the individual MOS and PN annuli.  
 In Figs.~\ref{fig:Ring_0} through~\ref{fig:Ring_7} we show the best-fit fluxes and significances (times the sign of the excess or deficit) for the UXLs for all of the annuli and for both MOS and PN.  Note that the shaded grey regions denote the masks that we use to avoid searching for UXLs in the direct vicinity of background lines included in the analyses.

\begin{figure*}[htb]
\includegraphics[width = .95\textwidth]{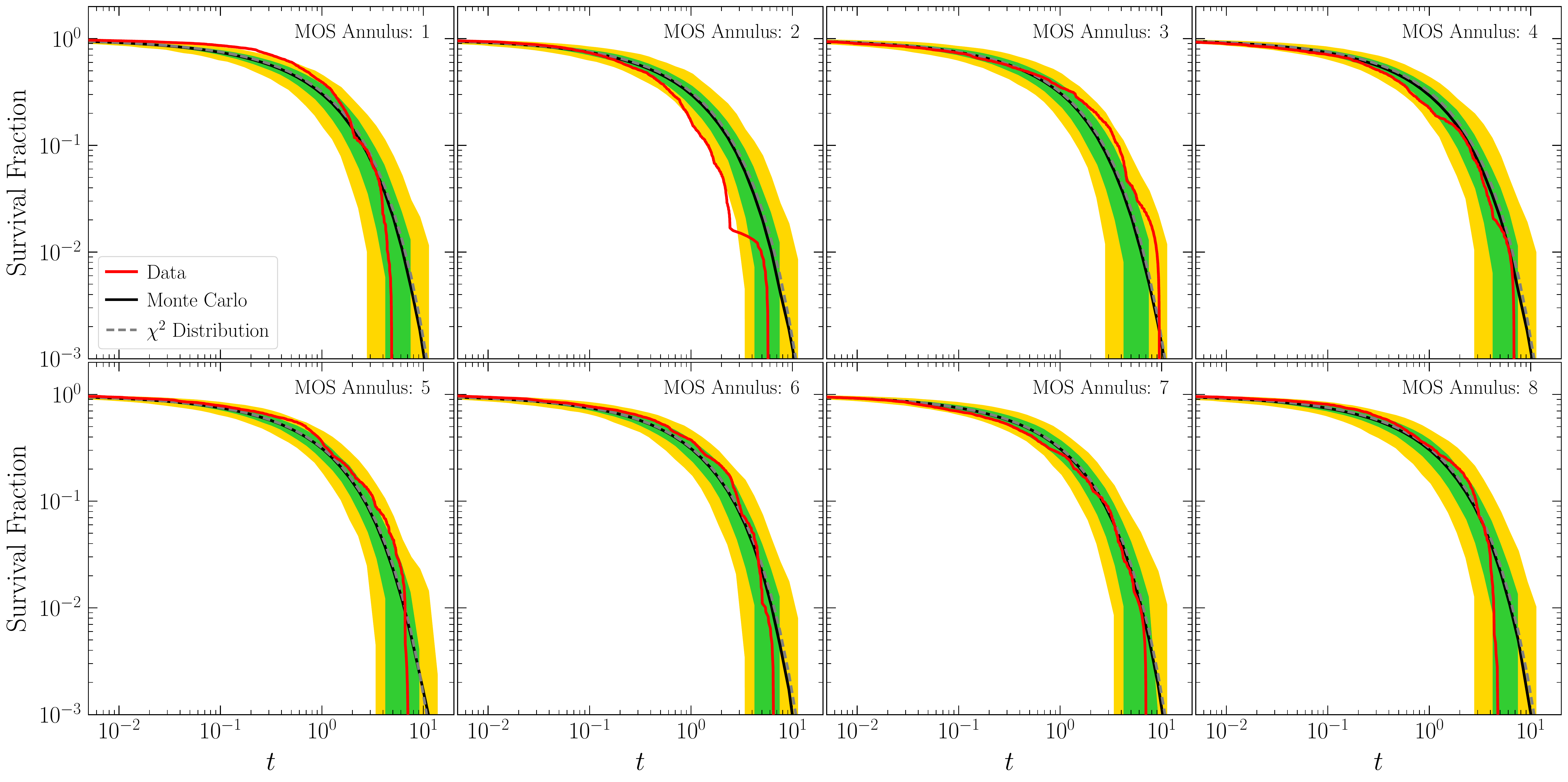}
\caption{As in Fig.~\ref{fig:Joined_Survival} but for the individual MOS annuli.  Note that the systematic nuisance parameter has not been applied since that is only incorporated in the joint likelihood that combines the results from the individual annuli.}
\label{fig:MOS_Ring_Survival}
\end{figure*}

\begin{figure*}[htb]
\includegraphics[width = .95\textwidth]{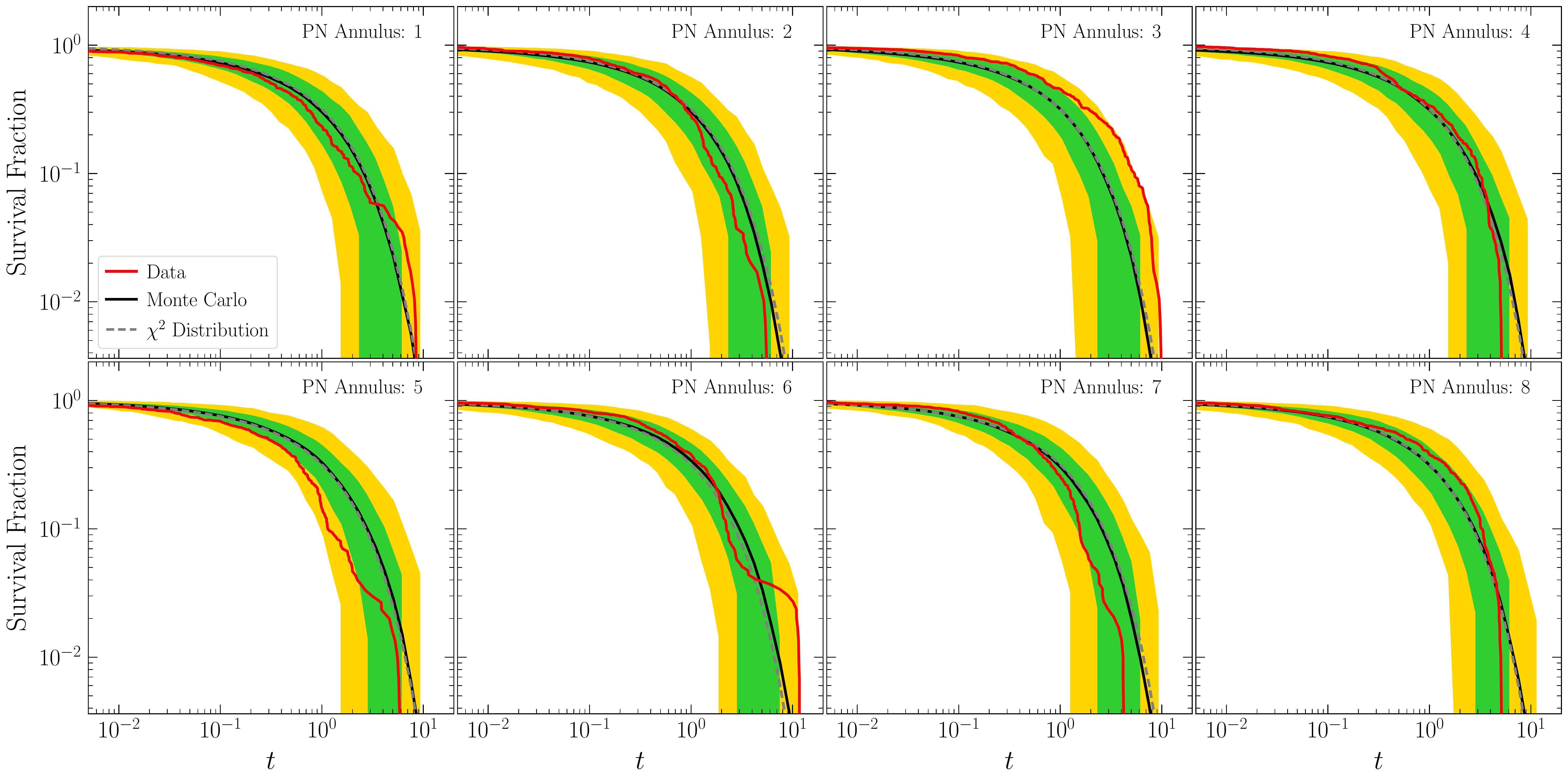}
\caption{As in Fig.~\ref{fig:MOS_Ring_Survival} but for the PN data sets.}
\label{fig:PN_Ring_Survival}
\end{figure*}

The distribution of discovery TSs that we observe in the individual annuli all appear consistent with expectations from MC, as illustrated in Fig.~\ref{fig:MOS_Ring_Survival} for MOS and Fig.~\ref{fig:PN_Ring_Survival} for PN.  These figures illustrate the survival fractions of TSs, as in Fig.~\ref{fig:Joined_Survival}, but at the level of the individual annuli instead of the joint analysis. Note that the MC expectations are constructed independently for each annulus and each data set.  These results do not include the systematic nuisance parameter since that is only included at the level of the joint likelihood, after combining the results from all of the individual annuli.  

\section{Synthetic signal tests for the fiducial analysis}
\label{app:injection}

In this section we verify that our analysis framework has the ability to discover real signals if they are present in the data.  We do so by injecting a synthetic signal into the real data and analyzing the hybrid data set with our full analysis. We also demonstrate the full analysis as applied to fully synthetic data generated with varying injected signal strengths.

\subsection{Signal injection in real data}
\label{sec:sig_inj_real}

\begin{figure*}[htb]
\includegraphics[width = .95\textwidth]{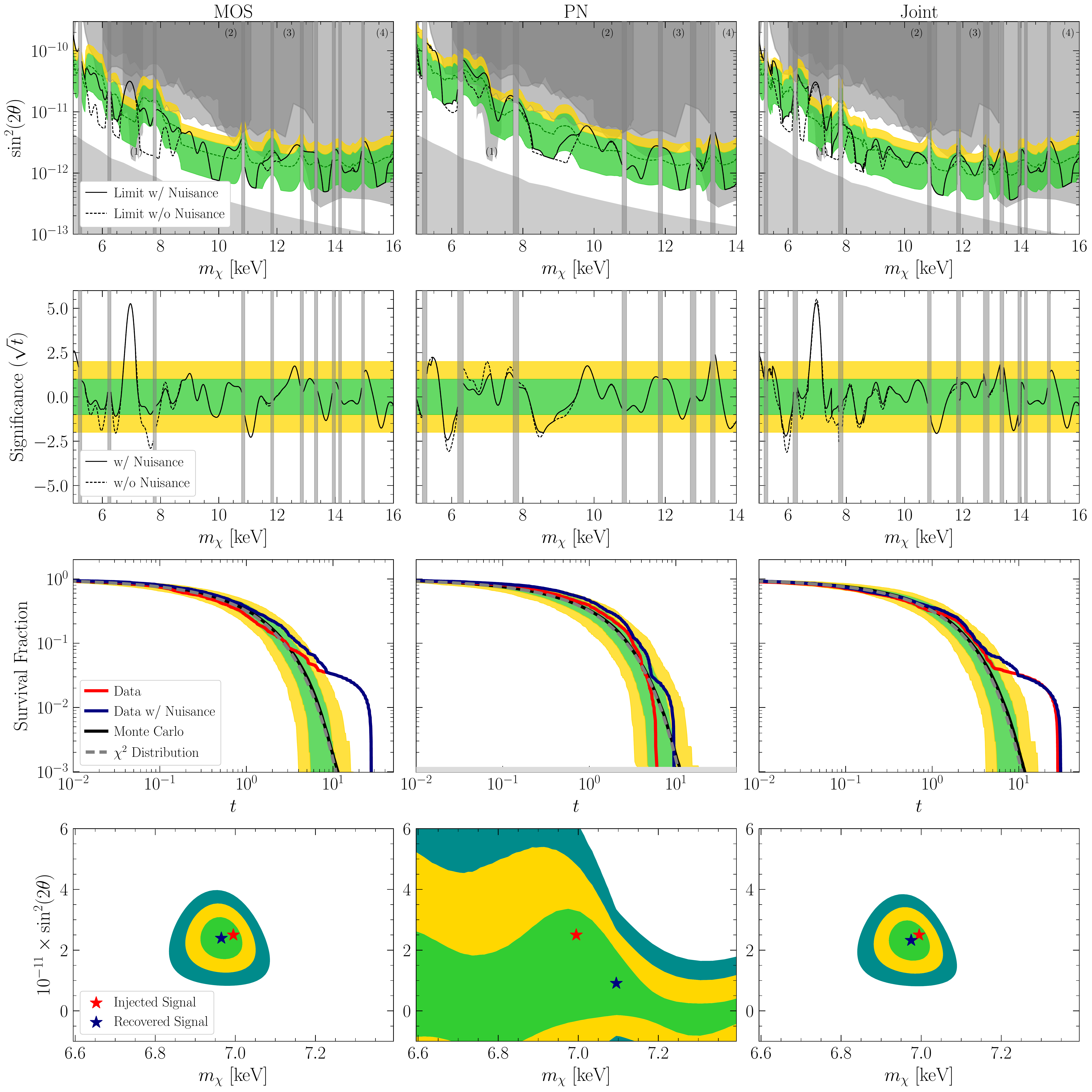}
\caption{The results of the analysis of the hybrid data that consists of the real MOS and PN data plus a synthetic DM signal.  The DM signal is generated with mass $m_\chi = 7.0$ keV and mixing angle $\sin^2(2\theta) = 2.5 \times 10^{-11}$ as described in the text.  The top, middle, and third rows are analogous to Figs.~\ref{fig:Joined_Survival} and~\ref{fig:final_limit_nuis}, but for the hybrid data set.  The last row shows the 1, 2, and 3 $\sigma$ recovered parameter space for the signal in the mass and mixing angle plane.  The best-fit recovered signal is indicated in dark blue, while the red star denotes the true value injected.  The synthetic signal is appropriately recovered, adding confidence that our analysis procedure has the ability to detect real DM signals if present in the data. }
\label{fig:Real_Injection}
\end{figure*}

For injection tests in the real data, we chose a DM mass $m_\chi = 7.0$ keV and a mixing angle $\sin^2(2\theta) = 2.5 \times 10^{-11}$.  We chose this mixing angle because we expect such a signal to be detected at approximately $5 \sigma$ significance.  We forward model this signal through the appropriate MOS and PN detector responses, draw Poisson counts, and then add these counts to the actual data sets.  The results of the data analysis of the hybrid data are illustrated in Fig.~\ref{fig:Real_Injection}. In the top panel we show the 95\% upper limits for MOS, PN, and the joint analysis, with and without the systematic nuisance parameter.  Note that the injected signal is indicated by the red star.  The upper limits weaken at the injected signal point, as expected, and do not exclude the injected signal coupling.  In the second row we show the corresponding detection significances.  The signal is detected at nearly $5\sigma$ in MOS alone and at around 2$\sigma$ in PN.  The systematics nuisance parameter slightly reduces the significance of the discovery, but by a minimal amount since we mask a 0.4 keV window around the test mass when determining the systematics nuisance parameter.  In the third row we show how the discovery of the injected signal extends the survival function to higher TS values.  Lastly, in the bottom row we show the 1, 2, and 3 $\sigma$ best-fit regions in the $m_\chi$-$\sin^2(2 \theta)$ plane for the DM candidate.  In red we mark the location of the injected signal, which is recovered appropriately.

\subsection{Signal injection in synthetic background data}

\begin{figure*}[htb]
\includegraphics[width = .95\textwidth]{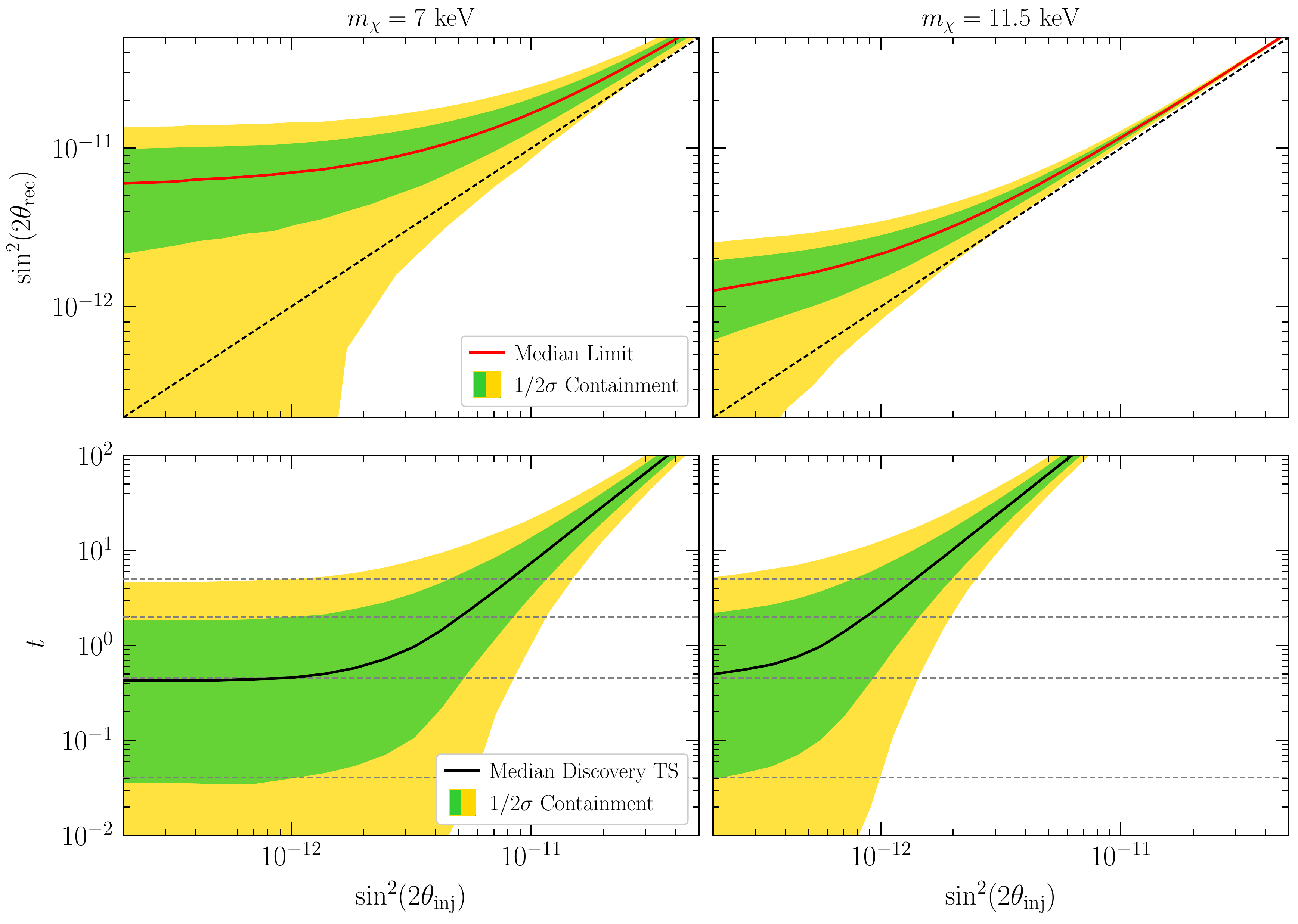}
\caption{(\textit{Top Row}) In red, the median 95$^\mathrm{th}$ percentile upper limit on the recovered signal as a function of the injection signal strength at two neutrino masses evaluated on synthetic data. We additionally indicate the $1$ and $2\sigma$ containment intervals for the ensemble of upper limits realized at each injected signal strength.  Note that these upper limits are not power constrained. These results demonstrate that our analysis framework places robust upper limits that do not rule out an injected signal. (\textit{Bottom Row}) In black, the median recovered detection test statistic for a signal injected in the synthetic data as a function of the injected signal strength, with the $1$ and $2\sigma$ containment intervals also indicated. Under the null hypothesis, the detection test statistic should follow a $\chi^2$-distribution; the median and $1\sigma$ and $2\sigma$ percentile values of the $\chi^2$-distribution are indicated by dashed grey lines. These results demonstrate that our detection test statistic follows its theoretically expected distribution under the null hypothesis ($\sin^2(2\theta_{\rm inj}) = 0$) and that our analysis framework can robustly identify a signal which is present in the data. The results are smoothed with a Savitzky–Golay filter for clarity.}
\label{fig:Synthetic_Injection}
\end{figure*}

For injection tests on the real data, we first generate synthetic data according to the best-fit null models for each of the eight rings studied in MOS and PN data sets. We then inject a synthetic signal at a specified value of $\sin^2(2 \theta)$ on top of the null-model realizations using the same procedure as applied for the signal injection on the real data and repeat our full analysis procedure in search of the injected signal with the exception that we do not apply a nuisance parameter tuning and correction. We perform 1000 independent realizations and analyses for each value of $\sin^2(2 \theta$, and we repeat this procedure for 30 values of $\sin^2(2 \theta)$ between $10^{-13}$ and $10^{-10}$ for two different neutrino masses: 7.0 keV and 11.5 keV. The results of the data analysis of the hybrid data are illustrated in Fig.~\ref{fig:Synthetic_Injection}. In the top row, we show ensemble statistics for the 95\% upper limits as a function of injected signal strength for the two neutrino masses studied in this test. In the bottom row, we show the ensemble statistics of the recovered detection test statistic as a function of the injected signal strength. The upper limits weaken with increasing injected signal strength without excluding the true value of the injected signal. Moreover, the detection test statistic smoothly increases as a function of increasing injected signal strength. Critically, at large injected signal strength, the test statistic safely exceeds $TS \approx 30$, which is the approximate threshold for a $5\sigma$ detection after correcting for the look-elsewhere effect.

\section{Systematic Analysis Variations}
\label{app:syst}

Our fiducial result, which is illustrated in Fig.~\ref{fig:final_limit}, made a number of physics-level and analysis-level choices.  These choices are justified in the main text and the supplementary results of the proceeding sections of the SM.  Still, it is worthwhile to consider how our results change for different physics and analysis assumptions and choices, as this gives an indication of the robustness of the limits and significances shown in Fig.~\ref{fig:final_limit}.

\subsection{Alternate DM Density Profiles}

\begin{figure*}[htb]
\includegraphics[width = .65\textwidth]{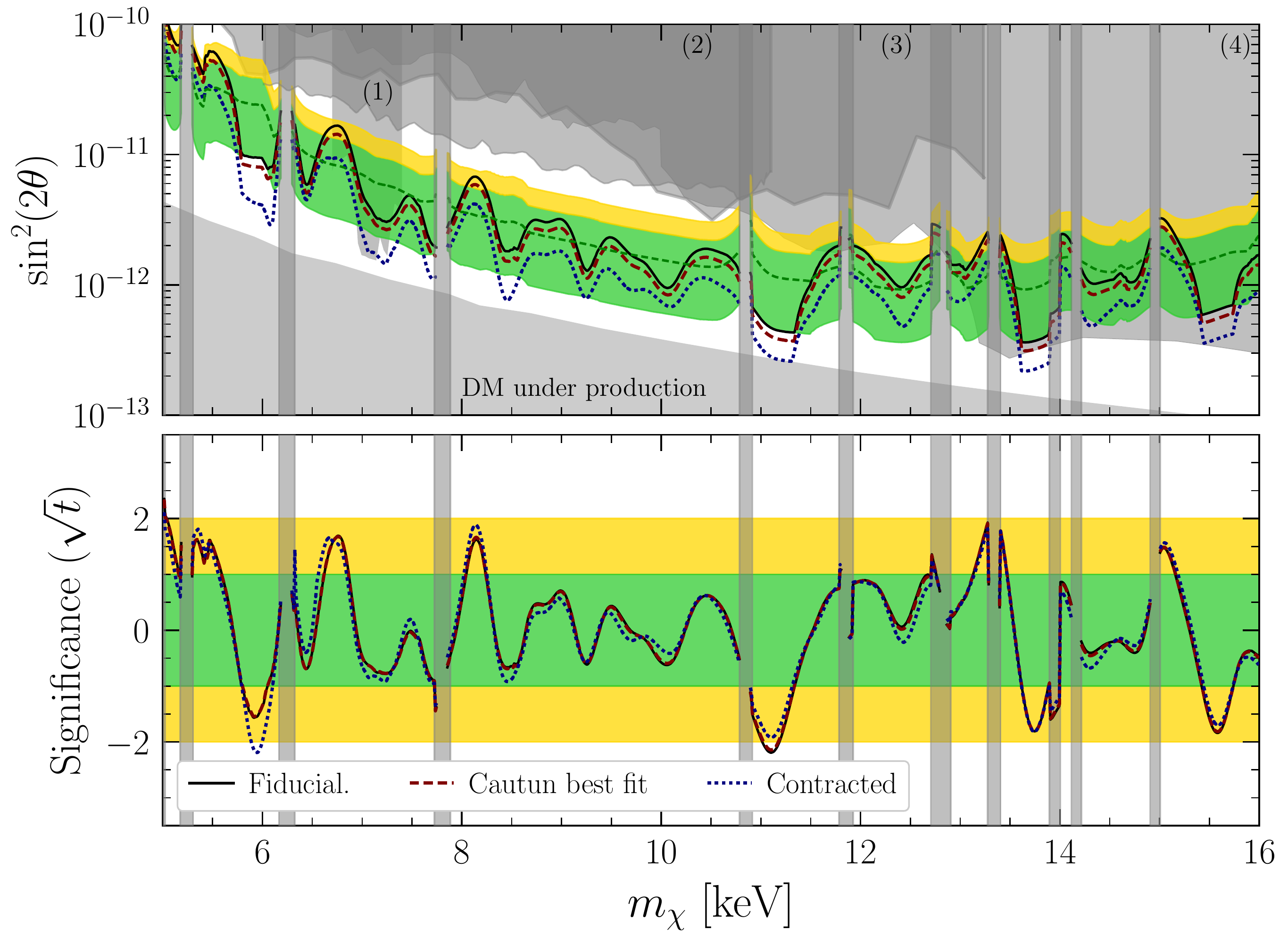}
\caption{As in Fig.~\ref{fig:final_limit}, but for three different DM density profiles, all based upon Ref.~\cite{2020MNRAS.494.4291C}.
In solid curve we show our fiducial results, corresponding to the uncontracted NFW profile with a conservative density.
The dashed curve then shows our results using the best fit NFW profile, whereas in dashed we show the stronger limits that would be obtained with a contracted DM distribution. Details of the distributions are provided in the text.}
\label{fig:All_Profiles}
\end{figure*}

In the main text, and in particular in Fig.~\ref{fig:final_limit}, we adopted the conservative DM profile that was shown in Fig.~\ref{fig:ROI_ill}.
As already described, the present expectation is that in the absence of baryons, the DM halo is well described by an NFW profile. Baryons are then expected to contract this profile, increasing the DM density towards the GC, and potentially also introducing a core on top of this. For our fiducial analysis we conservatively assumed an uncontracted NFW halo, using the most conservative parameters determined within the 68\% best fit region of \textcite{2020MNRAS.494.4291C}. In particular, we used an NFW profile with $r_s = 19.1$ kpc and normalized to a local DM density of $\rho_{\rm DM} = 0.29$ GeV/cm$^3$.

In Fig.~\ref{fig:All_Profiles}, we show our main results if instead we repeat the analysis for the best fit NFW profile determined in Ref.~\cite{2020MNRAS.494.4291C}, which corresponds to $r_s = 15.6$ kpc and $\rho_{\rm DM} = 0.31$ GeV/cm$^3$, as well as showing results for the more realistic contracted profile. There is not a parametric form for the contracted profile, however, Ref.~\cite{2020MNRAS.494.4291C} provides a best fit model for the DM mass distribution, which we use to infer the density and then $D$-factor. The model only provides an estimate down to 1 kpc from the GC, within which we conservatively assume the density profile is completely cored.

As the figure demonstrates, adopting a more realistic contracted DM profile strengthens our limits by roughly a factor of 2. Importantly, however, changing the profile does not appreciably change the distribution of significance, and we continue to see no clear evidence for an UXL.

\subsection{Dependence on the GP model}

For our fiducial analysis we use the GP kernel given in~\eqref{eq:kernel} with the choice $\sigma_E = 0.3$.  This choice was made so that the residual background model has the ability to adjust on scales around one order of magnitude larger in scale than the energy resolution of the detectors, which are $\delta E / E \sim 0.03$.  In this section we verify that our results do not depend in detail upon the particular value chosen.

\begin{figure*}[htb]
\includegraphics[width = .95\textwidth]{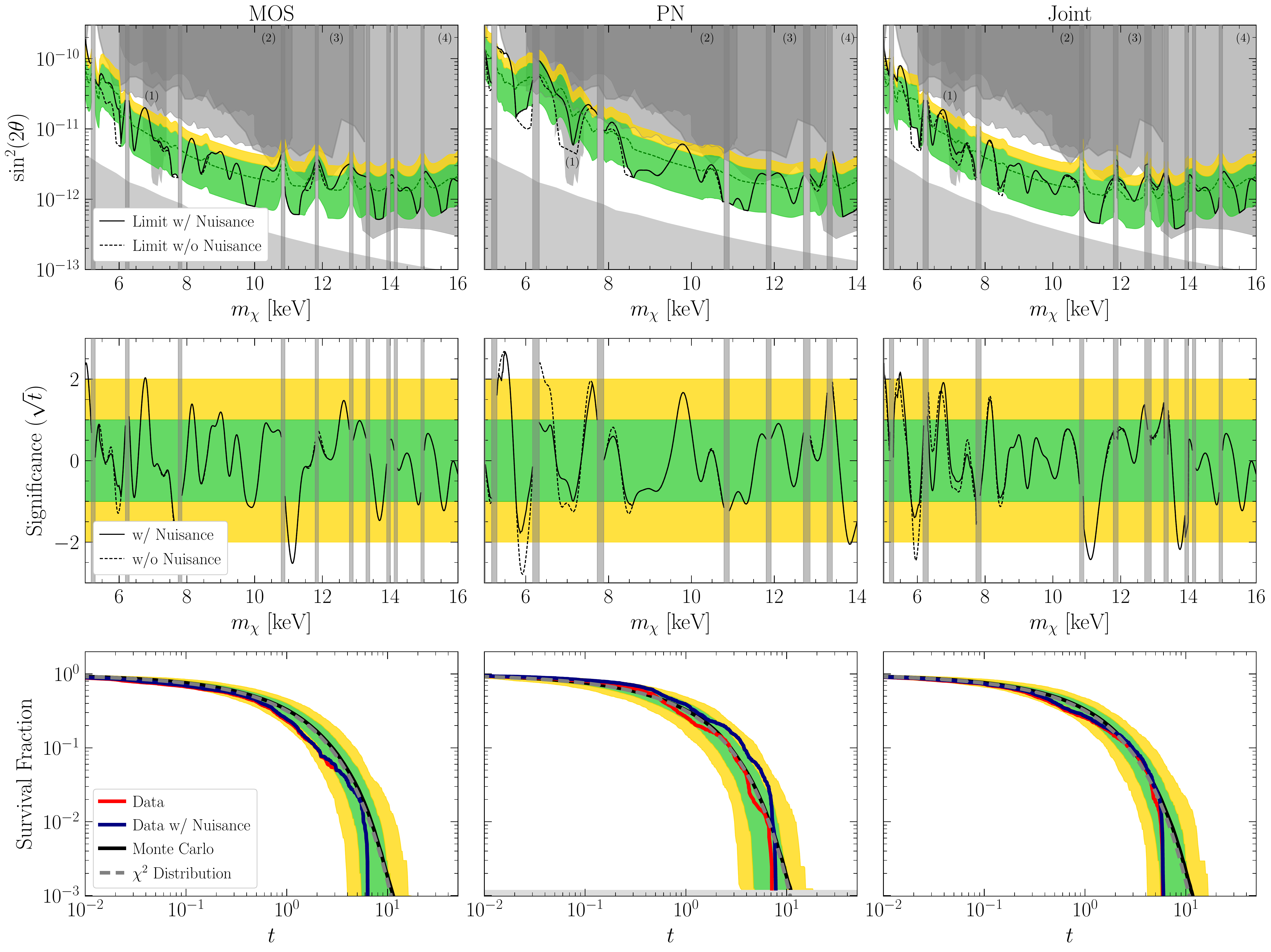}
\caption{The analogues of Figs.~\ref{fig:Joined_Survival} and~\ref{fig:final_limit_nuis}, but changing the kernel correlation length to $\sigma_E = 0.2$ (c.f. our fiducial value of $\sigma_E =0.3$).  The differences between the $\sigma_E = 0.2$ and $0.3$ results are minor.}
\label{fig:Relative_2e-1}
\end{figure*}

\begin{figure*}[htb]
\includegraphics[width = .95\textwidth]{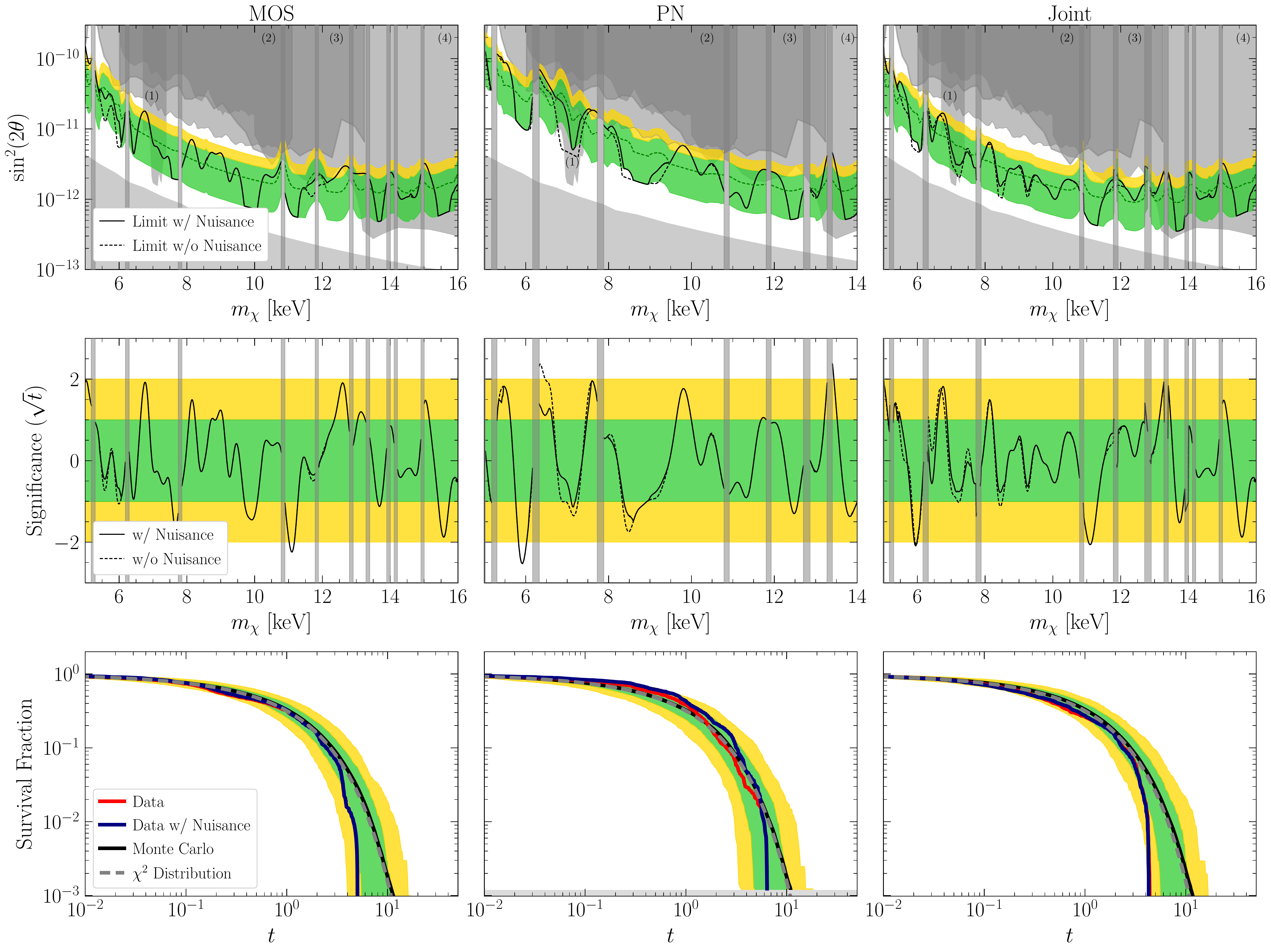}
\caption{As in Fig.~\ref{fig:Relative_2e-1} but with $\sigma_E = 0.4$. The limit is slightly strengthened, although again the differences are not significant.}
\label{fig:Relative_4e-1}
\end{figure*}

\begin{figure*}[htb]
\includegraphics[width = .95\textwidth]{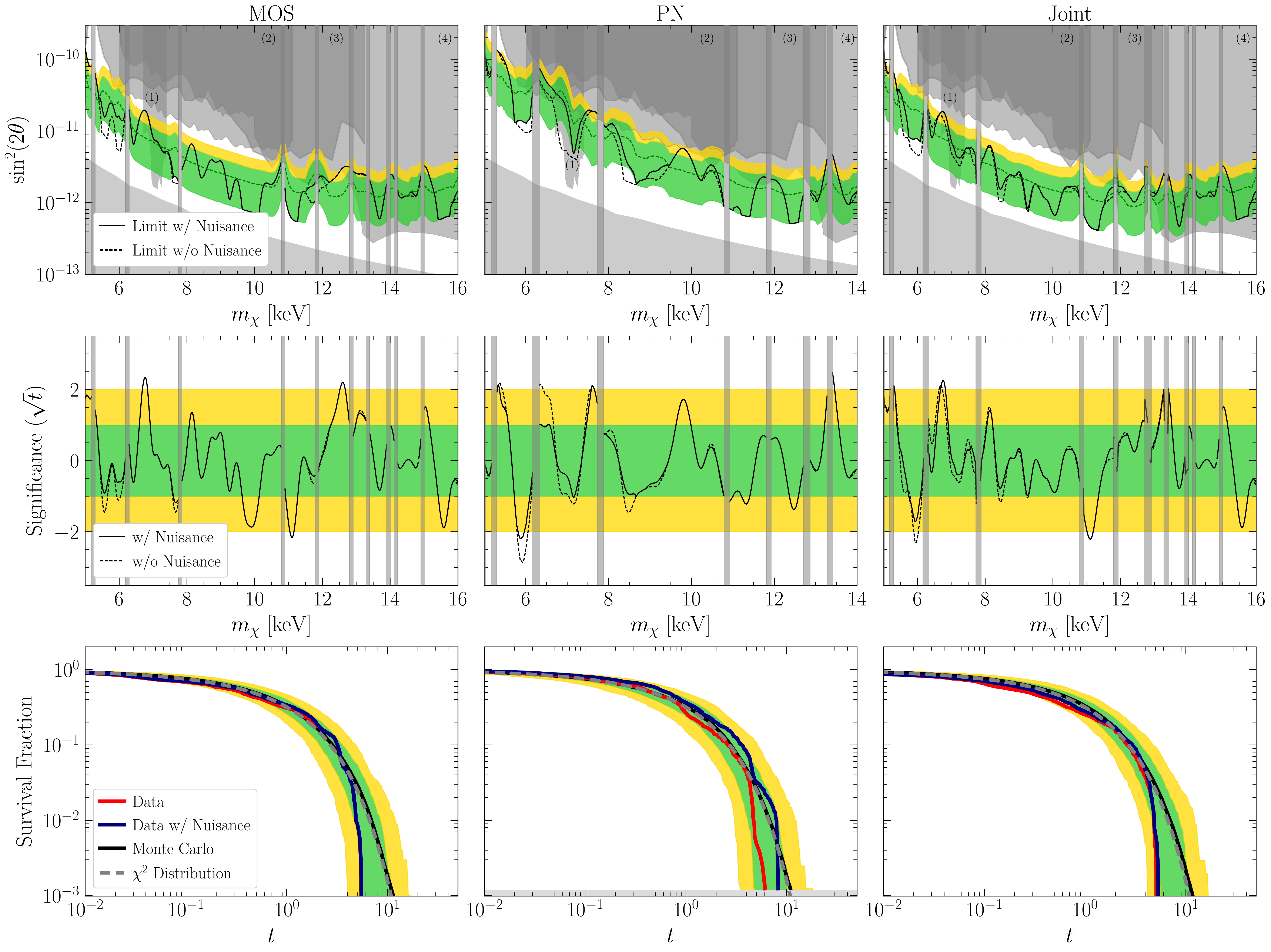}
\caption{{As in Fig.~\ref{fig:Relative_2e-1} but with $\sigma_E$ treated as a profiled nuisance parameter. The results demonstrate that even providing our background model this additional freedom does not have a significant impact on the limit.}}
\label{fig:Profiled_Scale}
\end{figure*}

First, we consider a small modification to our default analysis by taking  $\sigma_E = 0.2$ and $\sigma_E = 0.4$.  The results of these analyses are shown in Figs.~\ref{fig:Relative_2e-1} and~\ref{fig:Relative_4e-1}.  {As a further modification of our GP modeling procedure, we repeat our analysis with the relative scale of our kernel promoted to a nuisance parameter that we independently profile in each annulus in both instruments between the range of $0.15$ and $0.9$. We report the resulting best-fit GP scales in Tab.~\ref{tab:GPScale}. Results for this analysis are shown in Fig.~\ref{fig:Profiled_Scale}.}  In those figures we show the 95\% upper limits (upper panel), significances (middle panel), and survival fraction of significances (bottom panel).  We give the results both with an without nuisance parameters.  There is a slight trend where increasing $\sigma_E$ leads to a corresponding strengthening of the sensitivity, though this difference is minor compared to other choices in the analysis. In general, the results appear robust to the choice of $\sigma_E$.

\begin{table}[htb]
\ra{1.3}
\centering
\begin{tabularx}{\textwidth}{p{0.15\textwidth}*{8}{P{0.095\textwidth}}}
\hline
  Instruments & Ring 1 & Ring 2 & Ring 3 & Ring 4 & Ring 5 & Ring 6 & Ring 7 & Ring 8  \\ \hline
  MOS & 0.60 & 0.90 & 0.90 & 0.34 & 0.81 & 0.90& 0.42 & 0.90\\ \hline
  PN & 0.77 & 0.84  & 0.59 & 0.90 & 0.66 & 0.28 & 0.90& 0.54  \\ \hline
\end{tabularx}
\caption{{The best-fit scale $\sigma_E$, determined under the null model, when this scale is treated as profiled nuisance parameter. In all cases except Ring 6 of PN data, the best-fit scale is larger than the scale of the kernel used in our fiducial analysis, indicating that our fiducial choice of $\sigma_E$ = 0.3 was conservative and endowed the GP model with sufficient flexibility.}}
\label{tab:GPScale}
\end{table}

\begin{figure*}[htb]
\includegraphics[width = .95\textwidth]{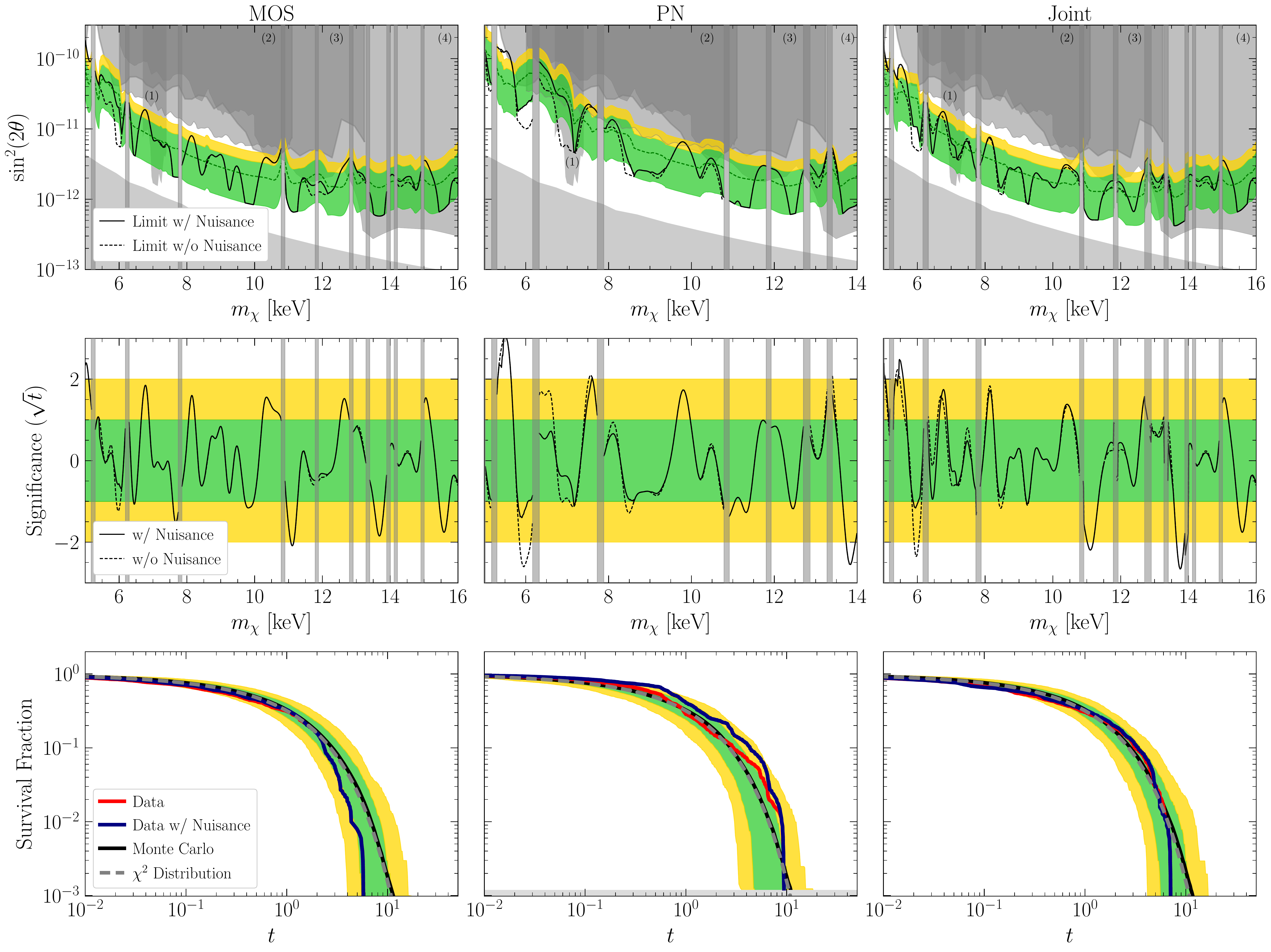}
\caption{As in Fig.~\ref{fig:Relative_2e-1} but with the alternate GP kernel, in~\eqref{eq:kernel_DE}, with {$\sigma^2 = 0.5$ keV$^2$.}}
\label{fig:Stationary_5e-1}
\end{figure*}

\begin{figure*}[htb]
\includegraphics[width = .95\textwidth]{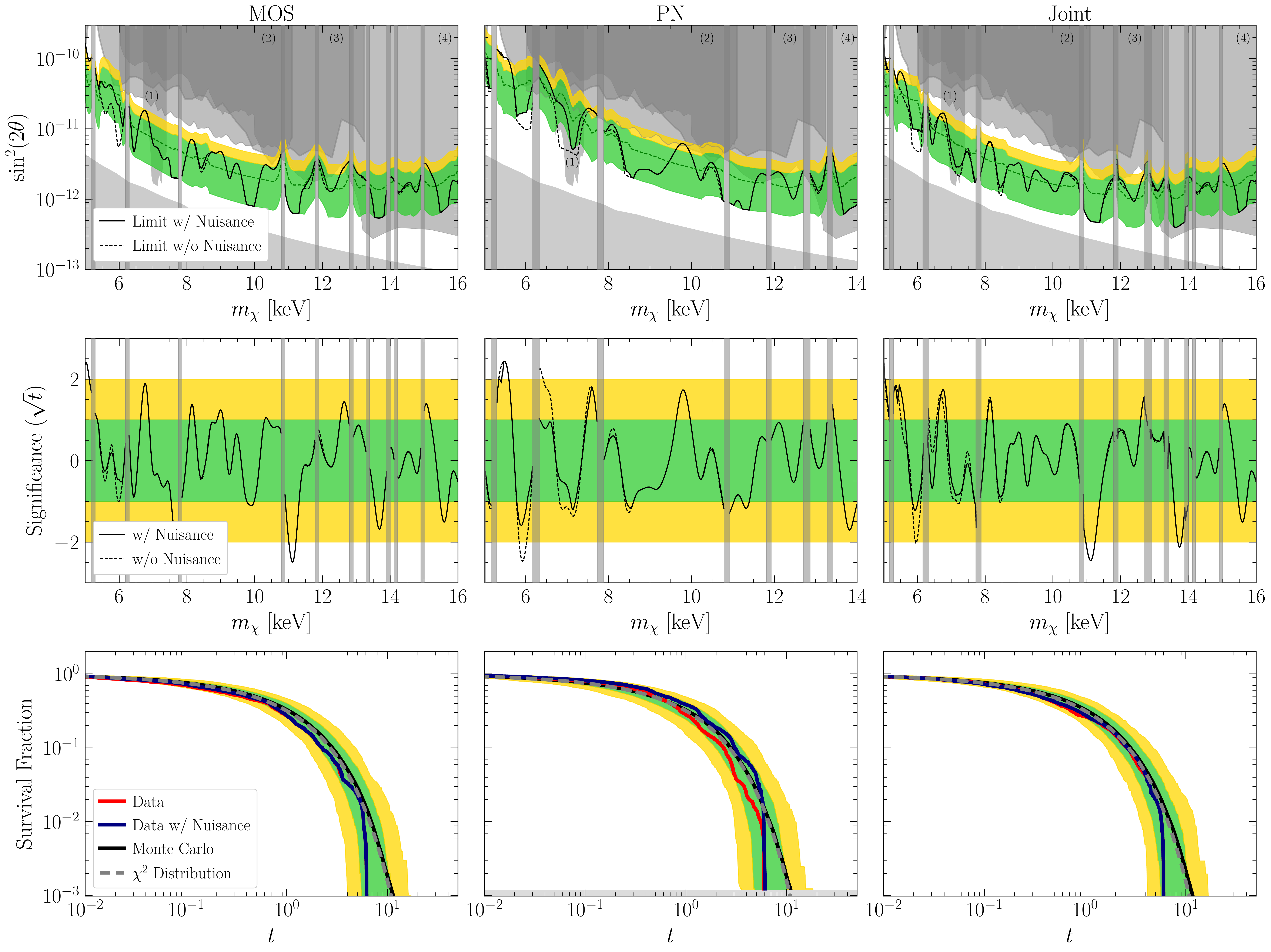}
\caption{As in Fig.~\ref{fig:Stationary_5e-1} but with {$\sigma^2 = 1.0$ keV$^2$}. Adopting a large scale length again slightly strengthens the limits, although again the systematic variation of our results with the kernel is relatively small.}
\label{fig:Stationary_1e0}
\end{figure*}

Next, we consider changing the GP modeling more significantly by adopting an alternate kernel. In particular, we consider the standard (and stationary) double exponential kernel
\es{eq:kernel_DE}{
K(E,E') = A_{\rm GP}  \exp \left[{- {(E-E')^2 \over  2 \sigma^2 }} \right]\,,
}
which has the hyperparameter $\sigma^2$.  Note that our fiducial kernel, given in~\eqref{eq:kernel}, has the property whereby the correlation length increases with the energy resolution of the detector.  The kernel in~\eqref{eq:kernel_DE}, on the other hand, has a fixed correlation length as a function of energy.  In Figs.~\ref{fig:Stationary_5e-1} and~\ref{fig:Stationary_1e0} we show the results of using the double exponential kernel with scale length {$\sigma^2 = 0.5$ keV$^2$} and {$\sigma^2 = 1.0$ keV$^2$}, respectively.  As with our fiducial kernel, in this case we also find that increasing $\sigma$ slightly increases the limits.  However, the differences between the double-exponential kernel results and our fiducial results are minor and most evident at high DM masses, $m_\chi$, where the two kernels predict the largest differences.  In particular, we find no evidence for decaying DM with the alternate kernels and similar 95\% upper limits. The systematic uncertainty associated with this choice is generally less than other aspects of the analysis such as our assumptions regarding the DM density profile.

\begin{figure*}[htb]
\includegraphics[width = .7\textwidth]{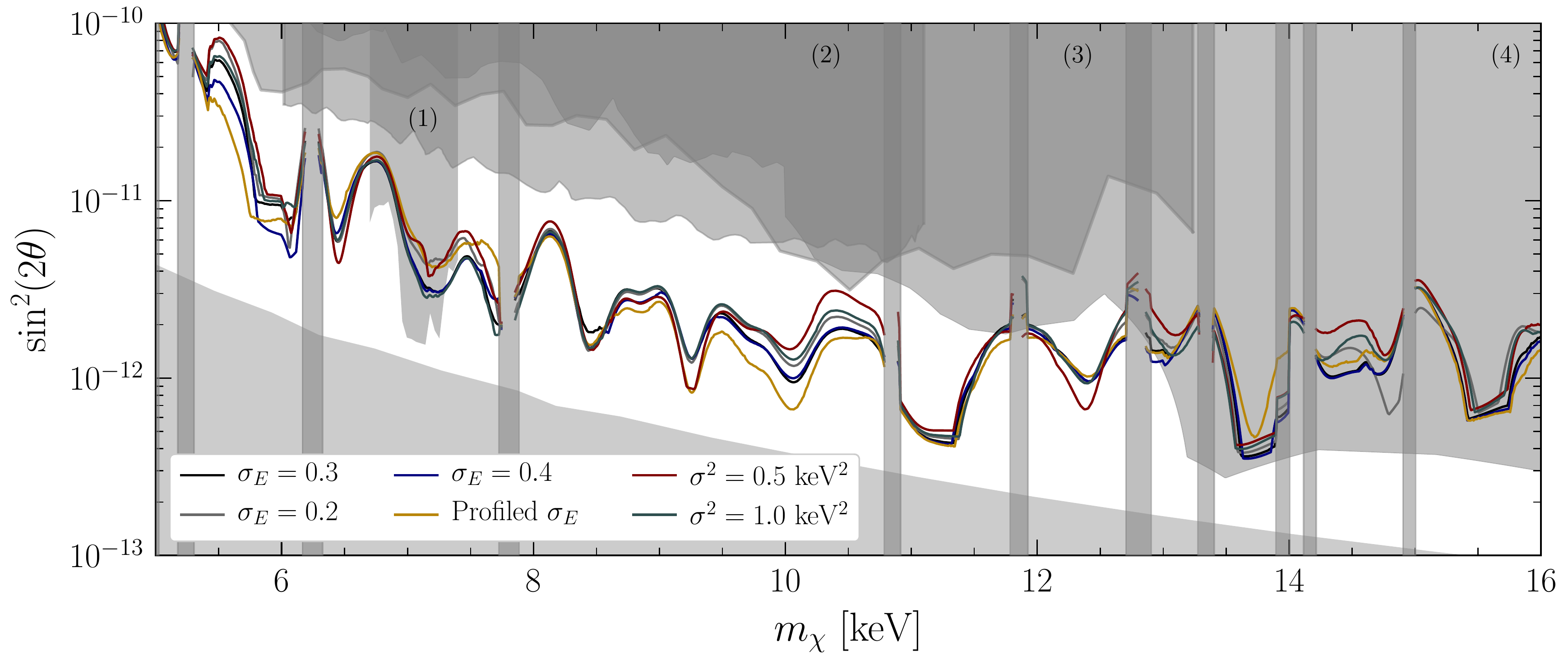}
\caption{{A comparison of the limits obtained across the full mass range for each variation of the GP correlation-length hyperparameter considered.
In particular we show results for variations of the relative-scale and fixed-scale kernels (denoted $\sigma_E$ and $\sigma^2$ respectively), as well as the relative-scale kernel where the scale profiled independently in each annulus.}}
\label{fig:ScaleComparison}
\end{figure*}

{A full comparison of the limits obtained under all the described GP kernel modeling choices is presented in Fig.~\ref{fig:ScaleComparison}. The choice of GP kernel and treatment of its scale as a fixed or profiled parameter is shown to have a marginal impact on our limit-setting procedure.}

\subsection{Unmodeled lines in the vicinity of 3.5 keV}

{
Our results have a significant impact on the decaying DM interpretation of the previously-observed 3.5 keV line from nearby galaxies and galaxy clusters~\cite{Bulbul:2014sua,Boyarsky:2014jta,Urban:2014yda,Jeltema:2014qfa,Cappelluti:2017ywp}.  Ref.~\cite{Dessert:2018qih} used a similar method to that presented in this work to argue that the non-observation of the UXL in {\it XMM-Newton} BSOs excluded the decaying DM origin of the 3.5 keV line.  However, subsequent works~\cite{Boyarsky:2020hqb,Abazajian:2020unr} questioned the validity of the results in~\cite{Dessert:2018qih} for three primary reasons: (i) the use of a narrow energy range, (ii) the possible importance of instrumental or astrophysical lines in the analysis region, (iii) the $D$-factor profile chosen with a local DM density of $0.4$ GeV/cm$^3$. These points were addressed extensively in the response~\cite{Dessert:2020hro}, and we do not review the arguments here for how these points are addressed within the context of the analysis in~\cite{Dessert:2018qih}. } 

{
Here we point out that the analysis in this work provides a probe of the decaying DM origin of the 3.5 keV line that is more robust to systematic uncertainties than~\cite{Dessert:2018qih} and that the null results from this work strongly disfavor the decaying DM interpretation of the 3.5 keV line.  Ref.~\cite{Dessert:2018qih} performed a similar analysis to this work, but the analysis focused on the limited mass range from 6.7 to 7.4 keV.  As in this work~\cite{Dessert:2018qih} used {\it XMM-Newton} blank sky data, with a comparable exposure time within the signal ROI to that in this work.  As mentioned in the main text, Ref.~\cite{Dessert:2018qih} used a joint likelihood over individual exposures, as opposed to this work where we stack the data in rings and construct the joint likelihood in individual rings.  Use of the ringed data facilitates our background subtraction and GP modeling procedures, in part because the number of counts in each ring is large enough that we may make the Gaussian approximation to the Poisson likelihood.  This work also performs a more systematic accounting of astrophysical and instrumental lines that are not fully removed by the background subtraction process.  Because we analyze a wide energy range in this analysis, we are able to use energy side-bands to determine the hyperparameter for the spurious-signal contribution to the likelihood, which accounts for residual mismodeling.  Thus while the limit presented in this work in Fig.~\ref{fig:final_limit} is slightly weaker than the fiducial limit from~\cite{Dessert:2018qih}, it is more robust to mismodeling.  Furthermore, we use a more conservative $D$-factor profile in this work, though astrophysical uncertainties on the DM density profile are not sufficiently large to explain why a decaying DM signal would have appeared in nearby galaxies and clusters but not in this work (see~\cite{Dessert:2020hro} for a discussion of this point).}

{Still, in this section we investigate the potential for mismodeling in the vicinity of 3.5 keV.  In particular,~\cite{Boyarsky:2018ktr} argued that lines may be present near 3.32 and 3.68 keV in both the MOS and PN data.  Note that in~\cite{Dessert:2018qih} these possible lines were tested for and their inclusion did not change the central conclusion of that work. Moreover, there is no robust evidence to-date for these lines in the MOS and PN data sets.  For completeness, however, we investigate how the inclusion of these lines affects the results of the analysis in this work.  Importantly, following our normal line-dropping procedure neither the 3.32 nor the 3.68 keV lines meet our criterion for inclusion in any of the rings for either MOS or PN.  This itself serves as evidence for the non-importance of these line candidates on our conclusions. However, as a systematic test we perform an analysis where we include these two lines in all of our rings for both MOS and PN, while performing the normal line-dropping procedure for the rest of the background lines. 
We treat the amplitude of these lines as a nuisance parameter which is allowed to take arbitrarily large positive or negative values.
} 

\begin{figure*}[htb]
\includegraphics[width = .95\textwidth]{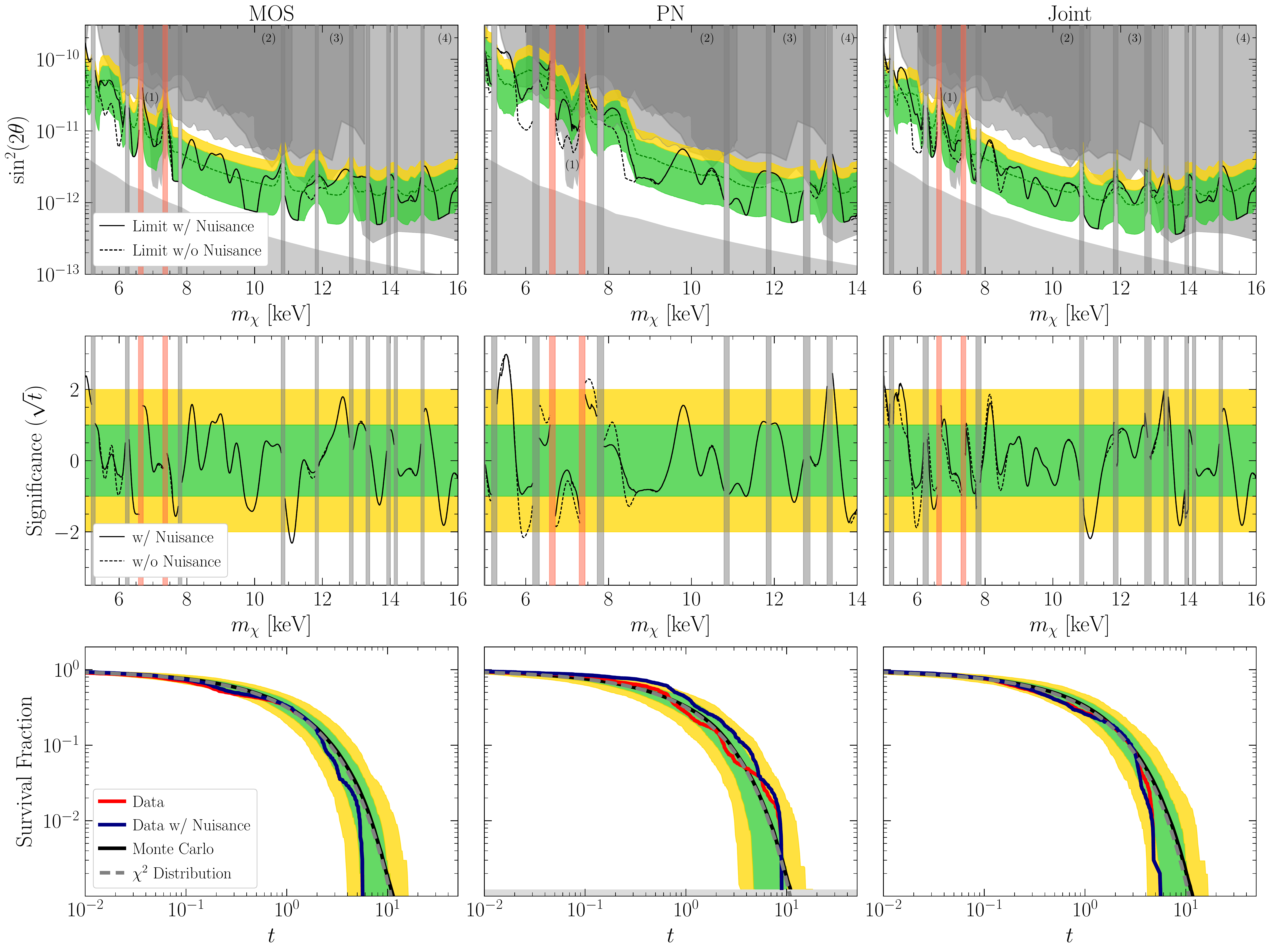}
\caption{{As in Fig.~\ref{fig:Relative_2e-1}, but with the fiducial GP kernel at $\sigma_E = 0.3$ and the inclusion of 3.32 and 3.68 keV lines in all analyzed annuli. The newly masked region associated with these two lines is highlighted in light red.}}
\label{fig:Extra_Lines_Summary}
\end{figure*}

\begin{figure*}[htb]
\includegraphics[width = .65\textwidth]{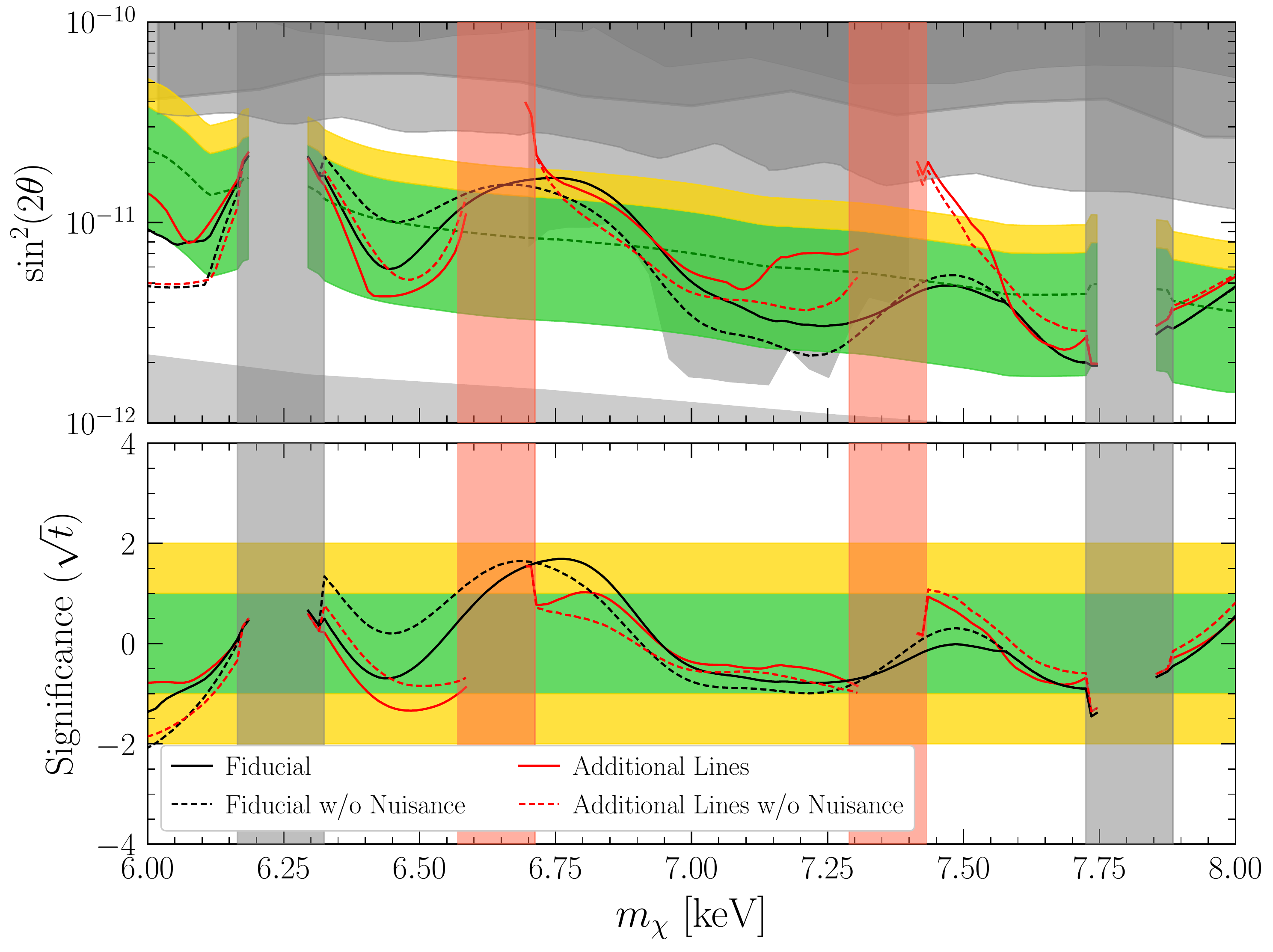}
\caption{{A close inspection of the limits set in our fiducial analysis and the modified analysis that includes a 3.32 and 3.68 keV line in each annulus. We compare the limits set in these two analyses both with (solid lines) and without (dashed lines) the inclusion of our systematic nuisance parameter designed to test for and correct possible mismodeling.}}
\label{fig:Extra_Lines_3.5}
\end{figure*}

{A summary of the full results of the analysis which includes these additional lines is provided in Fig.~\ref{fig:Extra_Lines_Summary}. No new detections are made. The limits obtained by this analysis in the 6-8 keV range are compared with the results obtained in our fiducial analysis in Fig.~\ref{fig:Extra_Lines_3.5}, which reveals small but unimportant changes in our limits.}

\subsection{Analysis of Fully Stacked Data}

In the main body, we divided our signal ROI into rings and modeled the flux independently in each ring.
The motivation behind this choice was to incorporate spatial information into the analysis, particularly as we expect the flux of an actual DM decay signal to steadily increase towards the GC.
Here we show the results of an alternative approach where instead of modeling the data ring-by-ring, we instead combine the data in the innermost three rings of the signal ROI and model that directly.
We effectively are then left with a single combined ring, which we analyze using our fiducial procedure.

\begin{figure*}[htb]
\includegraphics[width = .95\textwidth]{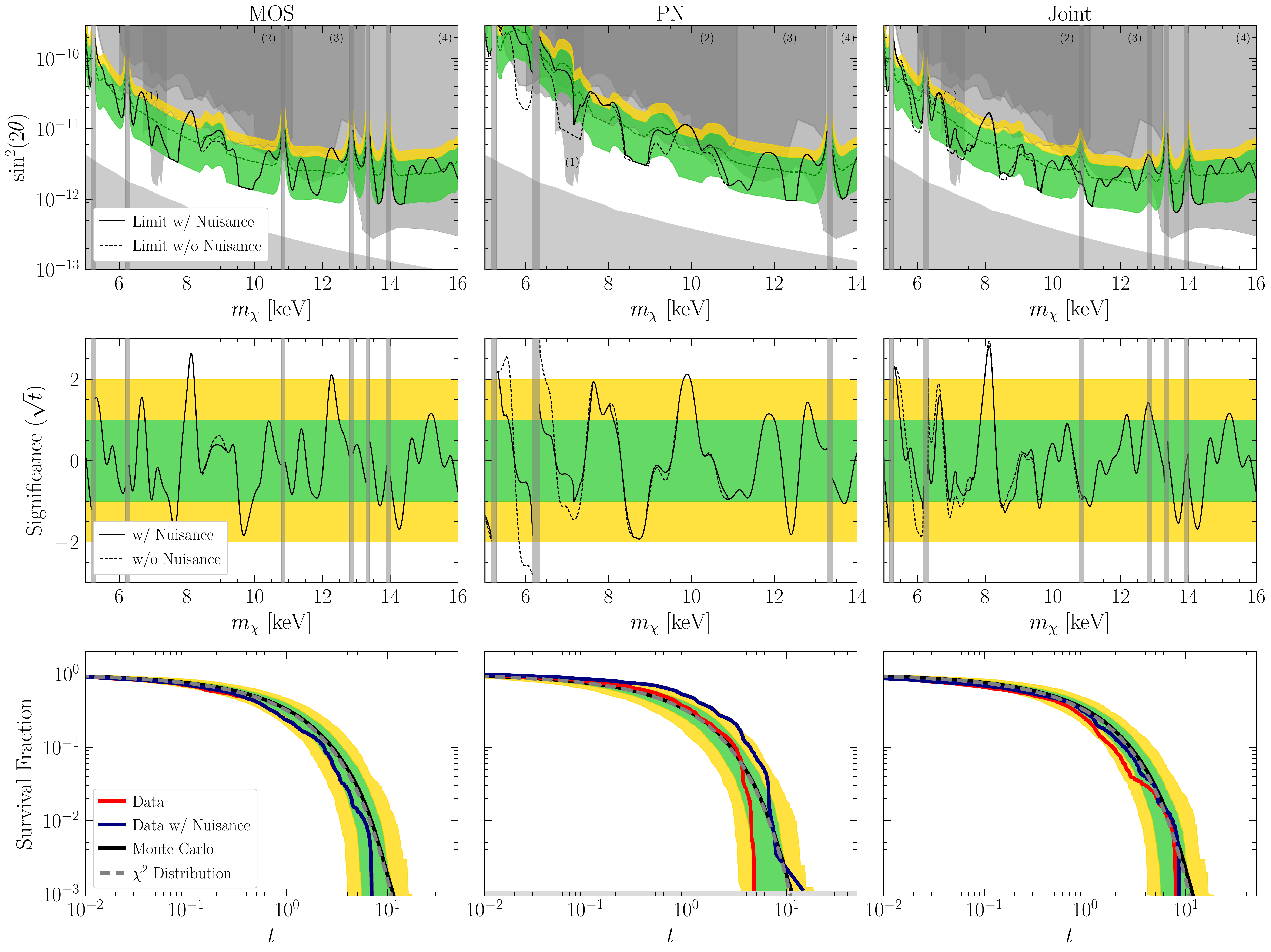}
\caption{The same results as presented in Figs.~\ref{fig:Joined_Survival} and~\ref{fig:final_limit_nuis}, however on a modified data set where instead of analyzing the signal ROI divided into eight individual rings, we stack the inner three rings into a single annulus.
As in our primary approach, we subtract the background ROI flux from the signal-region data.
The results are comparable to, although slightly weaker than, those from our fiducial approach, consistent with the reduced information available.
}
\label{fig:StackedSub}
\end{figure*}

In Fig.~\ref{fig:StackedSub} we show the resulting limit in the case where we also subtract the background-ROI flux from the stacked signal region data. While there are small differences, the resulting sensitivity and limits from this simpler approach are in good qualitative agreement to those of our default analysis. In detail, the result here are slightly weaker, which is as expected because there is less information in the signal ROI (we use fewer rings and by stacking the spatial information is partially erased).

\begin{figure*}[htb]
\includegraphics[width = .95\textwidth]{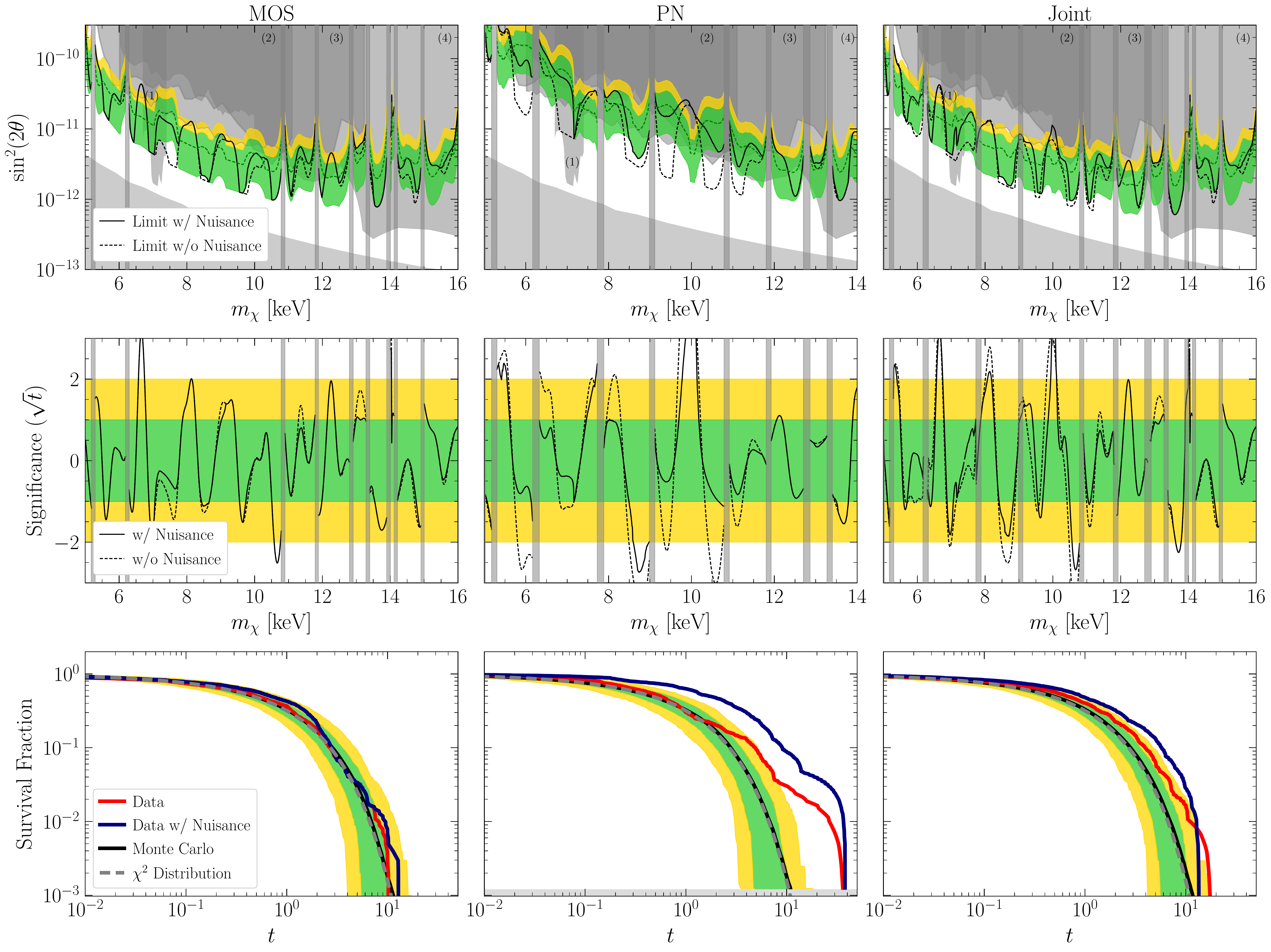}
\caption{As in Fig.~\ref{fig:StackedSub}, however considering the stacked signal ROI without subtracting the background. The limit is noticeably worse, and several excesses appear, highlighting the importance of the background subtraction.}
\label{fig:StackedNoSub}
\end{figure*}

Next, in Fig.~\ref{fig:StackedNoSub} we repeat this procedure but without subtracting the background flux. The differences are now more noticeable - the expected and resulting limit undergoes larger fluctuations and there are several mildly significant excesses. This emphasizes the importance of the background subtraction procedure in simplifying the data, particularly around bright instrumental lines. 

\subsection{Parametric Modeling without Background Subtraction}

In this section, we detail an alternate analysis to the one presented in the main body of this paper and provide a comparison between the fiducial and alternate analysis in a representative example over the 8-9 keV mass range. The alternate framework uses the same data as used in our primary analysis. However, a more traditional approach is adopted for the background modeling. Firstly, we consider the unsubstracted data in each ring within the signal ROI. The flux within each annulus is then modeled as follows. The background and putative signal lines are treated identically to our fiducial approach, but the smooth background contribution is modeled parametrically using an unfolded second order polynomial, rather than with a GP model. The three parameters that define the quadratic background component are treated as nuisance parameters and profiled over. As the quadratic background has less freedom than the GP model, we restrict to a smaller energy range. Specifically, we determine the energy range by fitting a Gaussian to the detector response at a given putative signal energy, and we define our energy range to extend 5 standard deviations out from the signal energy in either direction. In the 8-9 keV DM mass range, this corresponds to an approximate energy range of 0.60 keV. Furthermore, background lines within 7 standard deviations of the signal energy are included in the model. Thus, in the 8-9 keV mass range, the only line included is the 4.52 keV instrumental line for all PN annuli. We do not include the systematic nuisance parameter modeling for this example. 

\begin{figure*}[htb]
\includegraphics[width = .65\textwidth]{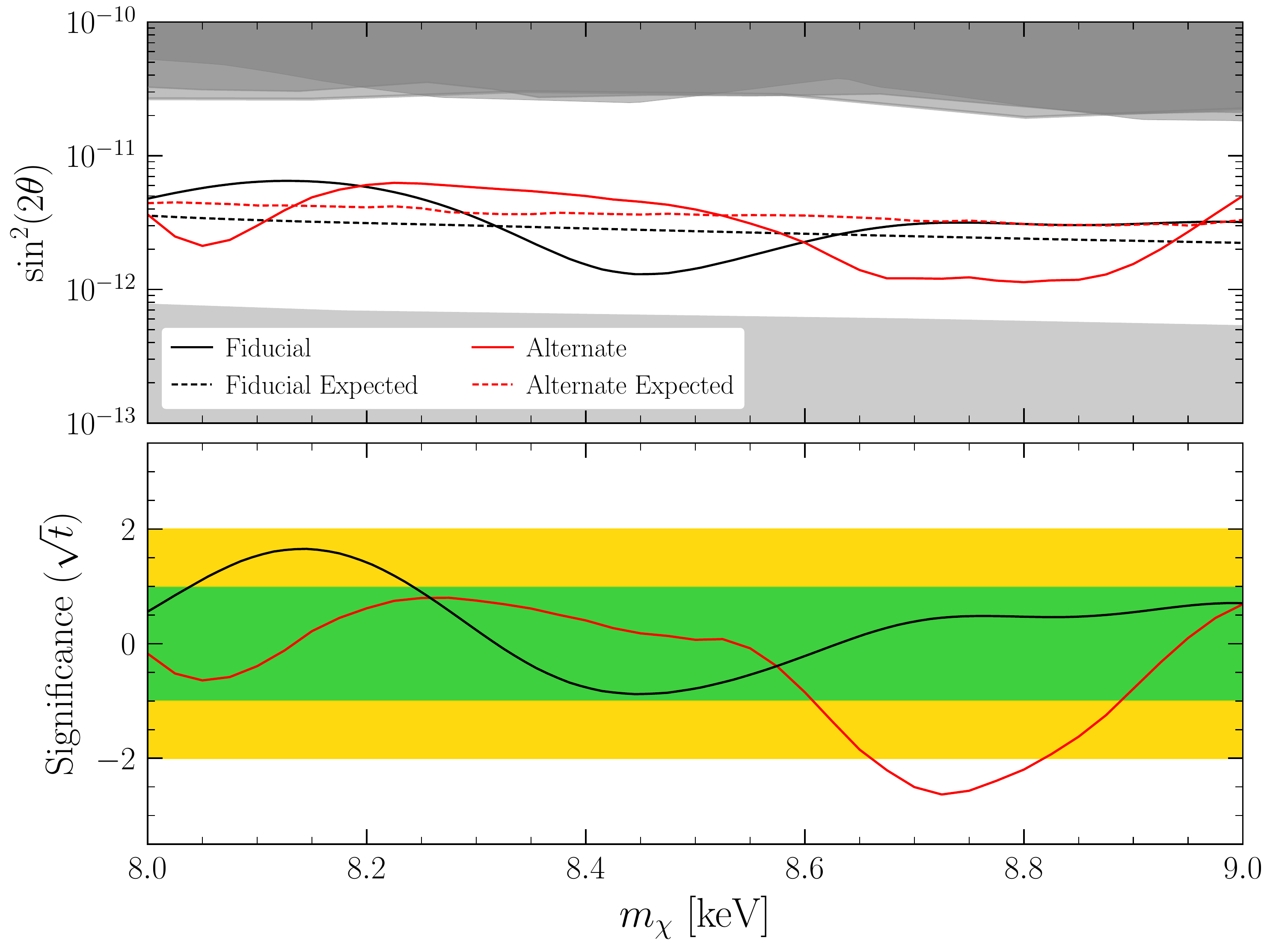}
\caption{Here we compare our fiducial results using a GP model, shown in black, to the result of an approach where the continuum background contribution is modeled with a second order polynomial, shown in red, as described in the text.
Both results are shown without imposing a systematic nuisance parameter.  While our fiducial approach uses the background-subtracted signal-ROI data, the alternate approach uses the un-subtracted data.
We see that in both cases the expected and resulting limits are in qualitative agreement, demonstrating that our choice of GP modeling in our fiducial analysis does not drive the sensitivity of our results.}
\label{fig:Compare_Results}
\end{figure*}

While our background modeling is significantly different in this case, we find again that our results are qualitatively unchanged compared to our fiducial analysis.
To provide a representative example, in Fig.~\ref{fig:Compare_Results} we show the comparison between our fiducial analysis (without the systematics nuisance parameter to facilitate the comparison) and this alternate approach over the mass range 8-9 keV. 
As can be seen, the expected sensitivity of the two approaches is almost identical. This is a significant further demonstration that our specific choice of background model is not underpinning our sensitivity.

\end{document}